\documentclass[11pt]{article}
\usepackage{hyperref}

\usepackage[utf8]{inputenc}
\usepackage{amsmath,amssymb}

\usepackage{amsthm}
\usepackage{comment}

\usepackage{graphicx}
\usepackage{xcolor}
\newcommand{\orange}{\textcolor{orange}}
\newcommand{\blue}{\textcolor{blue}}

\usepackage[bbgreekl]{mathbbol}
\usepackage{bbm}

\usepackage{epsfig}
\usepackage{subcaption}		   % for nicely aligned subfigures
\graphicspath{{pics/}}		   % for pdf_latex files

\usepackage{placeins}
\usepackage{float}
\usepackage{authblk}
\usepackage[shortlabels]{enumitem}

\graphicspath{{pics/}}					% for pdf_latex files

\def\bN{\mathbb{N}}
\def\bZ{\mathbb{Z}}
\def\bR{\mathbb{R}}
\def\bC{\mathbb{C}}

\newtheorem{definition}{Definition}[section]
\newtheorem{lemma}{Lemma}[section]

\newtheorem{remark}[lemma]{Remark}

\def\Re{{\rm Re}}

\def\tr{{\rm tr}}

\def\cL{\mathcal{L}}
\def\cN{\mathcal{N}}

\def\O{\mathcal{O}}

\begin{document}

%% Title, authors and addresses

\title{The Emergence of Spatial Patterns for Compartmental Reaction Kinetics Coupled by Two Bulk Diffusing Species with Comparable Diffusivities}

\date{\today} 

\author{Merlin Pelz\footnote{merlinpelz@math.ubc.ca} \text{ and }\addtocounter{footnote}{1}Michael J. Ward\footnote{ward@math.ubc.ca (corresponding author)}}

\affil[]{Dept. of Mathematics, University of British Columbia, Vancouver, BC, Canada}

\maketitle

\let\thefootnote\relax\footnote{Submitted to Proceedings of the Royal Society A.}

\begin{abstract}

Originating from the pioneering study of Alan Turing, the
  bifurcation analysis predicting spatial pattern formation from a
  spatially uniform state for diffusing morphogens or chemical species
  that interact through nonlinear reactions is a central problem in
  many chemical and biological systems. From a mathematical viewpoint,
  one key challenge with this theory for two component systems is that
  stable spatial patterns can typically only occur from a spatially
  uniform state when a slowly diffusing ``activator'' species reacts
  with a much faster diffusing ``inhibitor'' species. However, from a
  modeling perspective, this large diffusivity ratio requirement for
  pattern formation is often unrealistic in biological settings since
  different molecules tend to diffuse with similar rates in
  extracellular spaces. As a result, one key long-standing question is
  how to robustly obtain pattern formation in the biologically
  realistic case where the time scales for diffusion of the
  interacting species are comparable. For a coupled 1-D
  bulk-compartment theoretical model, we investigate the emergence of
  spatial patterns for the scenario where two bulk diffusing species
  with comparable diffusivities are coupled to nonlinear reactions
  that occur only in localized ``compartments'', such as on the
  boundaries of a 1-D domain. The exchange between the bulk medium and
  the spatially localized compartments is modeled by a Robin boundary
  condition with certain binding rates.  {As regulated by these binding
  rates, we show for various specific nonlinearities that our 1-D
  coupled PDE-ODE model admits symmetry-breaking  bifurcations, leading to
  linearly stable asymmetric steady-state patterns,
  even when the bulk diffusing species have equal
  diffusivities.} Depending on the form of the nonlinear kinetics,
  oscillatory instabilities can also be triggered. Moreover, the
  analysis is extended to treat a periodic chain of compartments.
\vspace{2mm}\\


\end{abstract}

\section{Introduction}\label{sec:intro}

In a pioneering study \cite{turing}, Alan Turing showed that a
spatially homogeneous steady-state solution of a reaction-diffusion
(RD) system that is stable in the absence of diffusion can be
destabilized in the presence of diffusing morphogens that have
different diffusivities. {From an application of this Turing
  stability theory to two-component activator-inhibitor systems, it is
  well-known from a mathematical viewpoint that, for any
  activator-inhibitor reaction-kinetics admitting a stable fixed
  point, the ratio of the diffusivities of the inhibitor and activator
  species must be sufficiently large, i.e.  above a threshold larger
  than unity, in order for spatially homogeneous patterns to emerge
  from the destabilization of a spatially uniform state
  (cf.~\cite{pearson1}, \cite{baker}, \cite{diambra}). Although this
  theoretically-predicted large diffusivity ratio requirement for
  pattern formation may be feasible to achieve in certain macroscale
  chemical systems, such as those involving a chemical species that
  can bind to a substrate (cf.~\cite{epstein}, \cite{french}), a large
  diffusivity ratio threshold in cellular biology is often unrealistic
  in applications where freely diffusing morphogen molecules have
  similar sizes and, consequently, comparable diffusivities.} In
scenarios where a large diffusivity ratio is a realistic assumption,
two-component RD systems modeling chemical and biological systems
typically admit a wide range of spatially localized patterns and
instabilities that occur in the fully nonlinear regime
(cf.~\cite{vanag}, \cite{ward}, \cite{frey1}, \cite{frey2}).

From a theoretical viewpoint, there has been considerable focus on
modifying the conventional two-component RD paradigm in order to
circumvent the requirement of a large diffusivity ratio for the
emergence of symmetry-breaking patterns. One such direction has
involved extending the two-component RD system to include additional
non-diffusible components (cf.~\cite{pearson2}, \cite{klika},
\cite{korv}), which (roughly) models either membrane-bound proteins or
other immobile chemically active substrates. This ``$2+1$'' extension
can generate stable spatial patterns even when the diffusible species
have a common diffusivity. In another direction, graph-theoretic
properties such as network topology have been used for predicting
Turing pattern formation for multi-component RD systems in the
presence of immobile species (cf. \cite{macron}, \cite{macron1},
\cite{landge}). More recently, it was shown in \cite{goldstein} using
a probabilistic approach that as the number of species in a
multi-component RD system increases, the required diffusivity ratio
threshold for pattern formation is typically reduced.

A new direction for theoretical RD modeling, largely initiated in
\cite{gomez2007} and \cite{levine2005}, involves exploring
pattern-forming properties associated with coupling dynamically active
spatially localized compartments via a bulk diffusion field. In a 1-D
context, and with one bulk diffusing species, this modeling paradigm
has been shown to lead to triggered oscillatory instabilities for
various reaction kinetics (cf.~\cite{gomez2007}, \cite{gou2015},
\cite{gou2016}, \cite{gou2017}), and whose bifurcation properties can
be characterized via amplitude equations obtained from a weakly
nonlinear analysis \cite{paquin_1d}.  This 1-D such framework was used in
\cite{xu} and \cite{xu2018} to model intracellular polarization and
oscillations in fission yeast.  Moreover, in multi-dimensional
spatial geometries, pattern-forming properties of bulk-membrane RD
systems have been studied both theoretically (cf.~(\cite{ratz2015},
\cite{elliott}, \cite{madzvamuse2015}, \cite{madzvamuse2016},
\cite{paquin_memb}), and for certain specific biological applications,
including cell polarization and Cdc42 protein pattern formation
(cf.~\cite{keshet}, \cite{ratz2012}, \cite{ratz2014}, \cite{holst},
\cite{paquin_model}).

The goal of this paper is to analyze symmetry-breaking pattern
formation for a 1-D coupled bulk-compartment RD system in which
spatially localized reaction compartments are coupled through a {\em
  two-component} linear bulk diffusion field with constant bulk
degradation rates. The two-component reaction kinetics inside each
compartment are assumed to be identical and the chemical exchange
between the bulk medium and the spatially localized reaction
compartments is modeled by a Robin boundary condition with certain
binding rates. We refer to this modeling framework as a {\em
  compartmental-reaction diffusion system}. {Although this
  coupled system does not admit spatially homogeneous steady-states,
  it does allow for the existence of steady-state solutions that are
  symmetric about the domain mid-line when the reaction compartments
  are identical. This symmetric steady-state solution is the {\em base
    state} for our analysis. From a linear stability analysis, and as
  confirmed by full PDE simulations, we will show for various specific
  activator-inhibitor intra-compartmental reaction kinetics that this
  coupled system admits symmetry-breaking pitchfork bifurcations of
  the symmetric base state, leading to {\em stable asymmetric
    steady-state patterns}, even when the bulk diffusing species have
  comparable or even {\em equal diffusivities}. We emphasize that the
  linear stability analysis predicting these symmetry-breaking
  bifurcations is more intricate than in \cite{turing} as it is based
  on a linearization around a {\em spatially non-uniform} symmetric
  steady-state, rather than around a spatially uniform steady-state as in
  \cite{turing}. In our coupled system, a key parameter regulating the
  emergence of these stable asymmetric steady-states is shown to be a
  sufficiently large binding rate ratio of the two species on the
  compartment boundary. This feature is in contrast to the usual large
  diffusivity ratio requirement that is a necessary condition for
  pattern formation from a spatially uniform state for typical
  two-component activator-inhibitor RD systems (cf.~\cite{nguyen},
  \cite{krause}).}

{A 2-D extension of the theoretical 1-D framework adopted in this
article can potentially be applied to characterizing symmetry-breaking
pattern-forming properties associated with biological or chemical modeling
scenarios that involve small dynamically active, but spatially
localized, discrete compartments or
``cells'' that are coupled through
multi-component linear bulk diffusion fields.} Examples of such
problems include the original 2-D agent-based model of Rauch and Millonas
\cite{rauch} describing Turing pattern formation for cell-cell
chemical signaling mediated by bulk diffusion, and the
experimental study of \cite{tompkins} of Turing patterns associated
with an emulsion of chemically loaded droplets dispersed in oil, in
which the interdrop coupling is purely diffusive. The full agent-based
numerical computations in \cite{rauch} showed that nonlinear kinetic
reactions restricted to lattice points on a 2-D lattice can generate
stable Turing-type spatial patterns when coupled through a spatially
discretized two-component bulk diffusion field where the diffusible
species have a comparable diffusivity. The formulation of our
theoretical model analyzed in a 1-D setting was inspired by this
observation in \cite{rauch}.

The outline of this paper is as follows. In \S \ref{2-cell system} we
formulate and analyze a compartmental-reaction diffusion system when
the reaction compartments are located on the boundaries of a 1-D
spatial domain. For this model, symmetric and asymmetric steady-state
solutions are determined for a class of reaction kinetics when the
bulk diffusing species have comparable or equal diffusivities. This
construction involves the solution of a nonlinear algebraic system,
whose bifurcation structure can be used to predict pitchfork
bifurcation points on the symmetric solution branch where asymmetric
patterns emerge.  Moreover, away from bifurcation points, the linear
stability problem for the symmetric steady-state is studied
numerically after first deriving a nonlinear matrix eigenvalue problem
for the eigenvalue parameter. In this way, symmetry-breaking
bifurcations of the symmetric steady-state, leading to stable
asymmetric patterns, as the binding rate ratio increases past a
threshold are illustrated for FitzHugh-Nagumo \cite{gomez2007},
Gierer-Meinhardt \cite{gm}, and Rauch-Millonas \cite{rauch} reaction
kinetics. For FitzHugh-Nagumo kinetics we show that oscillatory
instabilities of the symmetric steady-state are also possible in
certain parameter regimes.  In \S \ref{ring-cell system} we provide a
similar analysis for a compartmental-reaction diffusion system where
the localized reaction compartments are periodically spaced on a
ring. For this periodic problem, symmetric steady-states are
constructed analytically and, through a linear stability analysis
based on a Floquet-type approach, we derive a nonlinear matrix
eigenvalue problem that can be used to determine parameter values
where symmetry-breaking bifurcations occur due to multiple possible
modes of instability. The theory is illustrated for the three specific
reaction kinetics. The theoretical predictions in \S \ref{2-cell
  system} and \S \ref{ring-cell system} of symmetry-breaking
bifurcations and stable asymmetric steady-states are confirmed from
full time-dependent simulations of our PDE-ODE model.  Owing to the
novel form of the bulk-compartment coupling, off-the-shelf PDE
software was not available for the PDE numerical simulations. The
implicit-explicit (IMEX) numerical schemes and discretizations we
developed for the full PDE simulations in \S \ref{2-cell system} and
\S \ref{ring-cell system} are derived in Appendix \ref{CN-RK4 IMEX}
and \ref{num:ring}, respectively.

\section{Two Diffusion-Coupled Reaction Compartments} \label{2-cell system}

\begin{figure}[htbp]
\centering
    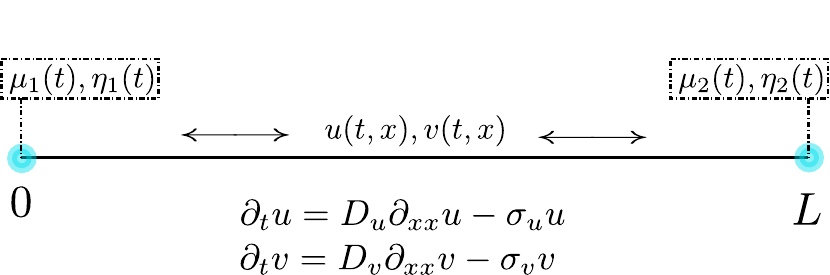
    \caption{The domain $[0,L]$ with two diffusion-coupled
      compartments (cells) and intracellular reaction kinetics on its
      boundary.}
    \label{fig:schematic}
%  \end{center}
\end{figure}

In our  modeling framework, we assume that there is a compartment or
``cell'' located at each boundary of a one-dimensional spatial bulk
domain $(0,L)$. These compartments are assumed to be reactive in the
sense that two species with concentrations $\mu(t)$ and $\eta(t)$ are
produced according to $\dot{\mu}(t) = f(\mu,\eta)$ and $\dot{\eta}(t) =
g(\mu,\eta)$, for some nonlinear reaction kinetics $f$
and $g$. Here the dot ``$\;\dot{}\;$" denotes the time derivative $d/dt$.
Further, we assume that these two compartmental species are
coupled to two bulk-defined species $u(t,x)$ and $v(t,x)$ on $(0,L)$
through a chemical exchange across the compartment boundary. The bulk
or extracellular species are assumed to undergo a linear diffusion
process with constant diffusivities $D_u$ and $D_v$ and constant
degradation rates $\sigma_u$ and $\sigma_v$ (see
Fig.~\ref{fig:schematic} for a schematic plot). This leads to a
coupled PDE-ODE system formulated as
\begin{subequations}\label{s:full}
\begin{eqnarray}
	\text{bulk} &&
	\begin{cases}
          \partial_t u = D_u\; \partial_{xx} u - \sigma_u u\,,
          \quad &x\in (0,L) \qquad \\
          \partial_t v = D_v\; \partial_{xx} v - \sigma_v v\,,
          \quad &x\in (0,L)
	\end{cases} \label{s:full_1} \\
	\text{reaction fluxes} &&
	\begin{cases}
          D_u \partial_xu(t,0)&=\beta_u\;(u(t,0)-\mu_1(t)) \; \,
          \qquad\text{(Robin boundary conditions)}\\
		D_v \partial_xv(t,0)&=\beta_v\;(v(t,0)-\eta_1(t)) \\
		-D_u \partial_xu(t,L)&=\beta_u\;(u(t,L)-\mu_2(t)) \\
		-D_v \partial_xv(t,L)&=\beta_v\;(v(t,L)-\eta_2(t)) \\
	\end{cases}\label{s:full_2}\\
	\text{compartments} &&
	\begin{cases}
          \dot{\mu}_1 = f(\mu_1,\eta_1) + D_u\; \partial_x u(t,0)
          \qquad \qquad\quad\text{(reaction kinetics at } x=0) \\
		\dot{\eta}_1 = g(\mu_1,\eta_1) + D_v\;\partial_x v(t,0) \\
		\dot{\mu}_2 = f(\mu_2,\eta_2) - D_u\; \partial_x u(t,L)
                \qquad \qquad \quad\text{(reaction kinetics at } x=L) \\
		\dot{\eta}_2 = g(\mu_2,\eta_2) - D_v\;\partial_x v(t,L)\,.
	\end{cases}\label{s:full_3}
\end{eqnarray}
\end{subequations}

To illustrate the analysis, in this section we will initially assume that the
common reaction kinetics in the two compartments is the FitzHugh-Nagumo
(FN) kinetics (with parameters $q>0,z>0,\varepsilon>0$), as considered
in \cite{gomez2007}, given by
\begin{equation}\label{s:FN}
	f(\mu_j,\eta_j) = \mu_j - q(\mu_j-2)^3 + 4 - \eta_j, \qquad
	g(\mu_j,\eta_j) = \varepsilon\mu_j z - \varepsilon\eta_j\,,
\end{equation}
for $j\in\{1,2\}$. We refer to the parameters $\beta_u$ and $\beta_v$
in (\ref{s:full_2}) as the binding rates, as they model the strength
of the exchange between the bulk and intra-compartmental species.

\subsection{Symmetry-breaking}

With this formulation, we will calculate a steady-state solution that
is symmetric in the sense that the concentrations in the two
compartments are identical. By analyzing the linear stability of this
steady-state, we will show that, depending on the parameters, it can
be destabilized through either an oscillatory instability due to a
Hopf bifurcation or from the emergence of a new stable steady-state
that is asymmetric in the two compartments. In contrast to the typical
Turing-type analysis for two component RD systems, this stable
asymmetric steady-state for our model, which arises from a
symmetry-breaking pitchfork bifurcation, occurs even when the bulk
diffusivities are comparable.

\subsubsection{Uncoupled system equilibrium}

In the absence of diffusion, the ODE system for the intra-compartmental
species is uncoupled from the bulk medium and reduces to
\begin{equation}\label{s:ode}
    \dot{\mu}(t) = f(\mu,\eta)\,, \qquad \dot{\eta}(t) = g(\mu,\eta)\,.
\end{equation}
Let $\mu_e$, $\eta_e$ be an equilibrium point for (\ref{s:ode}) and
label $F(\mu,\eta):=(f(\mu,\eta),g(\mu,\eta))$. For a specific
parameter set, the linear stability property of the equilibrium state
is characterized by whether the eigenvalues $\lambda$ of the Jacobian
$DF(\mu_e,\eta_e)$ have positive (unstable, exponentially growing
perturbations) or negative (stable, exponentially decaying
perturbations) real parts $\Re(\lambda)$.

\subsubsection{Coupled system equilibrium} \label{2cellcoupledsystem}

In the presence of bulk diffusion, the compartments are coupled to the bulk
medium. For this case, we first construct a steady-state solution for
(\ref{s:full}) without any symmetry assumptions.

In the bulk region $x\in (0,L)$, the steady-state solution for
(\ref{s:full_1}) has the form
\begin{equation}\label{ss:bulk}
  u_{e}(x)=A_1 \sinh(\omega_u(L-x)) + A_2 \sinh(\omega_u x) \,, \quad
  v_{e}(x)=B_1 \sinh(\omega_v(L-x)) + B_2 \sinh(\omega_v x) \,,
\end{equation}
where $\omega_u:=\sqrt{\sigma_u/D_u}$ and $\omega_v:=\sqrt{\sigma_v/D_v}$. Here
the constants $A_1, A_2, B_1$ and $B_2$ are to be determined. By employing the
boundary conditions in (\ref{s:full_2}) for $u$, we obtain 
\begin{subequations}\label{s:a1a2b1b2}
\begin{equation}
    \begin{pmatrix}
        A_1 \\
        A_2
    \end{pmatrix}
    = \frac{\beta_u}{D_u\omega_u(\gamma_u^2-1)}
    \begin{pmatrix}
        \gamma_u    &1 \\
        1 &\gamma_u
    \end{pmatrix}
    \begin{pmatrix}
        \mu_1^e \\
        \mu_2^e
    \end{pmatrix}
    \quad \text{where} \quad \gamma_u:= \cosh(\omega_u L) +
    \frac{\beta_u}{D_u\omega_u}\sinh(\omega_u L)\,,
  \end{equation}
while from the boundary conditions in (\ref{s:full_2}) for $v$ we get
\begin{equation}
    \begin{pmatrix}
        B_1 \\
        B_2
    \end{pmatrix}
    = \frac{\beta_v}{D_v\omega_v(\gamma_v^2-1)}
    \begin{pmatrix}
        \gamma_v    &1 \\
        1 &\gamma_v
    \end{pmatrix}
    \begin{pmatrix}
        \eta_1^e \\
        \eta_2^e
    \end{pmatrix}
    \quad \text{where} \quad \gamma_v:= \cosh(\omega_v L) + \frac{\beta_v}
    {D_v\omega_v}\sinh(\omega_v L)\,.
  \end{equation}
\end{subequations}

Next, from the compartmental reaction kinetics for the
$v$-species in (\ref{s:full_3}), and from use of (\ref{ss:bulk}), we get
\begin{equation}\label{ss:iv}
  \frac{\beta_v}{\gamma_v^2-1}\tilde{A} 
         \begin{pmatrix}
             \eta_1^e \\
             \eta_2^e
         \end{pmatrix}
         = 
         \begin{pmatrix}
             g(\mu_1^e,\eta_1^e) \\
             g(\mu_2^e,\eta_2^e)
         \end{pmatrix}
        \quad
        \text{where} \quad  \tilde{A} := 
        \begin{pmatrix}
        \gamma_v\cosh(\omega_v L) - 1 
        & \cosh(\omega_v L) - \gamma_v \\
        \cosh(\omega_v L) - \gamma_v
        & \gamma_v\cosh(\omega_v L) - 1 
        \end{pmatrix}\,.
\end{equation}
For the special case where $g(\mu, \eta)$ is linear in $\eta$, i.e.,
$g(\mu, \eta) = g_1(\mu) - g_2\eta$, as occurs for the FN kinetics in
(\ref{s:FN}), we can use (\ref{ss:iv}) to obtain 
\begin{equation}\label{ss:iv_new}
    A
    \begin{pmatrix}
        \eta_1^e \\
        \eta_2^e
    \end{pmatrix}
    =
    \begin{pmatrix}
        g_1(\mu_1^e) \\
        g_1(\mu_2^e)
    \end{pmatrix}
    \quad \text{with} \quad A := \frac{\beta_v}{\gamma_v^2-1}\tilde{A} +
    g_2 I\,,
\end{equation}
where $I$ is the identity matrix. 

Next, we use (\ref{ss:bulk}) in (\ref{s:full_3}) for the $u$-species. This
yields
\begin{equation}\label{ss:iu}
  \frac{\beta_u}{\gamma_u^2-1} B
         \begin{pmatrix}
             \mu_1^e \\
             \mu_2^e
         \end{pmatrix}
         = 
         \begin{pmatrix}
             f(\mu_1^e,\eta_1^e) \\
             f(\mu_2^e,\eta_2^e)
         \end{pmatrix}
        \quad
        \text{where} \quad
    B:=
    \begin{pmatrix}
        \gamma_u\cosh(\omega_u L) - 1 
        & \cosh(\omega_u L) - \gamma_u \\
        \cosh(\omega_u L) - \gamma_u 
        & \gamma_u\cosh(\omega_u L) - 1  
    \end{pmatrix}\,.
\end{equation}
Finally, by solving (\ref{ss:iv_new}) for $\eta_1^{e}$ and $\eta_{2}^{e}$,
we substitute the resulting expressions into (\ref{ss:iu}) to obtain a
nonlinear algebraic system for $\mu_1^{e}$ and $\mu_{2}^{e}$
\begin{equation}\label{s:iu_fin}
    \begin{pmatrix}
      f(\mu_1^e, (1,0)(\frac{\beta_v}{\gamma_v^2-1}\tilde{A} + g_2 I)^{-1}
      (g_1(\mu_1^e),g_1(\mu_2^e))^T) \\
      f(\mu_2^e, (0,1)(\frac{\beta_v}{\gamma_v^2-1}\tilde{A} + g_2 I)^{-1}
      (g_1(\mu_1^e),g_1(\mu_2^e))^T)
    \end{pmatrix}
    -\frac{\beta_u}{\gamma_u^2-1} B
    \begin{pmatrix}
        \mu_1^e \\
        \mu_2^e
    \end{pmatrix}
    = 0\,,
  \end{equation}
  where $^{T}$ denotes transposition.  In terms of a solution to
  (\ref{s:iu_fin}) we can determine the other steady-state values from
  (\ref{ss:iu}) and (\ref{s:a1a2b1b2}).  We emphasize that
  (\ref{s:iu_fin}) applies to all steady-states of (\ref{s:full})
  regardless of symmetry.

The matrices $A$ and $B$ that effectively couple the two compartments
are symmetric and circulant, and hence have the common eigenspace spanned by
$q_1:=(1,1)^T$ and $q_2:=(1,-1)^T$. The corresponding eigenvalues
$a_j$ and $b_j$ for $A$ and $B$ for $j=1,2$ are simply
\begin{equation}\label{ss:eig}
    \begin{array}{rclcl}
      A q_1 &=& (\frac{\beta_v}{\gamma_v^2-1}(\gamma_v(\cosh(\omega_v L)-1)
                + (\cosh(\omega_v L)-1)) + g_2)\;q_1 &=& a_1 q_1\,, \\
      A q_2 &=& (\frac{\beta_v}{\gamma_v^2-1}(\gamma_v(\cosh(\omega_v L)+1)
                - (\cosh(\omega_v L)+1)) + g_2) \;q_2 &=& a_2 q_2\,, \\
      B q_1 &=& (\gamma_u(\cosh(\omega_u L)-1) + (\cosh(\omega_u L)-1))\;q_1
                                                     &=& b_1 q_1\,, \\
      B q_2 &=& (\gamma_u(\cosh(\omega_u L)+1) - (\cosh(\omega_u L)+1))\;q_2
                                                     &=& b_2 q_2\,.
    \end{array}
  \end{equation}

To determine a symmetric solution for (\ref{s:iu_fin}) for which
$(\mu_{1}^{e},\mu_{2}^{e})=\mu_e q_1$, where $q_1=(1,1)^T$, we simply use
the eigenvalues $a_1$ and $b_1$ in (\ref{ss:eig}) to reduce
(\ref{s:iu_fin}) to 
\begin{equation}\label{ss:nonl_symm}
    f(\mu_e, \frac{g_1(\mu_e)}{a_1})-\frac{\beta_u}{\gamma_u^2-1} b_1\mu_e = 0\,.
\end{equation}
This scalar nonlinear algebraic problem for $\mu_e$ applies whenever
$g(\mu,\eta)$ has the specific form $g(\mu,\eta)=g_1(\mu) - g_2\eta$,
where $g_2$ is a constant. 

To detect any symmetry-breaking pitchfork bifurcation points of the
symmetric steady-state solution we can perform a linear stability
analysis of (\ref{s:full}) around the steady-state solution and seek
$\lambda=0$ eigenvalue crossings. More conveniently, to detect
zero-eigenvalue crossings for the linearization of the full system
(\ref{s:full}) we can more simply determine bifurcation points
associated with the linearization of the nonlinear algebraic system
(\ref{s:iu_fin}) around a symmetric steady-state.  To do so, we
introduce the perturbation
$(\mu_{1},\mu_{2})^T=\mu_eq_{1} +\delta\phi$ from the symmetric steady-state into
(\ref{s:iu_fin}) to obtain, for $\delta\ll 1$, the linearized problem
{\small
  \begin{equation}\label{ss:jac_1}
      \begin{pmatrix}
        \frac{d}{d\mu_1} f(\mu_1, (1,0)A^{-1}(g_1(\mu_1),g_1(\mu_2))^T)
        & \frac{d}{d\mu_2} f(\mu_1, (1,0)A^{-1}(g_1(\mu_1),g_1(\mu_2))^T) \\
        \frac{d}{d\mu_1} f(\mu_2, (0,1)A^{-1}(g_1(\mu_1),g_1(\mu_2))^T) &
        \frac{d}{d\mu_2} f(\mu_2, (0,1)A^{-1}(g_1(\mu_1),g_1(\mu_2))^T)
    \end{pmatrix}
    \phi - \frac{\beta_u}{\gamma_u^2-1} B \phi = 0 \,,
\end{equation}}
where the derivatives are to be evaluated at the symmetric state
$\mu_1=\mu_e$ and $\mu_2=\mu_e$. In this way, we obtain the linearized
problem
\begin{subequations}\label{ss:jacall}
\begin{equation}\label{ss:jac}
    \begin{pmatrix}
        J_{11} & J_{12} \\
        J_{21} & J_{22}
    \end{pmatrix}
    \phi - \frac{\beta_u}{\gamma_u^2-1} B \phi =0 \,,
\end{equation}
where the Jacobian matrix coefficients are given, in terms of
any solution $\mu_e$ of  (\ref{ss:nonl_symm}), by
\begin{eqnarray}\label{ss:jac_coeff}
  J_{11} &:=& \partial_\mu f(\mu_e,\frac{g_1(\mu_e)}{a_1}) + \partial_\eta
              f(\mu_e,\frac{g_1(\mu_e)}{a_1}) (1,0) A^{-1} (g_1^{\prime}
              (\mu_e),0)^T\,, \\
  J_{12} &:=& \partial_\eta f(\mu_e,\frac{g_1(\mu_e)}{a_1}) (1,0) A^{-1}
              (0,g_1^{\prime}(\mu_e))^T\,, \\
  J_{21} &:=& \partial_\eta f(\mu_e,\frac{g_1(\mu_e)}{a_1}) (0,1) A^{-1}
              (g_1^{\prime}(\mu_e),0)^T\,, \\
  J_{22} &:=& \partial_\mu f(\mu_e,\frac{g_1(\mu_e)}{a_1}) +
              \partial_\eta f(\mu_e,\frac{g_1(\mu_e)}{a_1}) (0,1) A^{-1}
              (0,g_1^{\prime}(\mu_e))^T\,.
\end{eqnarray}
\end{subequations}

From (\ref{ss:jacall}), a bifurcation with an in-phase perturbation
$\phi=q_1$, if it exists, must satisfy
\begin{equation}\label{+1}
  \partial_\mu f(\mu_e,\frac{g_1(\mu_e)}{a_1}) + \partial_\eta
  f(\mu_e,\frac{g_1(\mu_e)}{a_1}) \frac{g_1^{\prime}(\mu_e)}{a_1} -
  \frac{\beta_u}{\gamma_u^2-1} b_1 = 0 \,.
\end{equation}
Moreover, a bifurcation with an anti-phase perturbation $\phi=q_2$,
if it exists, has to fulfill
\begin{equation}\label{-1}
  \partial_\mu f(\mu_e,\frac{g_1(\mu_e)}{a_1}) + \partial_\eta
  f(\mu_e,\frac{g_1(\mu_e)}{a_1}) \frac{g_1^{\prime}(\mu_e)}{a_2} -
  \frac{\beta_u}{\gamma_u^2-1} b_2 = 0 \,.
\end{equation}

Below in \S \ref{s:FN_th} we will show for FN kinetics that the symmetric
steady-state is never destabilized through a zero-eigenvalue crossing
of the linearization with an in-phase perturbation. In contrast, to find a
bifurcation point for an anti-phase eigenperturbation, we must solve
the following coupled system for $\mu_e$ and a bifurcation parameter:
\begin{subequations}\label{ss:odd_FN}
\begin{eqnarray}
  f(\mu_e, \frac{g_1(\mu_e)}{a_1})-\frac{\beta_u}{\gamma_u^2-1} b_1\mu_e
  &=& 0\,, \\
  \partial_\mu f(\mu_e,\frac{g_1(\mu_e)}{a_1}) + \partial_\eta
  f(\mu_e,\frac{g_1(\mu_e)}{a_1}) \frac{g_1^{\prime}(\mu_e)}{a_2}
  - \frac{\beta_u}{\gamma_u^2-1} b_2 &=& 0\,.
\end{eqnarray}
\end{subequations}
The bifurcation parameter is taken either as a reaction kinetic
parameter or as the ratio $\rho := {\beta_v/\beta_u}$ of the binding
rates.  In (\ref{ss:odd_FN}) the explicit expressions for the
matrix eigenvalues $a_1$, $a_2$, $b_1$, and $b_2$ are defined in
(\ref{ss:eig}) in terms of the parameters.

\subsection{Compartments with FitzHugh-Nagumo kinetics} \label{2-cell FN system}

\subsubsection{Application of the theory}\label{s:FN_th}

\begin{figure}[htbp]
    \centering
    \includegraphics[width=0.9\textwidth]{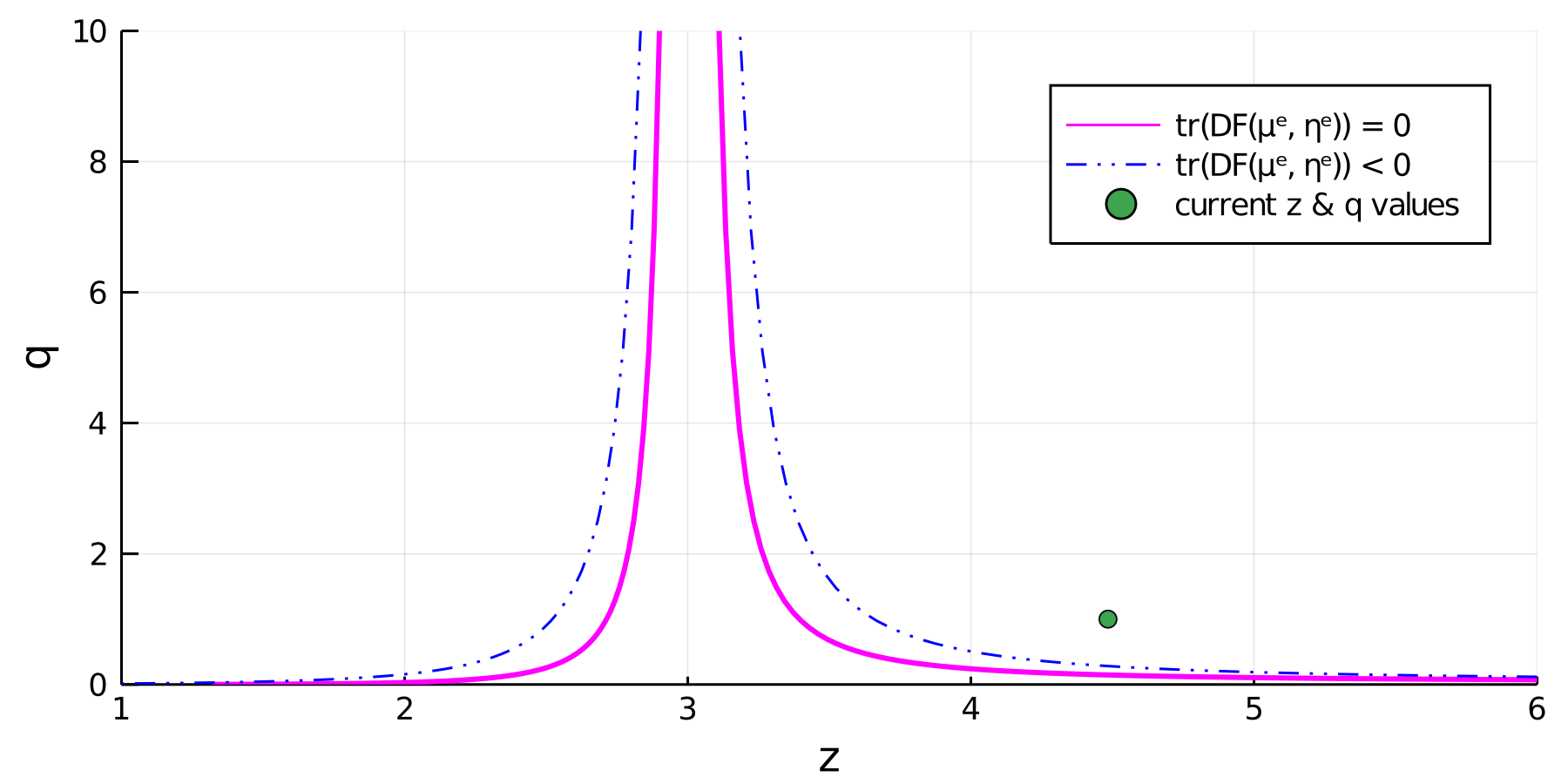}
    \caption{The instability boundary ${\rm tr}(DF(\mu_e,\eta_e))=0$
      (magenta curve) in the $(q,z)$-plane for a unique steady-state of
      (\ref{e:FN}) with $\varepsilon=0.7$, and a generic curve (blue
      curve) where the steady-state is linearly stable.  For $q=1$ and
      $z=4.48430$ (green dot) there is a unique stable steady state
      $\mu_e\approx 1.26293$ and $\eta_e=z\mu_e\approx 5.66336$ for
      which ${\rm tr}(DF(\mu_e,\eta_e))\approx -1.32981<0$.}
    \label{fig:tr-det-plot}
\end{figure}

When uncoupled from the bulk, the intra-compartmental dynamics with FN
kinetics is (see \cite{gomez2007})
\begin{equation}\label{e:FN}
  \dot{\mu} = f(\mu,\eta) := \mu - q(\mu-2)^3 + 4 - \eta\,,\qquad
   \dot{\eta} = g(\mu,\eta) := \varepsilon z \mu - \varepsilon \eta\,.
\end{equation}
As shown in \S 2 of \cite{gou2016}, there are up to three equilibria
for (\ref{e:FN}) that are determined by the cubic
\begin{equation*}
    \mu_e - q(\mu_e-2)^3 +4 - z \mu_e = 0 \,.
\end{equation*}
With $F(\mu,\eta):=(f(\mu,\eta),g(\mu,\eta))$, and for a specific parameter
set, the sign of the real parts of the eigenvalues of $DF(\mu_e,\eta_e)$
can be read off the $\tr$-$\det$-plot with, in this setting,
${\rm tr}(DF(\mu_e,\eta_e)) = 1-3q(\mu_e-2)^2 - \varepsilon$ and
$\det(DF(\mu_e,\eta_e)) = 3\varepsilon q(\mu_e-2)^2 + \varepsilon(z-1)$.
For $z>1$, for which $\det(DF(\mu_e,\eta_e))>0$, the linear stability of
$\mu_e$ is determined by the sign of ${\rm tr}(DF(\mu_e,\eta_e))$. We will
choose a parameter set for which there is a unique linearly stable
steady-state of the intra-compartmental dynamics (\ref{e:FN}) (see the
green dot in Fig.~\ref{fig:tr-det-plot}). 

To apply the steady-state theory of \S \ref{2cellcoupledsystem} for the
bulk-cell coupled system for FN kinetics we first identify that
$g(\mu,\eta)=g_1(\mu)-g_2\eta$, where
$g_1=\varepsilon \mu z$ and $g_2=\varepsilon$. From (\ref{ss:iv_new}) we
get
\begin{equation}\label{ssFN:iv_new}
    A
    \begin{pmatrix}
        \eta_1^e \\
        \eta_2^e
    \end{pmatrix}
    =\varepsilon z
    \begin{pmatrix}
        \mu_1^e \\
        \mu_2^e
    \end{pmatrix}
    \quad \text{with} \quad A := \frac{\beta_v}{\gamma_v^2-1}\tilde{A} +
    \varepsilon I\,,
\end{equation}
where $\tilde{A}$ was defined in (\ref{ss:iv}).  In terms of $A$, we
obtain from (\ref{s:iu_fin}) that the nonlinear algebraic system
characterizing both symmetric and asymmetric steady-states is
\begin{equation}\label{ssFN:iu_fin}
    \begin{pmatrix}
        \mu_1^e \\
        \mu_2^e
    \end{pmatrix}
    -q
    \begin{pmatrix}
        (\mu_1^e-2)^3 \\
        (\mu_2^e-2)^3
    \end{pmatrix}
    + 4
    \begin{pmatrix}
      1 \\
      1
    \end{pmatrix}
    - \varepsilon z A^{-1}
    \begin{pmatrix}
        \mu_1^e \\
        \mu_2^e
    \end{pmatrix}
    -\frac{\beta_u}{\gamma_u^2-1} B
    \begin{pmatrix}
        \mu_1^e \\
        \mu_2^e
    \end{pmatrix}
    = 0 \,.
\end{equation}
From (\ref{ss:nonl_symm}), the symmetric steady-state solution, for
which $\mu_1^{e}=\mu_{2}^{e}=\mu_e$, is obtained from
\begin{equation}\label{ssFN:symm}
  -q(\mu_e-2)^3+4 + \left(1-\frac{\varepsilon z}{a_1}-\frac{\beta_u}{\gamma_u^2-1}
  b_1\right)\mu_e = 0 \,,
\end{equation}
where $a_1$ is an eigenvalue of $A$ as  defined in (\ref{ss:eig}) with
$g_2=\varepsilon$.

Bifurcation points (if they exist) for either in-phase and anti-phase
perturbations of the symmetric steady-state, as obtained from
(\ref{ss:odd_FN}), satisfy
\begin{subequations}\label{ssFN:+-}
  \begin{align}
    - 3q(\mu_e-2)^2 + 1-\frac{\varepsilon z}{a_1}-\frac{\beta_u}{\gamma_u^2-1}
    b_1 & =  0 \,, \quad \mbox{(in-phase)} \label{+}\\
    - 3q(\mu_e-2)^2 + 1-\frac{\varepsilon z}{a_2}-
    \frac{\beta_u}{\gamma_u^2-1} b_2 &= 0 \,, \quad
                                       \mbox{(anti-phase)} \label{-}
  \end{align}
\end{subequations}
where $a_1,a_2,b_1,b_2$ are the eigenvalues of $A$ and $B$ as given in
(\ref{ss:eig}) with $g_2=\varepsilon$.

To show that there can be no bifurcation point for the in-phase perturbation
of the symmetric steady-state when $\mu_e\geq -1$, we use
(\ref{+}) in (\ref{ssFN:symm}) to eliminate ${\varepsilon z/a_1} +
{\beta_u/(\gamma_u^2-1)}$. This yields
\begin{align*}
 0 &= q(\mu_e-2)^3-4+ \left(\frac{\varepsilon z}{a_1}+\frac{\beta_u}
    {\gamma_u^2-1} b_1-1\right)\mu_e \\
  &= q(\mu_e-2)^2((\mu_e-2)-3\mu_e)-4 = -2q(\mu_e-2)^2(\mu_e+1)-4 \leq 0 \,,
\end{align*}
which shows that a bifurcation with an in-phase eigenperturbation is
impossible when $\mu_e\geq -1$.

To determine a bifurcation point associated with the anti-phase perturbation
of the symmetric steady-state we must solve the coupled system
(\ref{-}) and (\ref{ssFN:symm}) on the range where
${\varepsilon z/a_2} + {\beta_u b_2/(\gamma_u^2-1)}\leq 1$. By eliminating
$\mu_e$ between (\ref{-}) and (\ref{ssFN:symm}) we obtain 
\begin{equation}\label{ssFN:q}
  q - \frac{\left(1-\frac{\varepsilon z}{a_2} - \frac{\beta_u}{\gamma_u^2-1} b_2
      \right)}{27\left(6-\frac{2\varepsilon z}{a_1}-
    \frac{2\beta_u}{\gamma_u^2-1} b_1\right)^2}
\left(-2 + \varepsilon z \left(\frac{3}{a_1}-\frac{1}{a_2}\right) +
  \frac{\beta_u}{\gamma_u^2-1} (3b_1-b_2) \right)^2 = 0 \,.
\end{equation}
For a fixed parameter set, we view the parameter constraint
(\ref{ssFN:q}) as a nonlinear algebraic equation for either $z$ or for
the binding rate ratio $\rho := {\beta_v/\beta_u}$.

\begin{figure}[htbp]
    \centering
    \includegraphics[width=0.48\textwidth]{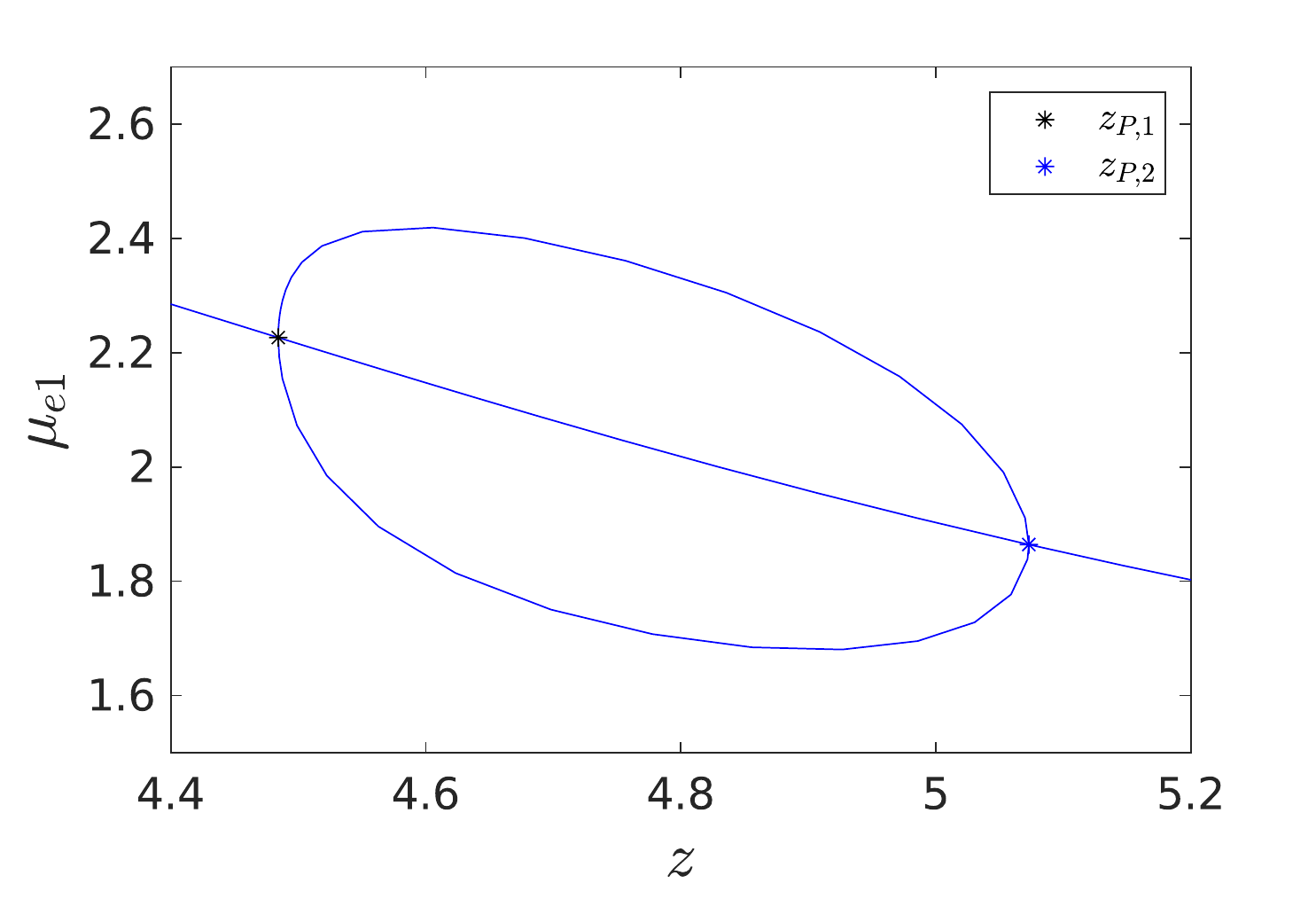}
    \includegraphics[width=0.48\textwidth]{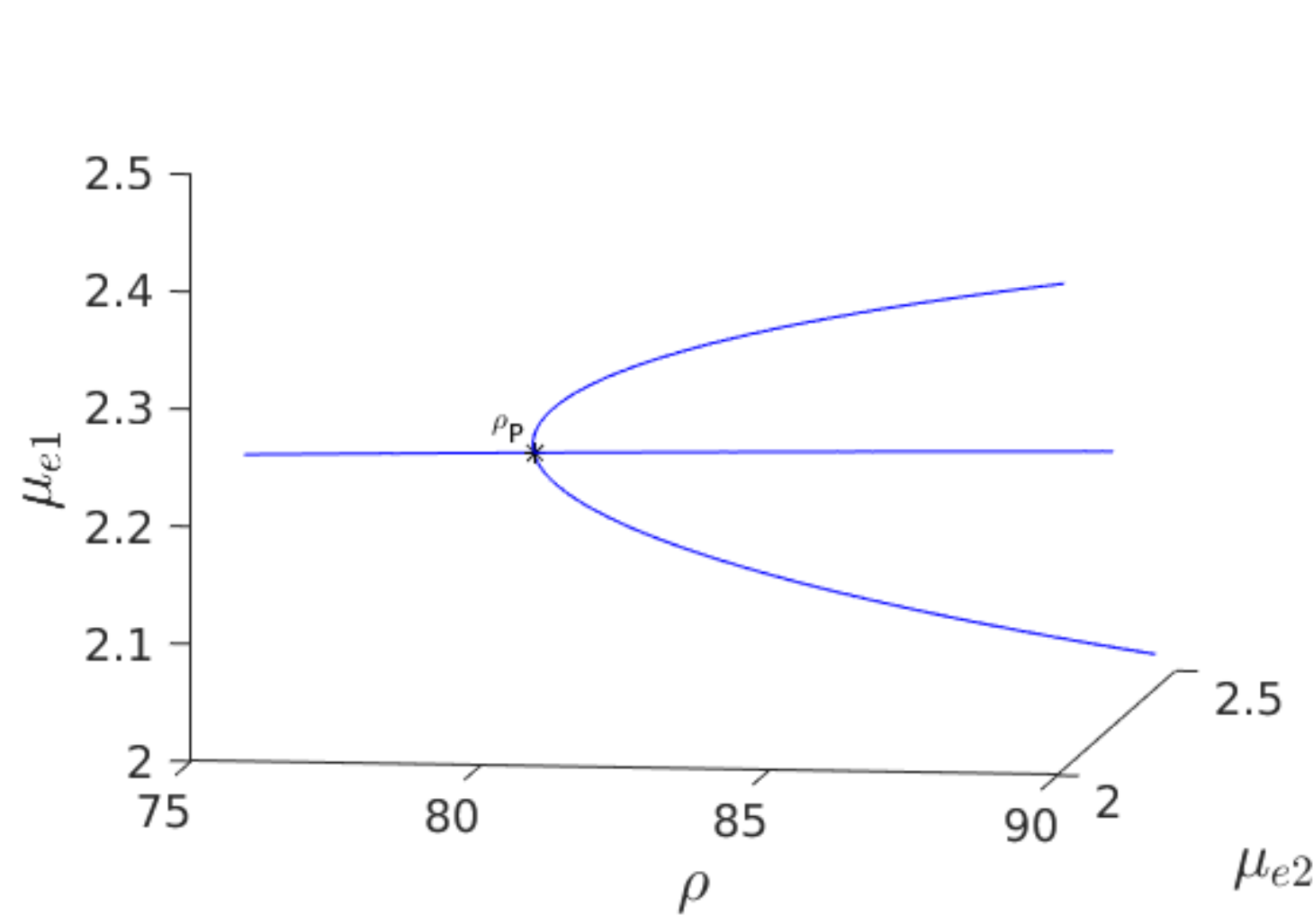}
    \caption{Left: $\mu_1^{e}$ versus $z$ showing that asymmetric
      equilibria exist inside a pitchfork bubble delimited by
      $z_{P,1}\approx 4.48430$ and $z_{P,2}\approx 5.07294$ when
      {$\rho={\beta_v/\beta_u}=80$.} Right: For $z=z_{P,1}$, there is a
      symmetry-breaking bifurcation of the symmetric steady-state as
      $\rho$ increases past the critical value $\rho_{p}=80$. Parameters:
      $D_u = 1, D_v=3, \sigma_u=\sigma_v=1, \varepsilon=0.7, q=1, L=1$,
      and $\beta_u = 0.1$.}
    \label{fig:FN_bubble}
\end{figure}

In terms of $z$, in Fig.~\ref{fig:FN_bubble}(left) we plot both
symmetric and asymmetric steady-state solution branches, as computed
from (\ref{ssFN:symm}) and (\ref{ssFN:iu_fin}) using MatCont
\cite{matcont} for the parameter set given in the figure caption. We
observe that asymmetric steady-state solution branches form a bubble
structure between two pitchfork bifurcation points $z_{P,1}$ and
$z_{P,2}$ associated with anti-phase perturbations of the symmetric
steady-state. For $z=z_{P,1}$, in Fig.~\ref{fig:FN_bubble}(right) we
plot symmetric and asymmetric solution branches in terms of
$\rho := {\beta_v/\beta_u}$ for the same parameter set in
Fig.~\ref{fig:FN_bubble}(left). This latter figure shows the existence
of a symmetry-breaking bifurcation as the binding rate ratio
$\rho$ increases past a critical value, and where the
bulk diffusivity ratio is fixed near unity.

\begin{figure}[]
    \centering
    \includegraphics[width=0.9\textwidth]{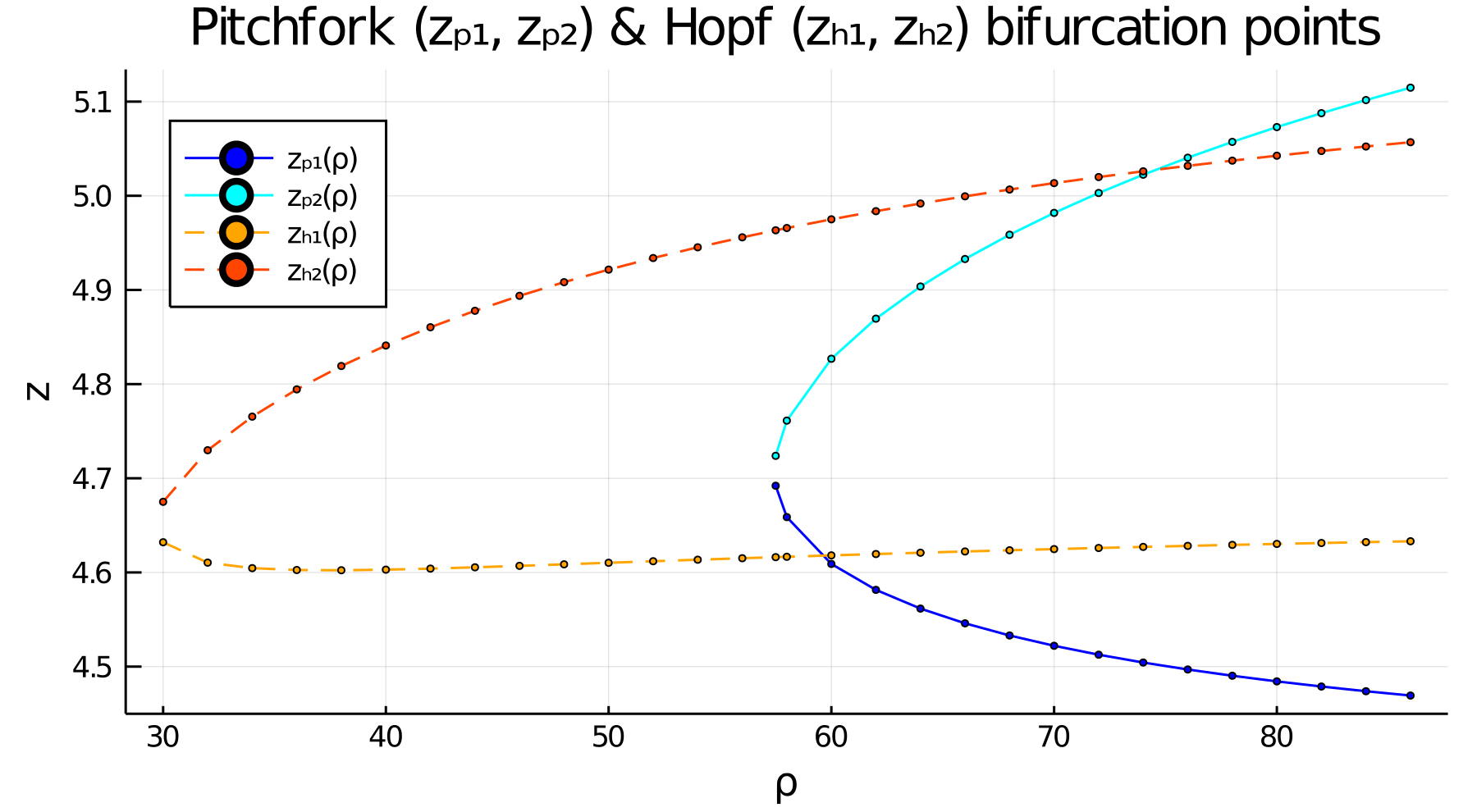}
    \caption{Symmetry-breaking pitchfork bifurcation thresholds
      $z_{P,1}$ and $z_{P,2}$ versus $\rho$ corresponding to
      zero-eigenvalue crossings for the anti-phase mode where the
      asymmetric steady-state bifurcates from the symmetric
      branch. The anti-phase mode is unstable on the range
      $z_{P,1}<z<z_{P,2}$.  The symmetric branch is unstable to an
      oscillatory instability for the in-phase mode when
      $z_{H,1}<z<z_{H,2}$, where $z_{H,1}$ and $z_{H,2}$ are Hopf
      bifurcation thresholds for the in-phase mode. On the range
      $\rho>75$, we observe the ordering
      $z_{P,1}<z_{H,1}<z_{H,2}<z_{P,2}$. For this range of $\rho$,
      where the unstable oscillatory range for the in-phase mode is
      contained within the pitchfork range, the asymmetric branches of
      equilibria that bifurcate from the symmetric branch are
      predicted to be locally linearly stable. Parameters:
      $D_u=1, D_v=3, \sigma_u=\sigma_v=1, \varepsilon=0.7, q=1, L=1$,
      $\beta_u=0.1$, and $\beta_v=\rho \beta_u$.}
    \label{fig:my_label}
\end{figure}

\begin{figure}[]
	\begin{subfigure}[b]{.49\textwidth}
	    \begin{subfigure}[b]{1.\textwidth}
    		\centering
			\includegraphics[width=1.\linewidth]{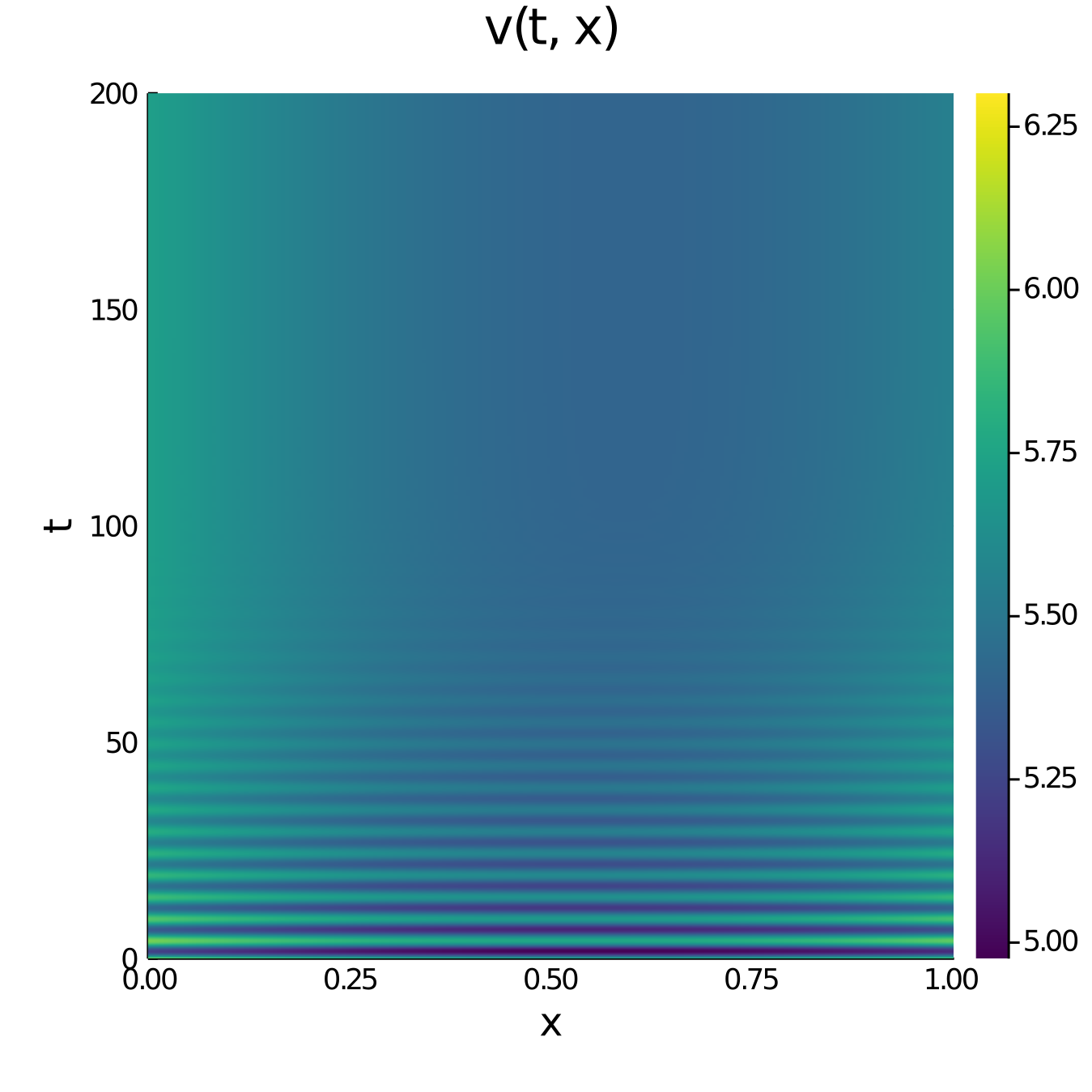}  
		\end{subfigure}
		\begin{subfigure}[b]{1.\textwidth}
    		\centering
			\includegraphics[width=1.\linewidth]{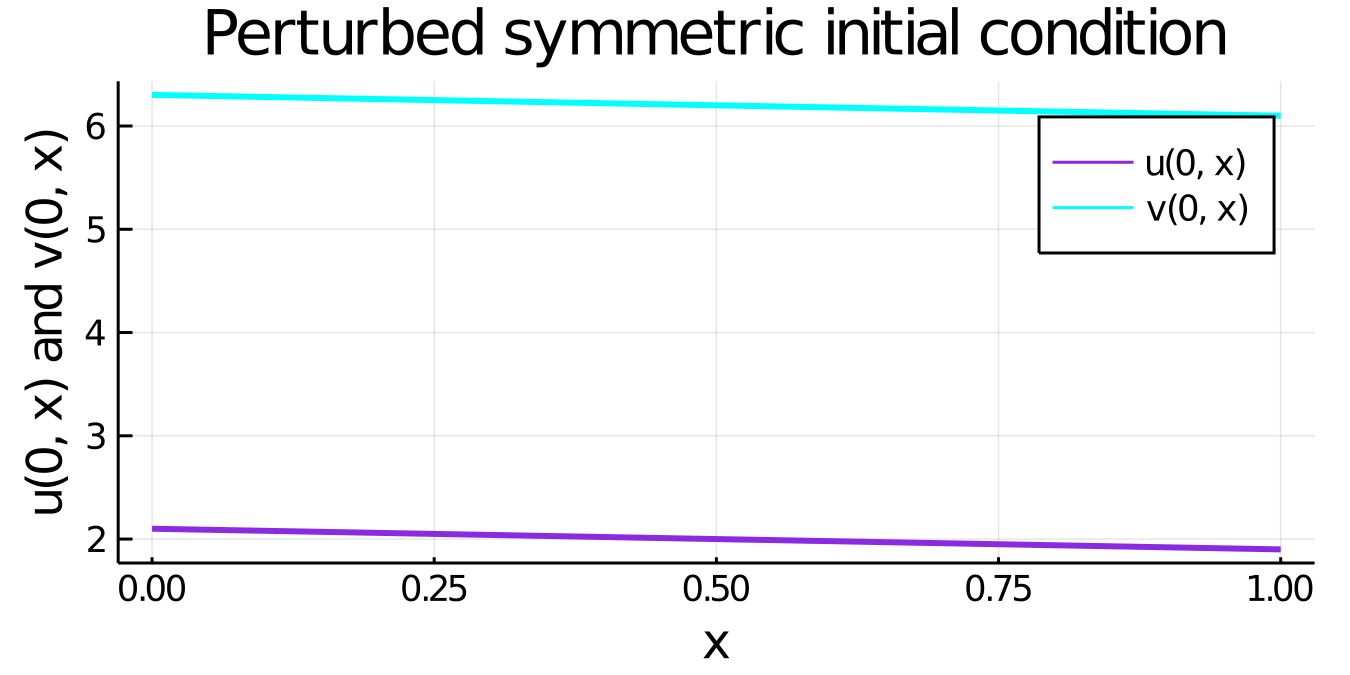}  
		\end{subfigure}
	\end{subfigure}
	\begin{subfigure}[b]{.50\textwidth}
		\begin{subfigure}[b]{1.\textwidth}
    		\centering
			\includegraphics[width=1.\linewidth]{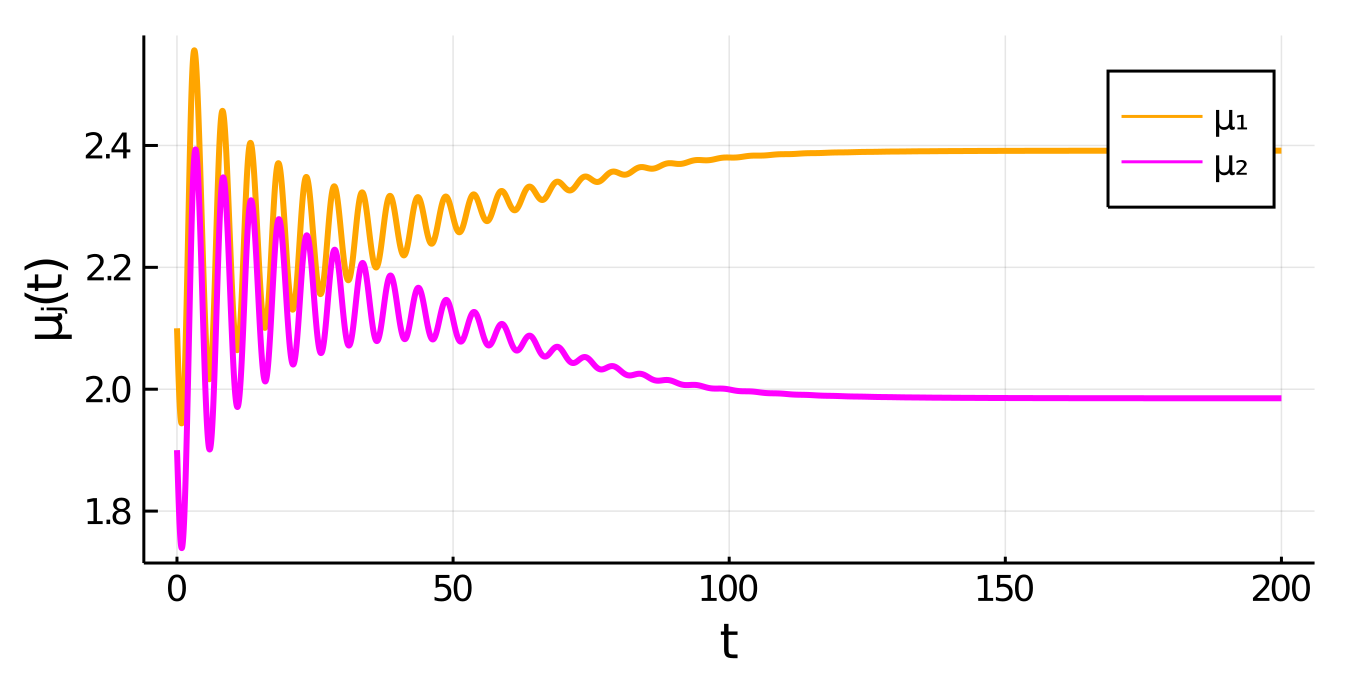}  
		\end{subfigure}
		\begin{subfigure}[b]{1.\textwidth}
			\centering
			\includegraphics[width=1.\linewidth]{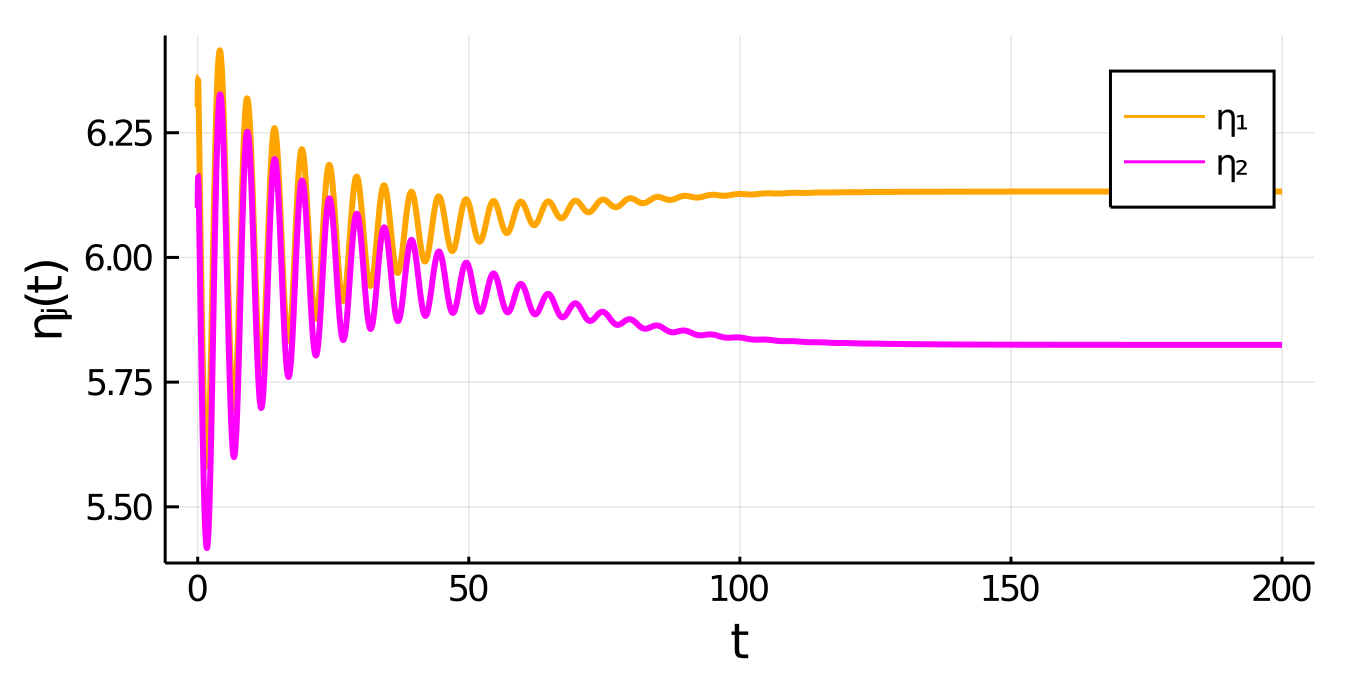}  
		\end{subfigure}
		\begin{subfigure}[b]{1.\textwidth}
			\centering
			\includegraphics[width=1.\linewidth]{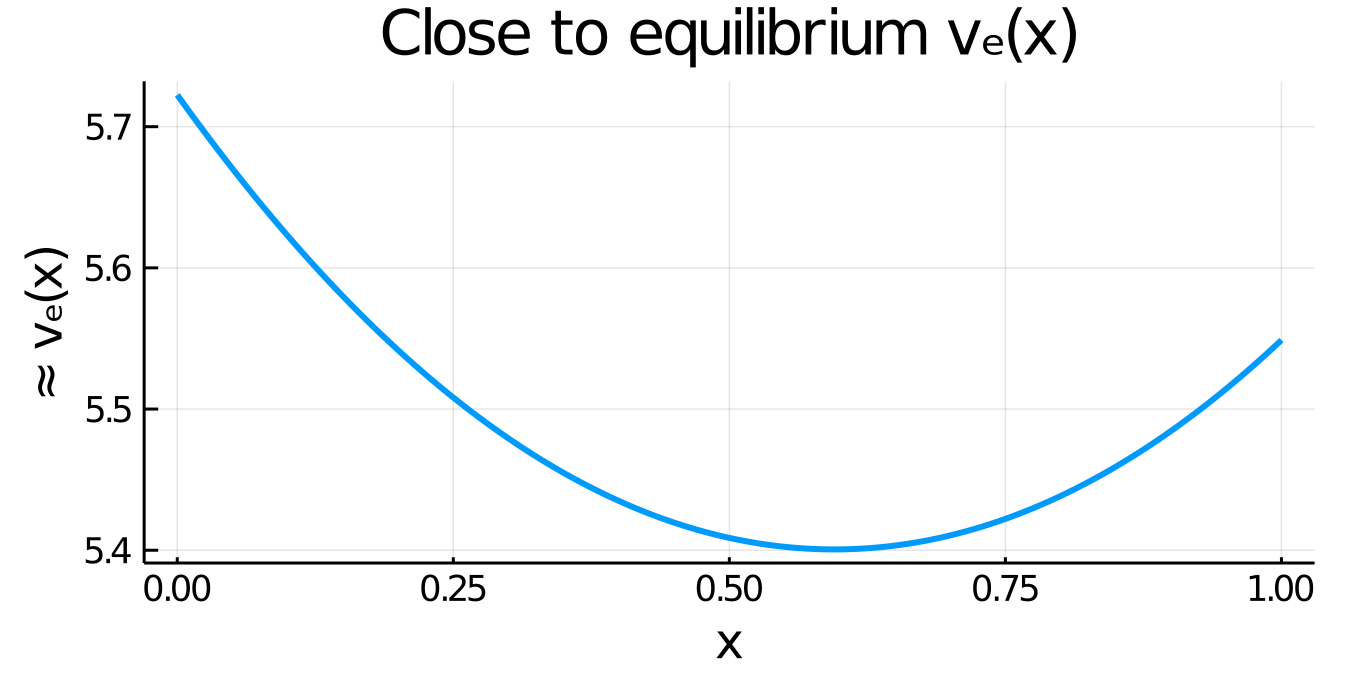}  
		\end{subfigure}
	\end{subfigure}
	\caption{For an initial condition near the unstable symmetric branch,
          and for $z=4.52211$ and $\rho=80$, we predict that the asymmetric
          solution branch is linearly stable since $z_{P,1}<z<z_{H,1}$ (see
          Fig.~\ref{fig:my_label}). The
          full PDE simulation of (\ref{s:full}) by the CN-RK4 IMEX method
          of Appendix \ref{CN-RK4 IMEX} supports this conjecture and
          shows the existence of a stable asymmetric steady-state. For
          $z=4.52211$, the $\rho$-pitchfork point was $\rho\approx 70$
          (see Fig.~\ref{fig:my_label}). Parameters:
          $D_u=1,D_v=3,\sigma_u=\sigma_v=1,\varepsilon=0.7,q=1,L=1,\beta_u=0.1,
          \rho={\beta_v/\beta_u}=80$.}\label{fig:fullFN}.
\end{figure}

\subsubsection{Hopf bifurcation bubble}\label{sec:stab_hopf}

The analysis in (\ref{s:FN_th}) has constructed both symmetric and
asymmetric equilibria of the coupled bulk-cell model with FN kinetics
and has detected symmetry-breaking bifurcations of the symmetric
solution branch $u_e,v_e,\mu_e,\eta_e$ that correspond to
zero-eigenvalue crossings of the linear stability problem.

We now formulate and study the linear stability problem for
the symmetric steady-state solution branch to determine whether
other instabilities, such as Hopf bifurcations, are possible. For the
linear stability analysis we perturb the symmetric steady-state by
introducing
$u(t,x)=u_e(x)+ \phi(x) e^{\lambda t}, v(t,x)=v_e(x)+\psi(x)
e^{\lambda t}, \mu_j(t)=\mu_e + \xi_j e^{\lambda t}, \eta = \eta_e +
\zeta_j e^{\lambda t}$, where $|\phi|\ll 1, |\psi|\ll 1, |\xi_j|\ll 1$
and $|\zeta_j|\ll 1$ for $j\in\{1,2\}$, into (\ref{s:full}) and
linearizing. This yields the linearized eigenvalue problem
\begin{subequations}\label{eig:all_new}
\begin{eqnarray}
	\text{bulk} &&
	\begin{cases}
  \partial_{xx}\phi - \Omega_u^2 \phi = 0\,, \quad &x\in(0,L) \\
  \partial_{xx}\psi - \Omega_v^2 \psi = 0\,, \quad &x\in(0,L)
	\end{cases} \label{stab:bulk}\\
	\text{reaction fluxes} &&
	\begin{cases}
          D_u \partial_x\phi(0)&=\beta_u\;(\phi(0)-\xi_1) \qquad\qquad \qquad \;
          \text{(Robin boundary conditions)}\\
		D_v \partial_x\psi(0)&=\beta_v\;(\psi(0)-\zeta_1) \\
		-D_u\partial_x\phi(L)&=\beta_u\;(\phi(L)-\xi_2) \\
		-D_v \partial_x\psi(L)&=\beta_v\;(\psi(L)-\zeta_2)
	\end{cases} \label{stab:bc}\\
	\text{compartments} &&
	\begin{cases}
          \lambda \xi_1 = \partial_\mu f_e \xi_1 + \partial_\eta f_e
          \zeta_1 + \beta_u\;(\phi(0)-\xi_1)  \qquad \,
          \text{(reaction kinetics at } x=0) \\
          \lambda \zeta_1 = \partial_\mu g_e \xi_1 +
          \partial_\eta g_e \zeta_1 + \beta_v\;(\psi(0)-\zeta_1) \\
          \lambda \xi_2 = \partial_\mu f_e \xi_2 + \partial_\eta
          f_e \zeta_2 + \beta_u\;(\phi(L)-\xi_2)  \qquad
          \text{(reaction kinetics at } x=L) \\
          \lambda \zeta_2 = \partial_\mu g_e \xi_2 +
          \partial_\eta g_e \zeta_2 + \beta_v\;(\psi(L)-\zeta_2)
	\end{cases}\,. \label{stab:jac}
\end{eqnarray}
\end{subequations}
In (\ref{stab:jac}) the partials of $f$ and $g$ evaluated at the
symmetric steady-state are denoted by $\partial_\mu f_e$,
$\partial_\eta f_e$, $\partial_\mu g_e$, and $\partial_\eta g_e$.  In
(\ref{stab:bulk}) we defined $\Omega_u:=\sqrt{(\lambda+\sigma_u)/D_u}$
and $\Omega_v:=\sqrt{(\lambda+\sigma_v)/D_v}$.

{Our goal is to determine all of the eigenvalues $\lambda$ 
  that are associated with nontrivial solutions to
  (\ref{eig:all_new}). The symmetric steady-state solution is linearly
  stable if all such eigenvalues satisfy $\mbox{Re}(\lambda)<0$, and
  it is unstable if for one such eigenvalue we have
  $\mbox{Re}(\lambda)>0$.  Although our eigenvalue problem is not
  self-adjoint, and as such there may be a short-term transient growth
  of initial perturbations (cf.~\cite{transient}), our numerical
  simulations below, confirming results from the linear stability
  theory, suggest that transient growth effects are unlikely to play a
  key role in the pattern-forming process.}

{Owing to the symmetry of the steady-state solution about the
  mid-line, the eigen-modes for (\ref{eig:all_new}) can be decomposed
  into two distinct classes: in-phase eigenperturbations (+) or
  anti-phase eigenperturbations (-), for which
  $\phi_+^{\prime}(\frac{L}{2})=0=\psi_+^{\prime}(\frac{L}{2})$ or
  $\phi_-(\frac{L}{2})=0=\psi_-(\frac{L}{2})$, respectively. We now
  derive the eigenvalue problem for these two possible eigenmodes.}

For the in-phase mode, we calculate from (\ref{stab:bulk}) and
(\ref{stab:bc}) that
\begin{align*}
  \phi_+(x) &= d_1^+ \frac{\cosh\left(\Omega_u(\frac{L}{2}-x)\right)}
                {\cosh(\Omega_u L)} \qquad \text{with} \qquad d_1^+ :=
      \frac{\beta_u}{\beta_u+D_u\Omega_u\tanh(\Omega_u \frac{L}{2})}\xi_1, \\
  \psi_+(x) &= d_2^+ \frac{\cosh\left(\Omega_v(\frac{L}{2}-x)\right)}
                {\cosh(\Omega_vL)}
                \qquad \text{with} \qquad d_2^+ := \frac{\beta_v}
                {\beta_v+D_v\Omega_v\tanh(\Omega_v \frac{L}{2})}\zeta_1.
\end{align*}
For this in-phase mode, we have $\xi_1=\xi_2:=\xi$ and
$\zeta_1=\zeta_2:=\zeta$. Then, by using (\ref{stab:jac}), we obtain the
$2\times 2$ nonlinear matrix eigenvalue problem
\begin{subequations}
  \begin{equation}\label{stab:m+1}
      M_{+}(\lambda)     \begin{pmatrix}
        \xi \\
        \zeta
    \end{pmatrix} =     \begin{pmatrix}
        0 \\
        0
    \end{pmatrix}\,,
\end{equation}
where
\begin{equation}\label{stab:m+2}
    M_{+} :=  
    \begin{pmatrix}
      \lambda - \partial_\mu f_e + \beta_u(1-\frac{\beta_u}
      {\beta_u+D_u\Omega_u\tanh(\Omega_u L/2)}) & - \partial_\eta f_e \\
      - \partial_\mu g_e & \lambda - \partial_\eta g_e +
      \beta_v(1-\frac{\beta_v}{\beta_v+D_v\Omega_v\tanh(\Omega_v L/2)})\,.
    \end{pmatrix}
  \end{equation}
\end{subequations}
Eigenvalues associated with the in-phase eigenperturbation are roots
$\lambda$ of $\det(M_+(\lambda))=0$.

Similarly, for the anti-phase mode, we calculate from
(\ref{stab:bulk}) and (\ref{stab:bc}) that
\begin{align*}
  \phi_-(x) &= d_1^- \frac{\sinh(\Omega_u(\frac{L}{2}-x))}{\sinh(\Omega_u L)}
     \qquad \text{with} \qquad d_1^- := \frac{\beta_u}{\beta_u+D_u\Omega_u
                \coth(\Omega_u \frac{L}{2})}\xi_1\,, \\
  \psi_-(x) &= d_2^- \frac{\sinh(\Omega_v(\frac{L}{2}-x))}{\sinh(\Omega_v L)}
   \qquad \text{with} \qquad d_2^- := \frac{\beta_v}{\beta_v+D_v\Omega_v
                \coth(\Omega_v \frac{L}{2})}\zeta_1\,,
\end{align*}
where we now label $\xi_1=-\xi_2:=\xi$ and
$\zeta_1=-\zeta_2:=\zeta$. From (\ref{stab:jac}), eigenvalues $\lambda$
associated with the anti-phase mode are roots of
$\det(M_-(\lambda))=0$, where
\begin{equation}\label{stab:m-2}
    M_{-} :=  
    \begin{pmatrix}
      \lambda - \partial_\mu f_e +
      \beta_u(1-\frac{\beta_u}{\beta_u+D_u\Omega_u\coth(\Omega_u L/2)}) &
      - \partial_\eta f_e \\
      - \partial_\mu g_e & \lambda - \partial_\eta g_e +
      \beta_v(1-\frac{\beta_v}{\beta_v+D_v\Omega_v\coth(\Omega_v L/2)})
    \end{pmatrix}\,.
\end{equation}

By a simple scaling, the eigenvalues for the in-phase and anti-phase mode are
equivalently written as the roots of $\mathcal{F}_\pm(\lambda)=0$, where
\begin{equation}\label{stab:roots}
  \mathcal{F}_\pm(\lambda) := \frac{\det(M_{\pm}(\lambda))}
  {\det(\lambda I - DF_e)}
  \quad \mbox{with} \quad DF_e :=
    \begin{pmatrix}
    \partial_\mu f_e & \partial_\eta f_e\\
    \partial_\mu g_e &  \partial_\eta g_e
    \end{pmatrix}\,.
\end{equation}
Roots of (\ref{stab:roots}) with $\mbox{Re}(\lambda)>0$ correspond to
unstable modes, while roots with $\lambda = i \lambda_i$ for $\lambda_i>0$
yield Hopf bifurcations.  To determine the number of zeroes
of $\mathcal{F}_\pm$ in the right-half
plane of $\bC$, denoted by $\mathcal{N}_+(\mathcal{F}_\pm)$, we
numerically implement the Argument Principle given by
\begin{equation}\label{stab:arg}
  \frac{1}{2\pi i} \int_{\Gamma_R} \frac{\frac{d}{d\lambda}
    \mathcal{F}_\pm(\lambda)}{\mathcal{F}_\pm(\lambda)}\;d\lambda =
  \mathcal{N}_+(\mathcal{F}_\pm) - \mathcal{P}_+(\mathcal{F}_\pm)\,,
\end{equation}
where $\mathcal{P}_+(\mathcal{F}_\pm)$ is the number of poles of
$\mathcal{F}_\pm$ in $\mbox{Re}(\lambda)>0$. Here we have labeled the
counterclockwise integration contour as
$\Gamma_R:=\{R \exp(i\varphi) |
\varphi\in[-\frac{\pi}{2},\frac{\pi}{2}]\} \cup \left[-i R, i
  R\right]$, with $R$ sufficiently large so that all zeroes 
$\mathcal{N}_+(\mathcal{F}_\pm)$ and poles $\mathcal{P}_+(\mathcal{F}_\pm)$
are in ${\rm int}(\Gamma_R)$. 

Our numerical results based on (\ref{stab:arg}) applied to
${\mathcal F}_{-}(\lambda)$ show for the parameter set in
Fig.~\ref{fig:FN_bubble}(left) that there is a unique real unstable
eigenvalue for the anti-phase mode on the range $z_{P,1}<z<z_{P,2}$
between the two pitchfork points. Moreover, we numerically verified that
the symmetric branch in Fig.~\ref{fig:FN_bubble}(right) is linearly
stable to the anti-phase mode only when $\rho$ is below the pitchfork
point.

In Fig.~\ref{fig:my_label}, the pitchfork points $z_{P,1}$ and
$z_{P,2}$ are plotted versus $\rho$. From a numerical root finding of
$\mathcal{F}_{+}(i\lambda_i)=0$ in (\ref{stab:roots}), together with a
numerical implementation of the Argument Principle (\ref{stab:arg}),
we identify two Hopf bifurcation thresholds $z_{H,1}$ and $z_{H,2}$
for the in-phase mode, with the symmetric branch being unstable to an
oscillatory instability only on the range $z_{H,1}<z<z_{H,2}$. In
Fig.~\ref{fig:my_label} we observe for $\rho>75$ that the unstable
oscillatory range is contained within the pitchfork range, in that the
ordering $z_{P,1}<z_{H,1}<z_{H,2}<z_{P,2}$ holds. For this range of
$\rho$, the symmetric branch is linearly stable to both in-phase and
anti-phase modes on $z<z_{P,1}$ and on $z>z_{P,2}$. As a result, we
predict that the symmetry-breaking bifurcation at $z_{P,1}$ and
$z_{P,2}$ leads to {\em linearly stable asymmetric solution branches}
at least locally near either $z_{P,1}$ or $z_{P,2}$.  As shown in
Fig.~\ref{fig:fullFN}, this conjecture is verified in the full PDE
simulations of (\ref{s:full}) by the CN-RK4 IMEX method of Appendix
\ref{CN-RK4 IMEX}.

\subsection{Compartments with Gierer-Meinhardt kinetics}

Next, we extend our analysis to a compartmental-reaction variant of the Gierer-Meinhardt (GM) model, where the reaction kinetics are now
confined to the boundaries. The original GM model, as
introduced in \cite{gm} and \cite{gierer} to model pattern formation
in biological morphogenesis, given by
\begin{equation*}
    \begin{array}{rcl}
      \partial_t u &=& \varrho_0(x) + c_u\varrho_u(x) \frac{u^2}{v} +
                       D_u \Delta u  - \sigma_u u\,,\\
         \partial_t v &=& c_v\varrho_v(x) u^2 + D_v \Delta v - \sigma_v v\,, 
    \end{array}
\end{equation*}
disregards that in biological tissues morphogen-producing reactions mostly
occur intracellularly and on the membranes of cells. For simplicity, we
illustrate our compartmental-reaction diffusion theory for the 1-D case
when $\varrho_0\equiv 0, c_u\varrho_u(x)\equiv 1$, and $c_v\varrho_v\equiv 1$.
In our model, we restrict the reaction kinetics to the boundaries so that
for the uncoupled bulk-cell model we have
\begin{equation}\label{cell:GM}
         \dot{\mu}(t) = f(\mu,\eta) := \frac{\mu^2}{\eta}\,, \qquad
        \dot{\eta}(t) = g(\mu,\eta) := \mu^2\,.
\end{equation}
Similar to the case with FN kinetics we observe that $\mu$
activates self-production and the production of $\eta$, whereas
the species $\eta$ only inhibits the production of
$\mu$. Hence, the GM model has a more indirect activator-inhibitor
structure than the FN model, in which $\eta$ also
inhibits directly its own production. The uncoupled equilibrium for
(\ref{cell:GM}) is non-hyberbolic in all directions, as
$DF(\mu_e,\eta_e)=0$, where $\mu_e = 0$ and $\eta$ is a constant.

To apply the bulk-cell steady-state analysis of
\S \ref{2cellcoupledsystem} we simply identify that $g(\mu,\eta)=g_1(\mu)-
g_2\eta$, where $g_1=\mu^2$ and $g_2=0$. From (\ref{ss:iv_new}) we
get
\begin{subequations}
\begin{equation}\label{ssGM:iv_new}
    A
    \begin{pmatrix}
        \eta_1^e \\
        \eta_2^e
    \end{pmatrix}
    =
    \begin{pmatrix}
        (\mu_1^e)^2 \\
        (\mu_2^e)^2
    \end{pmatrix}
    \quad \text{with} \quad A = \frac{\beta_v}{\gamma_v^2-1}\tilde{A} :=
    \begin{pmatrix}
        A_{11} &A_{12} \\
        A_{12} & A_{11}
    \end{pmatrix} \,,
  \end{equation}
  where, by using (\ref{ss:iv}) for $\tilde{A}$, we identify that
  \begin{equation}\label{ssGM:iv_new_b}
    A_{11} = \frac{\beta_v}{\gamma_v^2-1} \left( \gamma_v \cosh(\omega_vL)-1
    \right) \,, \qquad
    A_{12} = \frac{\beta_v}{\gamma_v^2-1} \left( \cosh(\omega_vL)-\gamma_v
    \right) \,.
  \end{equation}
\end{subequations}
In terms of $A_{11}$ and $A_{12}$, we obtain from (\ref{s:iu_fin}) after
calculating $A^{-1}$ that the nonlinear algebraic system for
$\mu_1^e$ and $\mu_2^e$ reduces to
\begin{equation}\label{ssGM:iu_fin}
  \begin{pmatrix}
        (\mu_1^e)^2 \\
        (\mu_2^e)^2
    \end{pmatrix}
    - \frac{\beta_u}{\gamma_u^2-1} \frac{1}{\left(A_{11}^2 - A_{12}^2\right)}
    \begin{pmatrix}
        A_{11} (\mu_1^e)^2 - A_{12} (\mu_2^e)^2 & 0 \\
        0 & -A_{12} (\mu_1^e)^2 + A_{11} (\mu_2^e)^2
    \end{pmatrix}
    B
    \begin{pmatrix}
        \mu_1^e \\
        \mu_2^e
    \end{pmatrix}
    =0\,.
\end{equation}
From (\ref{ss:nonl_symm}), or equivalently from (\ref{ssGM:iu_fin}),
the symmetric steady-state solution, for which $\mu_1^{e}=\mu_{2}^{e}=\mu_e$,
is given explicitly by 
\begin{equation}\label{ssGM:symm}
  \mu_e = \frac{a_1}{b_1}\frac{(\gamma_u^2-1)}{\beta_u}\,, \qquad
  \eta_e=\frac{\mu_e(\gamma_u^2-1)}{\beta_u b_1} =
    \frac{a_1}{b_1^2}\frac{(\gamma_u^2-1)^2}{\beta_u^2}\,,
  \end{equation}
  where $a_1=A_{11}+A_{12}$ and $b_1$ are the in-phase eigenvalues of $A$ and
  $B$ given in (\ref{ss:eig}) with $g_2=0$.

  The linear stability of the symmetric steady-state solution is
  studied numerically using the methodology in \S
  \ref{sec:stab_hopf}. In contrast to the case with FN kinetics, there
  are no Hopf bifurcation points associated with the in-phase mode for
  the GM model. As a result, the symmetric steady-state cannot be
  destabilized by the in-phase mode. {Moreover, the anti-phase
    mode for the linearization of the symmetric steady-state undergoes
    a zero-eigenvalue crossing only at any pitchfork bifurcation point
    $\rho=\rho_p$ that can be calculated from the nonlinear algebraic
    system (\ref{ssGM:iu_fin}).  This symmetric steady-state is
    unstable to the anti-phase mode only when $\rho>\rho_p$ (as
    suggested by the full numerical results shown below in
    Fig.~\ref{fig:symmhysteresis2cells} and
    Fig.~\ref{fig:asymmhysteresis2cells}).}
  
\begin{figure}[htbp]
    \centering
    \includegraphics[width=1.\textwidth]{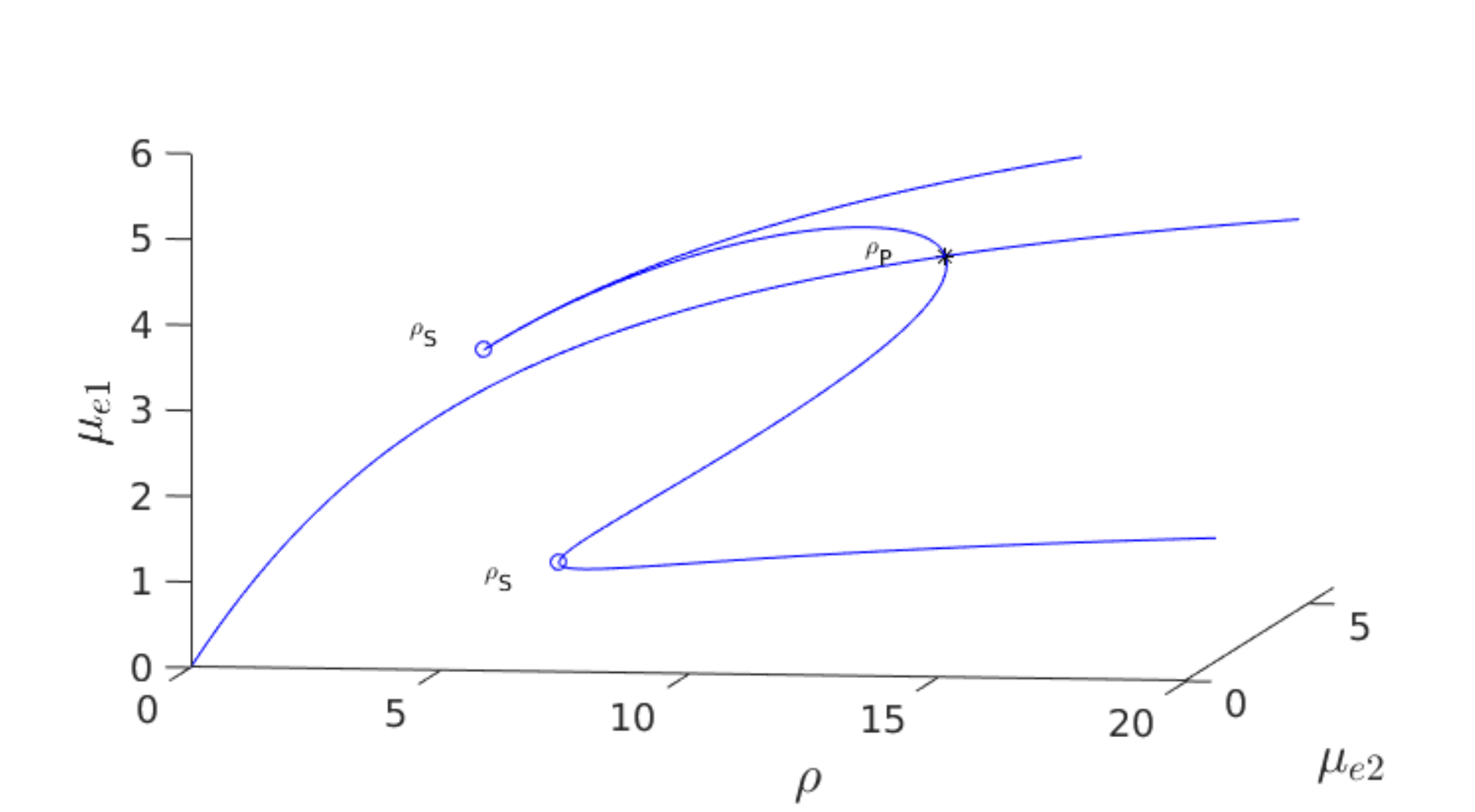}
    \caption{{3-D bifurcation diagram for $\mu_{e1}$ and
        $\mu_{e2}$ with bifurcation parameter $\rho={\beta_v/\beta_u}$
        for steady-states of the GM compartmental-reaction diffusion
        system as computed from (\ref{ssGM:iu_fin}) using MatCont
        \cite{matcont}.} There is a subcritical pitchfork bifurcation
        from the symmetric steady-state at $\rho_p \approx 13.09728$,
        with (secondary) fold point bifurcations occurring on the
        asymmetric branches at $\rho_f \approx 5.59447$. For
        $\rho<\rho_p$ the symmetric steady-state is linearly stable,
        while for $\rho>\rho_f$ the asymmetric branches are linearly
        stable. Parameters: $D_u=D_v=1, \sigma_u=\sigma_v=1, L=1$, and
        $\beta_u=0.1$. }
    \label{fig:bifdiagrhoGMiota0p1}
  \end{figure}

\begin{figure}[] 
    \centering
	\begin{subfigure}[b]{.45\textwidth}
	    \begin{subfigure}[b]{1.\textwidth}
	        \centering
	        \includegraphics[width=1.\linewidth]{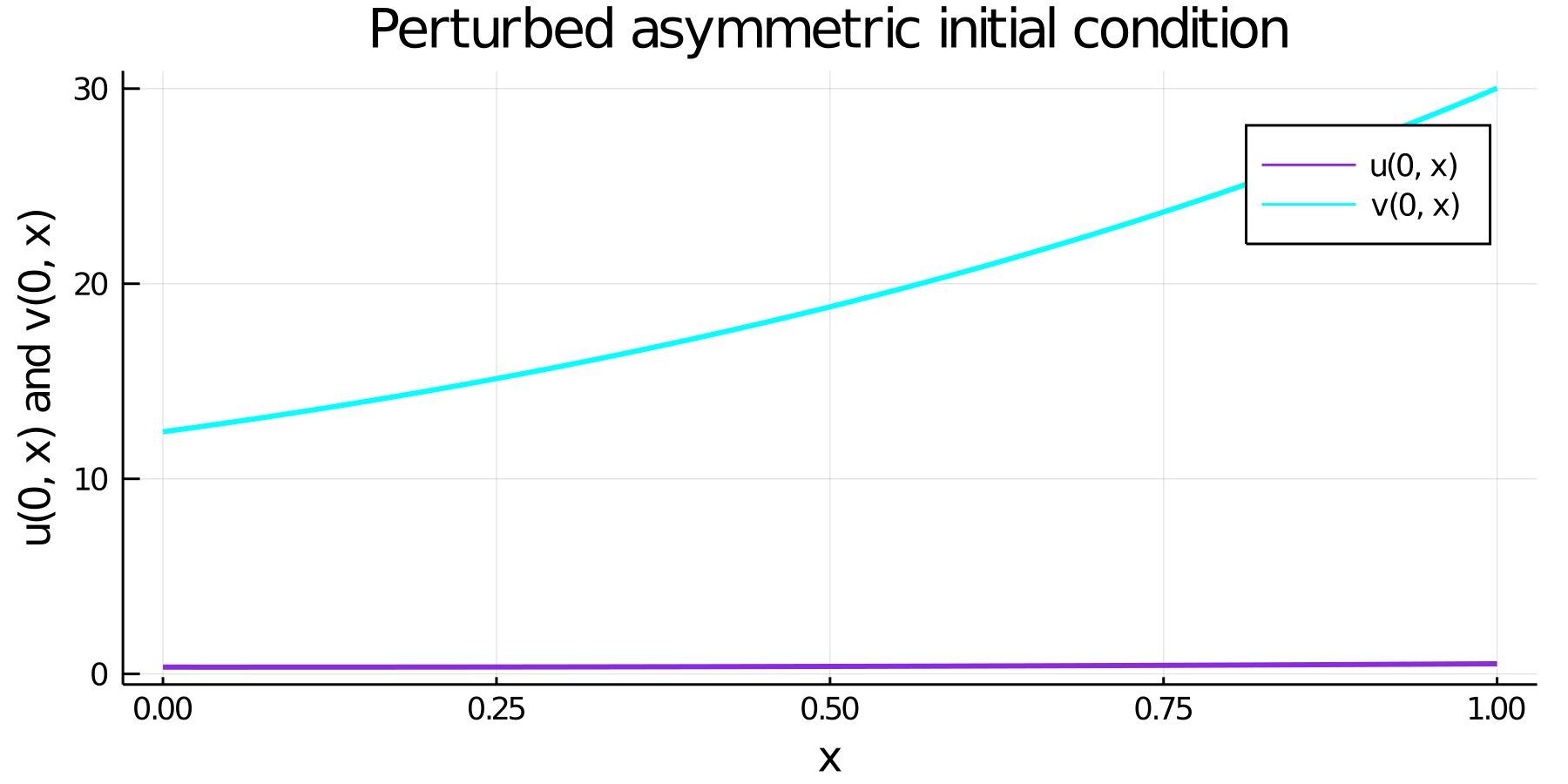}
	    \end{subfigure}
	    \begin{subfigure}[b]{1.\textwidth}
	        \centering
	        \includegraphics[width=1.\linewidth]{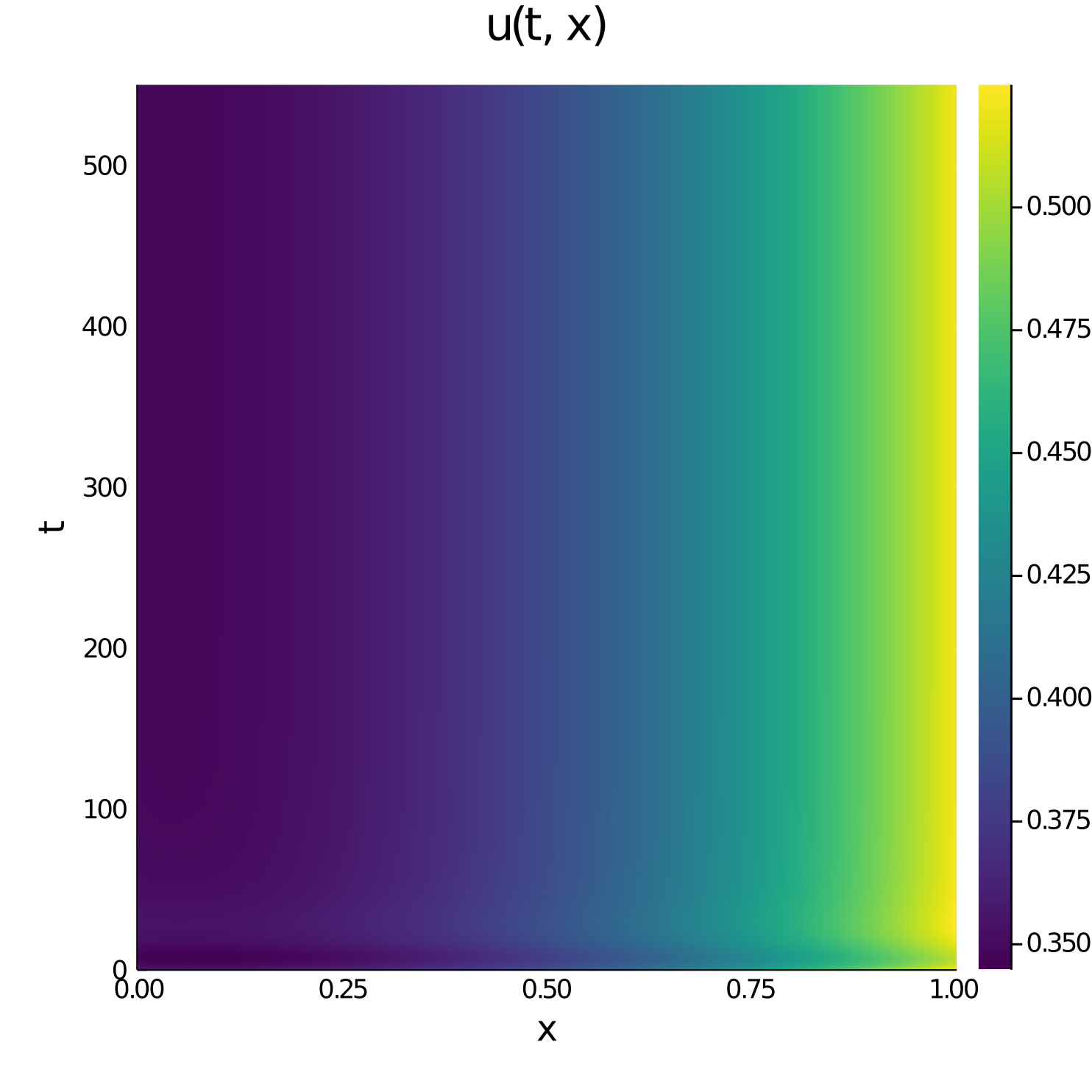}
	    \end{subfigure}
	    \begin{subfigure}[b]{1.\textwidth}
    		\centering
			\includegraphics[width=1.\linewidth]{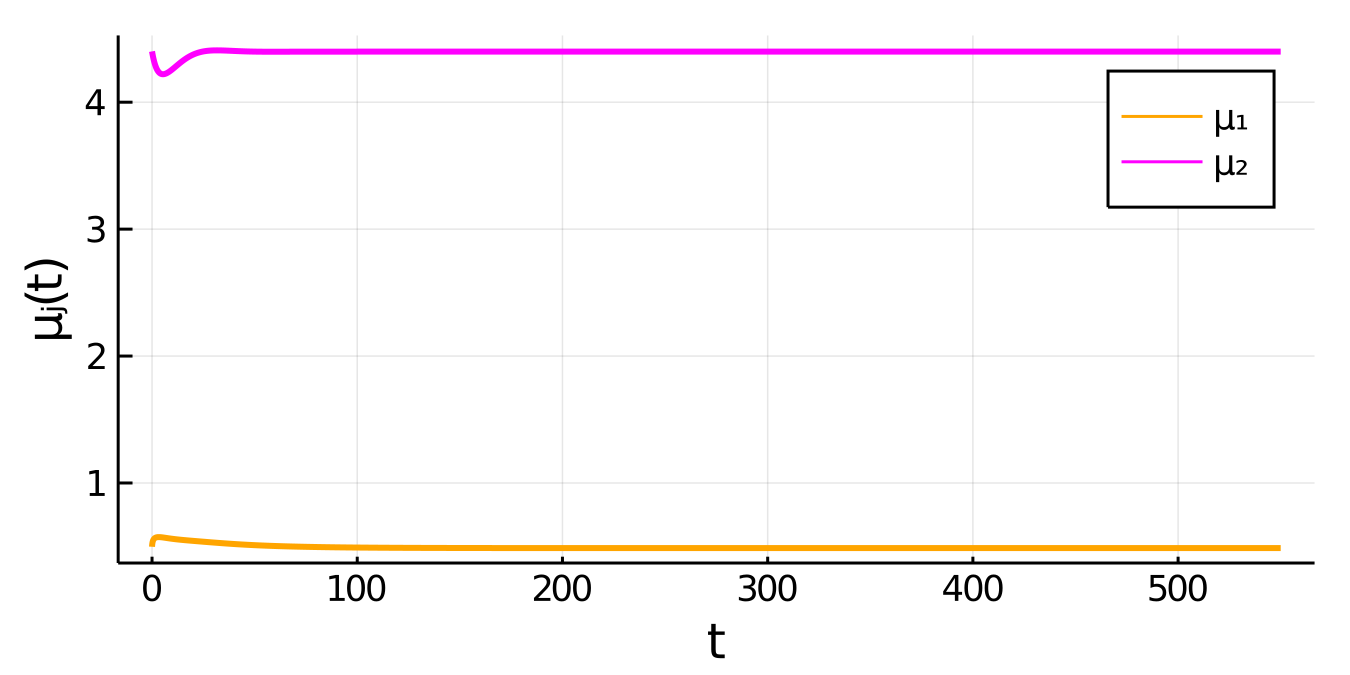}  
		\end{subfigure}
		\begin{subfigure}[b]{1.\textwidth}
			\centering
			\includegraphics[width=1.\linewidth]{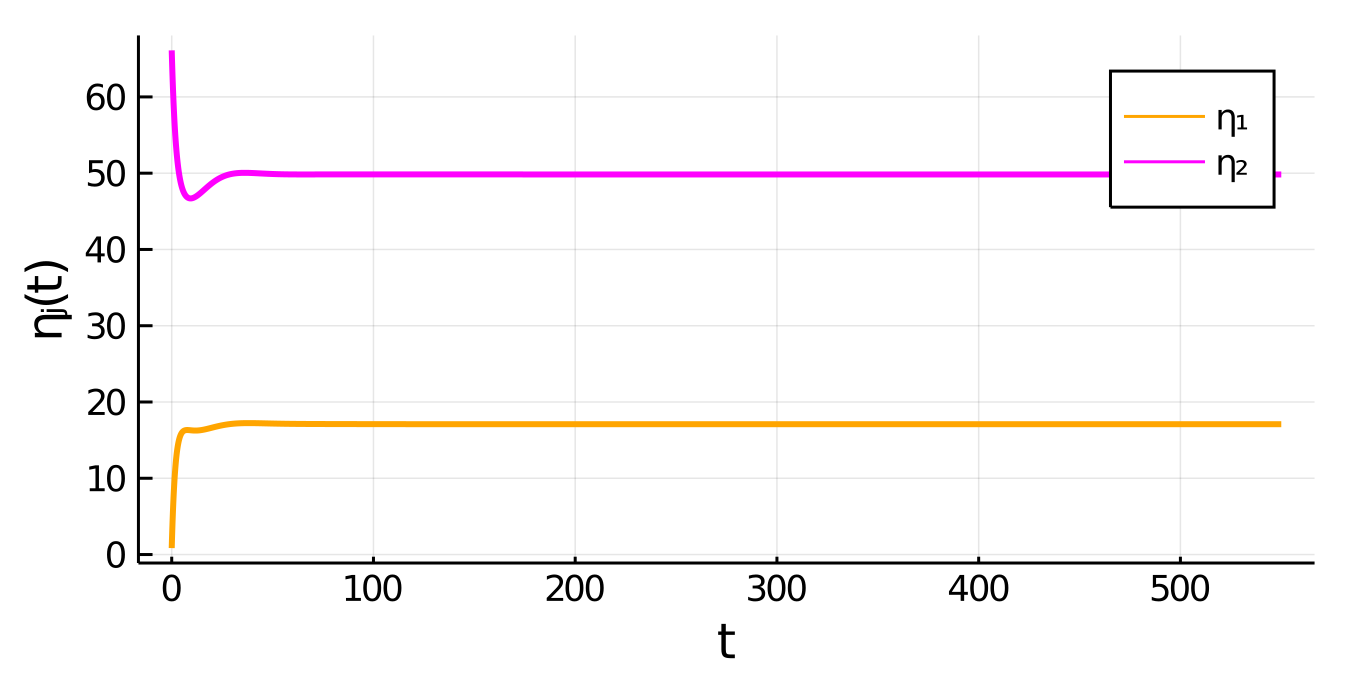}  
		\end{subfigure}
		\begin{subfigure}[b]{1.\textwidth}
			\centering
			\includegraphics[width=1.\linewidth]{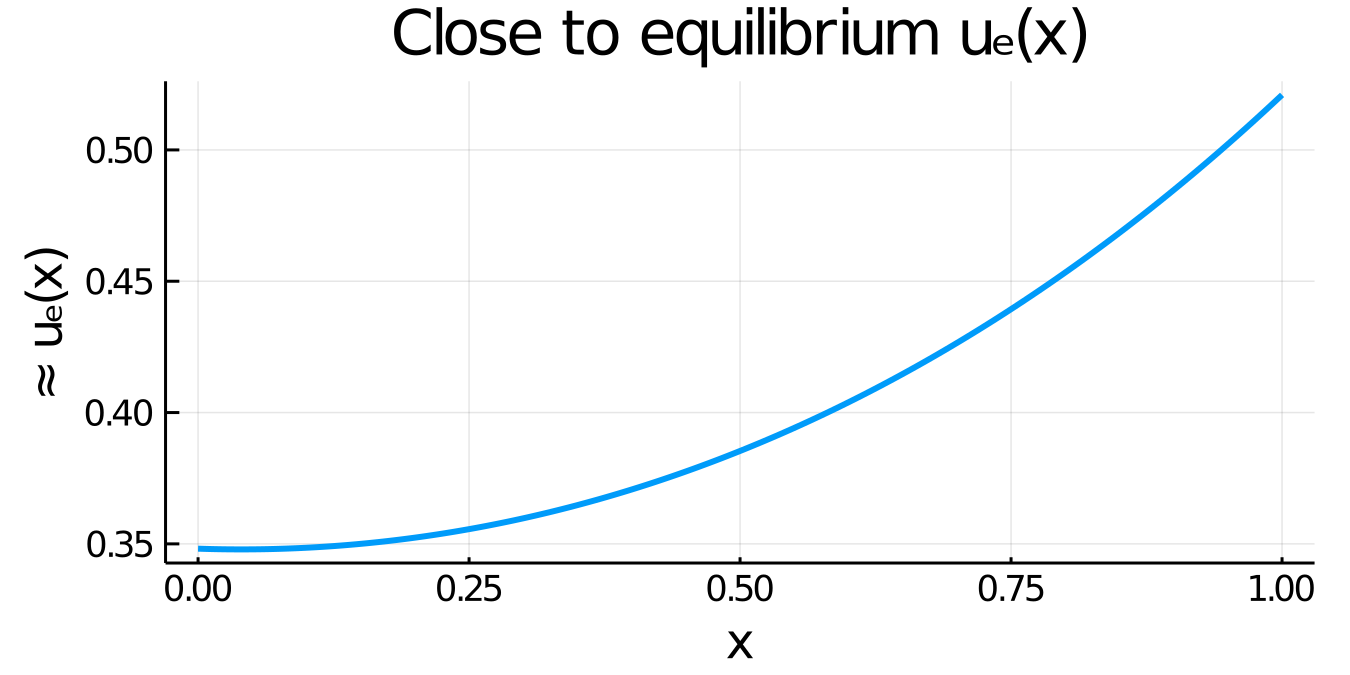}  
		\end{subfigure}
	\end{subfigure}
	\begin{subfigure}[b]{.45\textwidth}
	    \begin{subfigure}[b]{1.\textwidth}
	        \centering
	        \includegraphics[width=1.\linewidth]{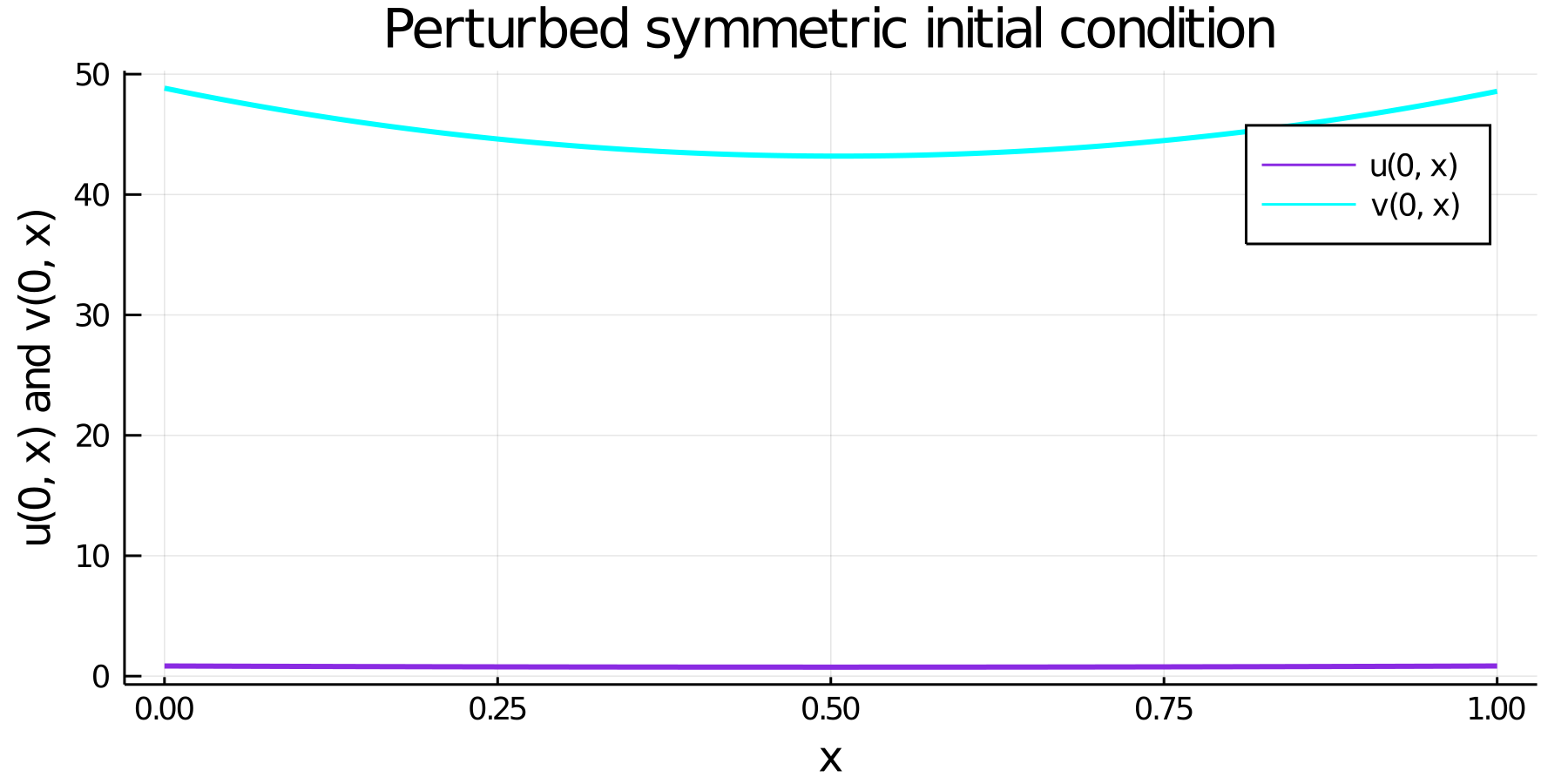}
	    \end{subfigure}
	    \begin{subfigure}[b]{1.\textwidth}
	        \centering
	        \includegraphics[width=1.\linewidth]{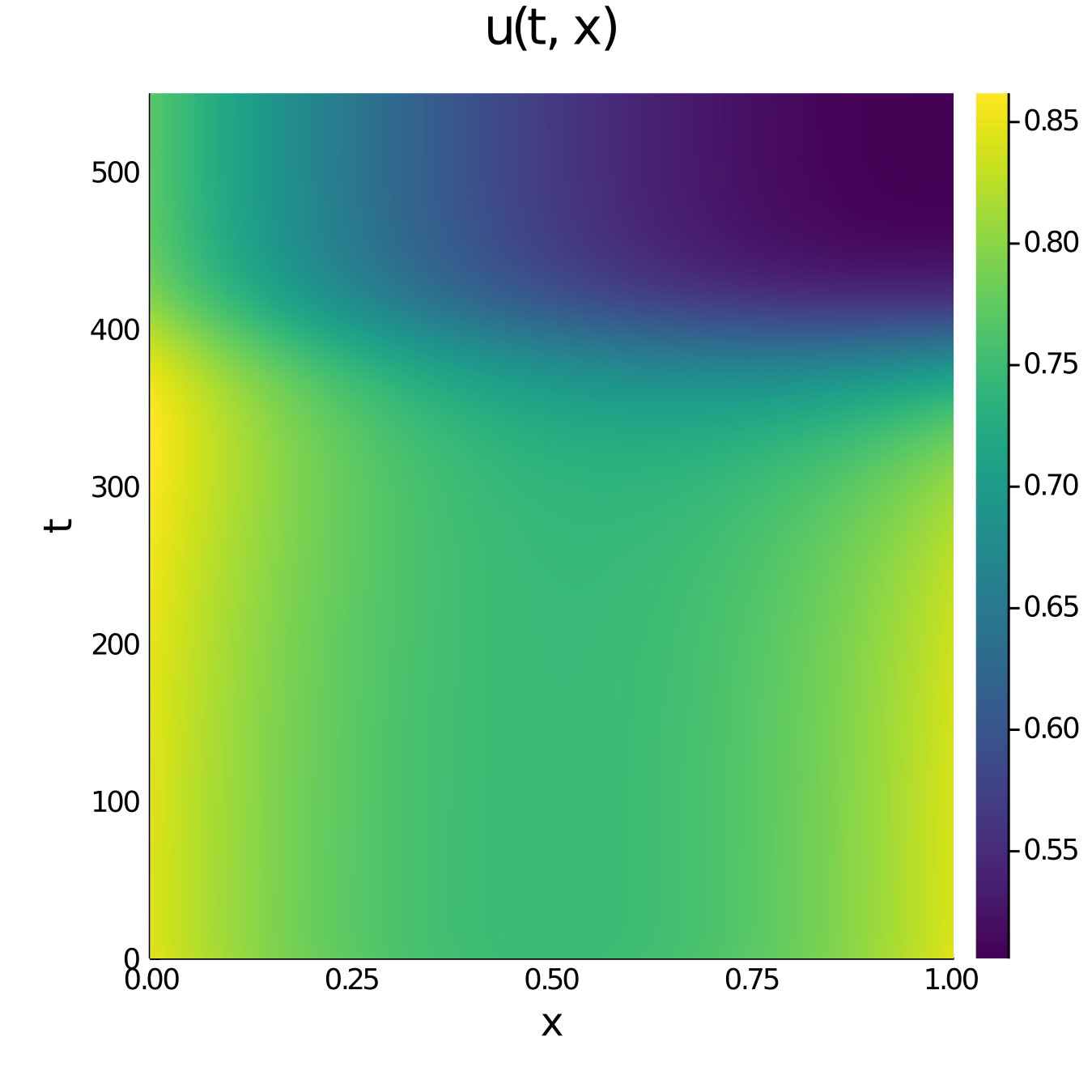}
	    \end{subfigure}
	    \begin{subfigure}[b]{1.\textwidth}
    		\centering
			\includegraphics[width=1.\linewidth]{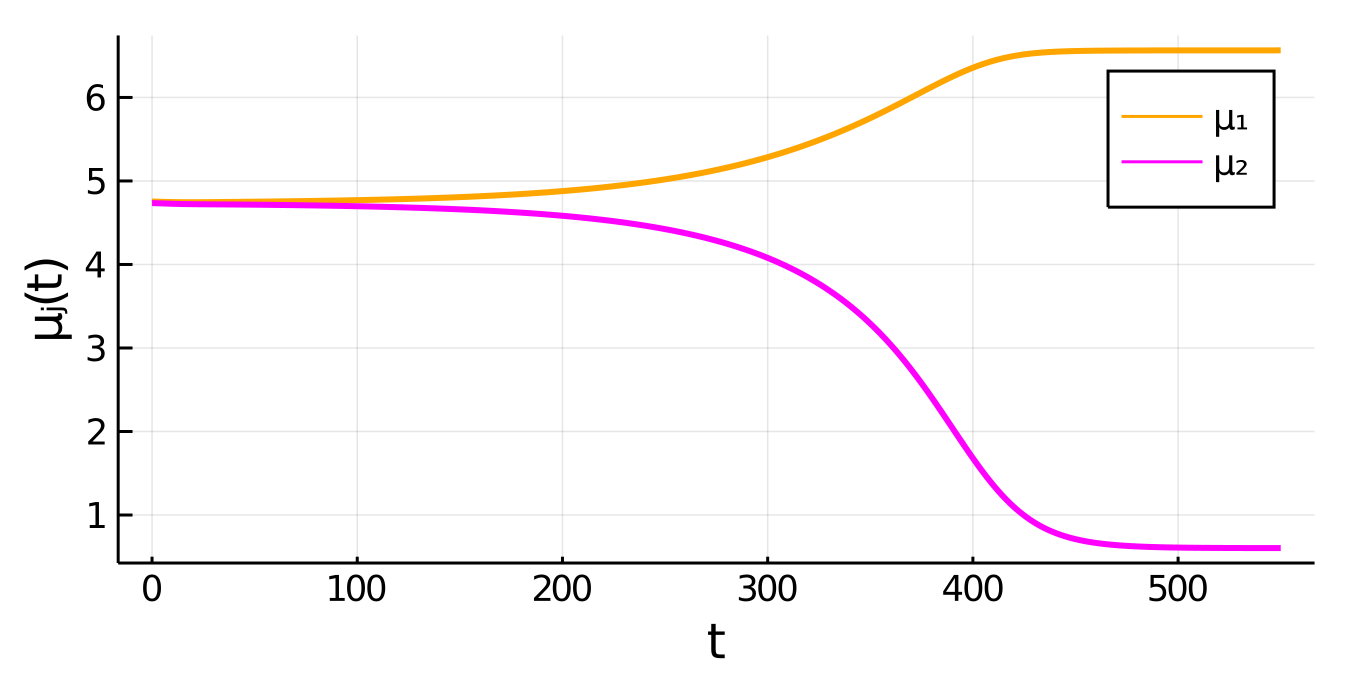}  
		\end{subfigure}
		\begin{subfigure}[b]{1.\textwidth}
			\centering
			\includegraphics[width=1.\linewidth]{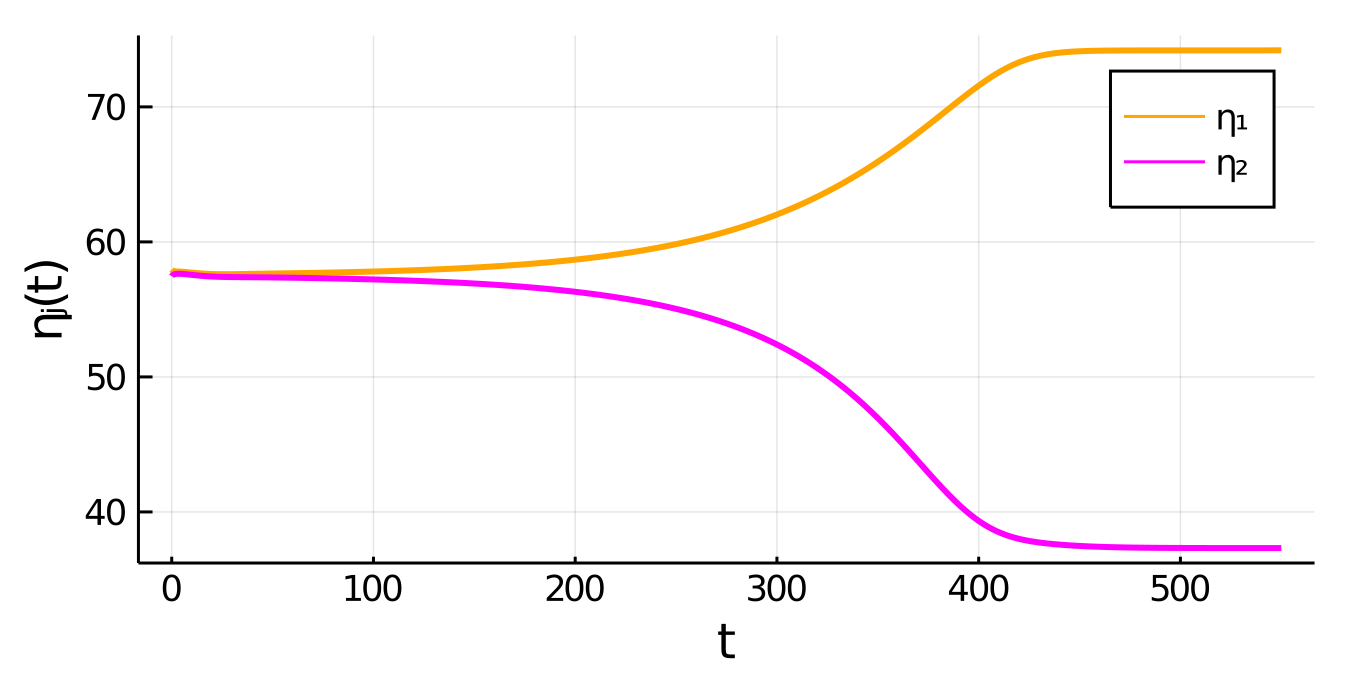}  
		\end{subfigure}
		\begin{subfigure}[b]{1.\textwidth}
			\centering
			\includegraphics[width=1.\linewidth]{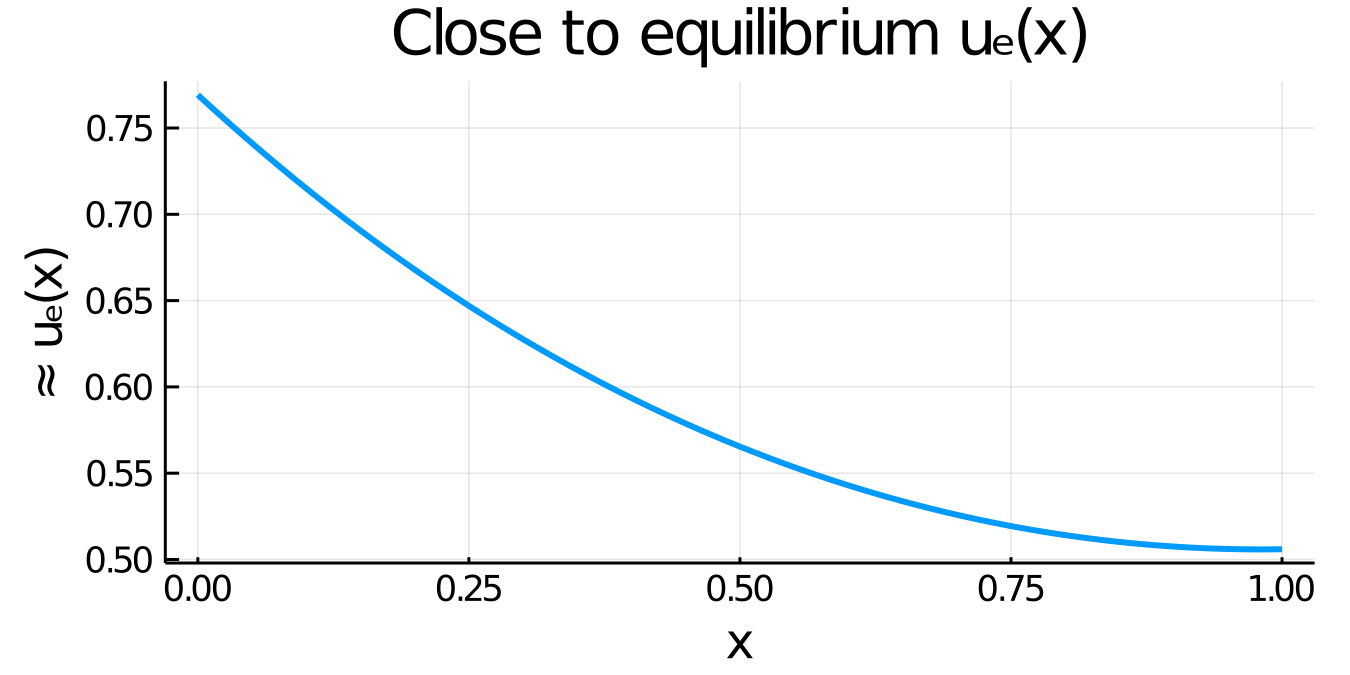}  
		\end{subfigure}
	\end{subfigure}
	\caption{Full numerical computations of (\ref{s:full}) by the
          CN-RK4 IMEX method for GM kinetics. Left: convergence to the
          asymmetric branch for $\rho=8$ starting near one of the
          stable asymmetric branches for $\rho>\rho_f$. Right:
          convergence to an asymmetric branch for $\rho=25$ starting
          close to the symmetric branch with perturbation toward the
          respective asymmetric branch. The remaining parameter values
          are as in the caption to
          Fig.~\ref{fig:bifdiagrhoGMiota0p1}.}
	\label{fig:asymmhysteresis2cells}
\end{figure}

  In Fig.~\ref{fig:bifdiagrhoGMiota0p1} we plot solution branches of
  (\ref{ssGM:iu_fin}) for $D_u=D_v=\sigma_u=\sigma_v=1$, $L=1$, and
  $\beta_u=0.1$, as a function of $\rho={\beta_v/\beta_u}$. We observe
  that the symmetric branch undergoes a subcritical pitchfork
  bifurcation at $\rho=\rho_p\approx 13.10$. The emerging unstable
  asymmetric equilibria undergo a secondary fold bifurcation at
  $\rho_f\approx 5.59$. The asymmetric branches are found to be
  linearly stable for $\rho>\rho_f$. As a result, we conclude that
  linearly stable asymmetric patterns occur even when $D_u=D_v$.
  
  From Fig.~\ref{fig:bifdiagrhoGMiota0p1}, the bifurcation structure
  of equilibria for (\ref{ssGM:iu_fin}) exhibits a hysteresis
  structure.  As shown in Fig.~\ref{fig:symmhysteresis2cells} for
  $\rho=4$ and $\rho=8$, for an initial condition near the symmetric
  branch when $\rho\in[1,\rho_p)$, the full PDE  simulations
  of (\ref{s:full}) by the CN-RK4 IMEX method of Appendix \ref{CN-RK4
    IMEX} verifies the convergence to the symmetric steady-state
  branch. In contrast, the full numerical results shown in
  Fig.~\ref{fig:asymmhysteresis2cells}(left) for $\rho>\rho_f$ confirms
  that for an initial condition near the stable asymmetric solution branch
  the time-dependent solution converges to the asymmetric
  branch. Moreover, in Fig.~\ref{fig:asymmhysteresis2cells}(right)
  when $\rho=25>\rho_p$ an initial condition near the unstable
  symmetric branch is shown to converge to the asymmetric solution
  branch.

  For the classical GM RD model without reaction compartments, it is
  well-known from a standard Turing analysis that stable spatially
  non-uniform solutions can only occur when the diffusivity ratio
  ${D_v/D_u}$ is sufficiently large. In our compartmental-reaction
  diffusion model, in Table \ref{tab:2-cell GM hysteresis} we show
  that asymmetric solution branches bifurcate from the symmetric
  branch even when $D_v < D_u$. As seen from this table, the required
  binding rate ratio threshold $\rho_p$ increases as ${D_v/D_u}$
  decreases. Furthermore, the extent of the hysteresis, as measured by
  the values of $\rho_p - \rho_f$ and $\mu_1^e - \mu_2^e$, becomes
  larger as ${D_v/D_u}$ decreases.  Regarding the dependence of the
  hysteresis on the binding rates $\beta_u$ and
  $\beta_v=\rho \beta_u$, Table \ref{tab:2-cell GM hysteresis iota}
  shows that if $\beta_u\geq 0.35$ there is now a supercritical
  pitchfork bifurcation from the symmetric branch and the hysteresis
  structure disappears.

\begin{table}[ht]
\centering
\begin{tabular}{|c||c|c|c|c|c|c|c|}
    \hline
    ${D_v/D_u}$   & 0.4   & 0.5       & 0.6       & 0.7       & 0.8       & 0.9       & 1\\
    \hline\hline
    $\rho_p$            & $>100$& 52.18853  & 26.86097  & 19.71589  & 16.34470  & 14.38188  & 13.09728 \\
    \hline
    $\mu_e$             &       & 4.83783   & 4.60295   & 4.43970   & 4.31969   & 4.22776   & 4.15511\\
    \hline\hline
    $\rho_f$            &       & 9.33646   & 7.71235   & 6.82347   & 6.26265   & 5.87653   & 5.59447 \\
    $\mu_1^e$           &       & 4.24548   & 4.03936   & 3.89609   & 3.79078   & 3.71011   & 3.64635 \\
    $\mu_2^e$           &       & 0.70991   & 0.67544   & 0.65148   & 0.63387   & 0.620385  & 0.60972 \\
    \hline
\end{tabular}

\begin{tabular}{|c||c|c|c|c|c|c|c|}
    \hline
    ${D_v/D_u}$   & 1.1       & 1.2       & 1.3       & 1.4       & 1.5       & 1.6 \\
    \hline\hline
    $\rho_p$            & 14.38188  & 11.51761  & 10.99734  & 10.58334  & 10.24607  & ...  \\
    \hline
    $\mu_e$             & 4.22776   & 4.04759   & 4.00670   & 3.97186   & 3.94181   & \\
    \hline\hline
    $\rho_f$            & 5.87653   & 5.20997   & 5.07306   & 4.96011   & 4.86535   &  \\
    $\mu_1^e$           & 3.71011   & 3.55200   & 3.51612   & 3.48554   & 3.45917   &  \\
    $\mu_2^e$           & 0.62038   & 0.59395   & 0.58795   & 0.58283   & 0.57842   &  \\
    \hline
\end{tabular}
\caption{Numerical values (rounded to 5th decimal place) of the subcritical
  pitchfork bifurcation point $\rho_p$ for the symmetric branch and
  the two fold bifurcation points $\rho_f$ along the asymmetric branches.
  Bifurcation values for the symmetric ($\mu_e$) and one of the
  asymmetric ($\mu_1^e$, $\mu_2^e$) solution branches are listed. As the ratio
  ${D_v/D_u}$ decreases the range of $\rho$ where hysteresis occurs
  increases. Parameters: $D_u=1,  \sigma_u = \sigma_v=1, L=1$, and
  $\beta_u=0.1$.}
\label{tab:2-cell GM hysteresis}
\end{table}

\begin{table}[ht]
\centering
\begin{tabular}{|c||c|c|c|c|c|c|c|}
    \hline
    $\beta_u$             & 0.05      & 0.1       & 0.15      & 0.2       & 0.25      & 0.3       & 0.35\\
    \hline\hline
    $\rho_p$            & 20.44213  & 13.09728  & 11.42015  & 11.59667  & 13.33412  & 17.91935  & 33.18867 \\
    \hline
    $\mu_e$             & 7.05336   & 4.15511   & 3.21380   & 2.76056   & 2.50167   & 2.33930   & 2.23156 \\
    \hline\hline
    $\rho_f$            & 4.69415   & 5.59447   & 6.82631   & 8.63523   & 11.58010  & 17.27125   & - \\
    $\mu_1^e$           & 3.82650   & 3.64635   & 3.45845   & 3.25882   & 3.03717   & 2.76171   & - \\
    $\mu_2^e$           & 0.32048   & 0.60972   & 0.87895   & 1.13950   & 1.40694   & 1.71692  & - \\
    \hline
\end{tabular}
\caption{Numerical values (rounded to 5th decimal place) of the subcritical (or
  supercritical) pitchfork bifurcation point $\rho_p$, the
  two fold bifurcation points $\rho_f$, and the corresponding values
  for the symmetric ($\mu_e$) and one of the asymmetric ($\mu_1^e$,
  $\mu_2^e$) solution branches. As $\beta_u$ increases from
  $0.05$, the range of $\rho$ where hysteresis occurs decreases, until
  a supercritical pitchfork bifurcation arises when $\beta_u\approx
  0.35$. Parameters: $D_u=D_v=1, \sigma_u =\sigma_v=1, L=1$.}
\label{tab:2-cell GM hysteresis iota}
\end{table}

\begin{figure}[]
    \centering
	\begin{subfigure}[b]{.45\textwidth}
	    \begin{subfigure}[b]{1.\textwidth}
	        \centering
	        \includegraphics[width=1.\linewidth]{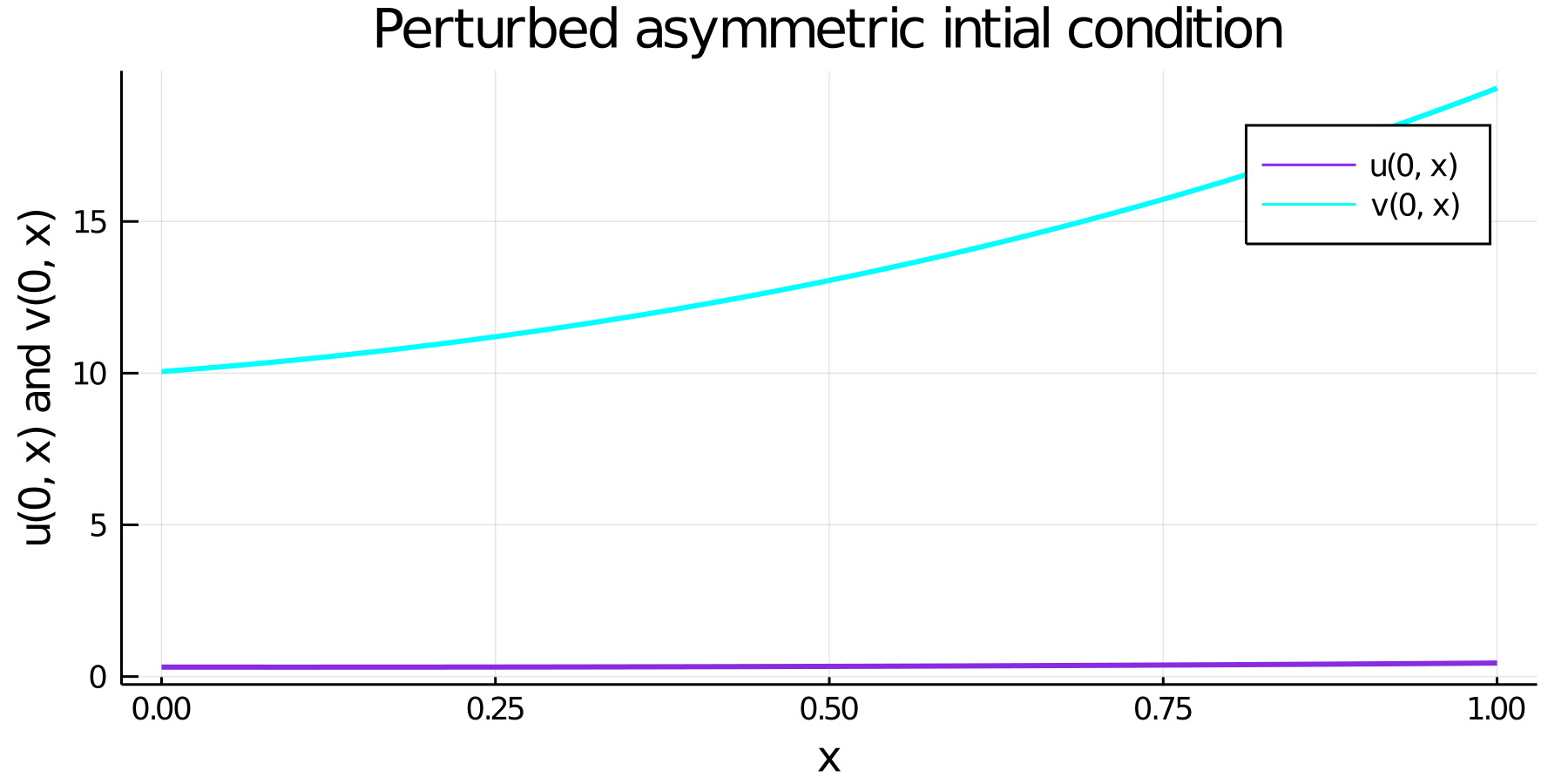}
	    \end{subfigure}
	    \begin{subfigure}[b]{1.\textwidth}
	        \centering
	        \includegraphics[width=1.\linewidth]{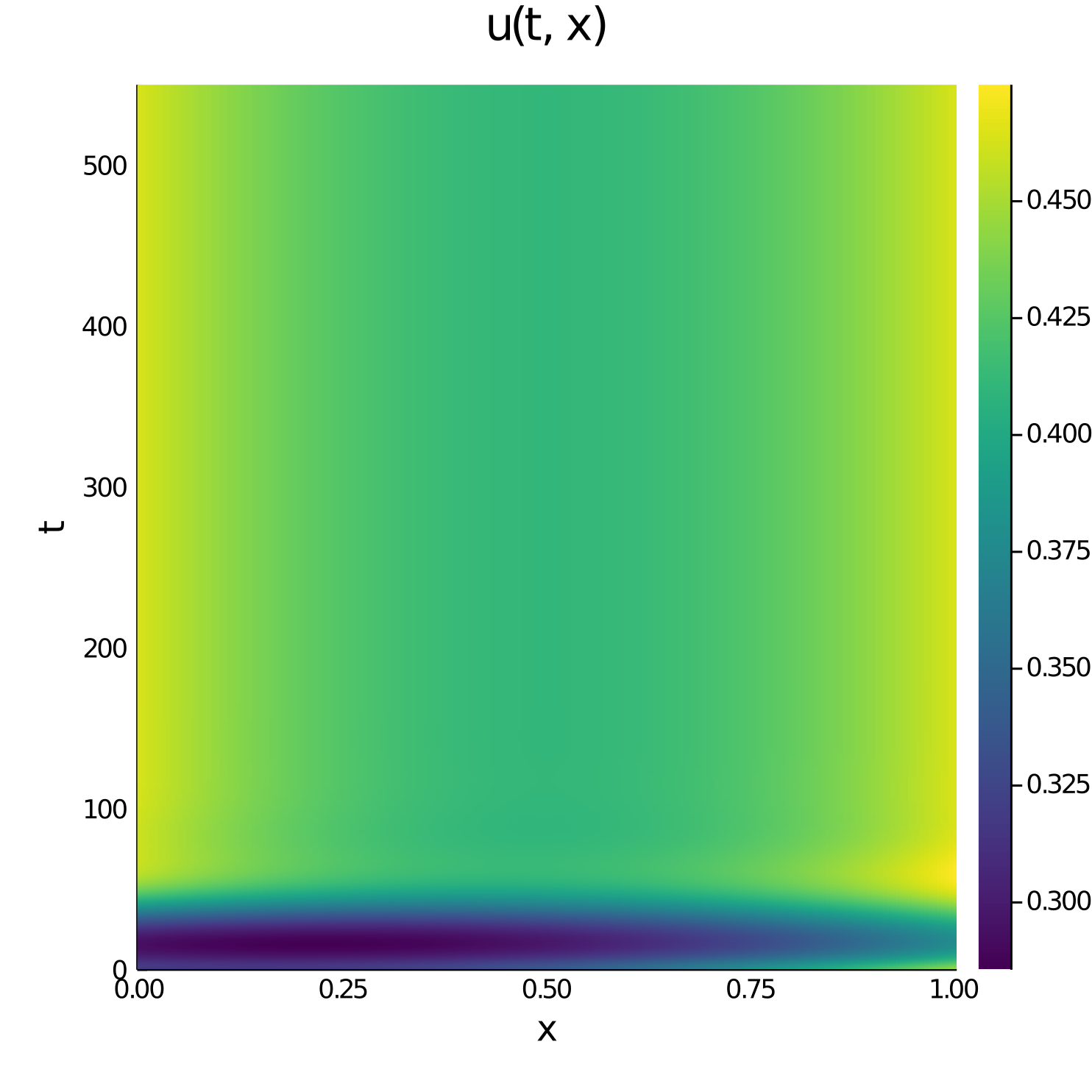}
	    \end{subfigure}
	    \begin{subfigure}[b]{1.\textwidth}
    		\centering
			\includegraphics[width=1.\linewidth]{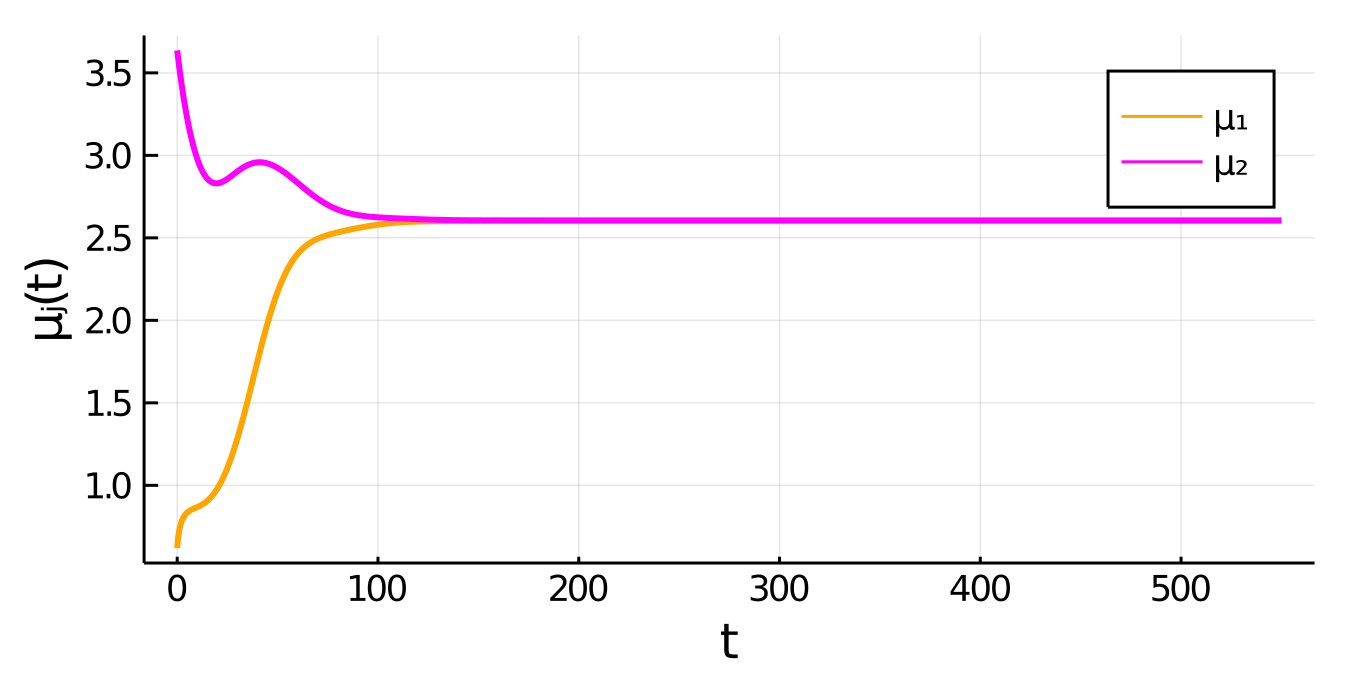}  
		\end{subfigure}
		\begin{subfigure}[b]{1.\textwidth}
			\centering
			\includegraphics[width=1.\linewidth]{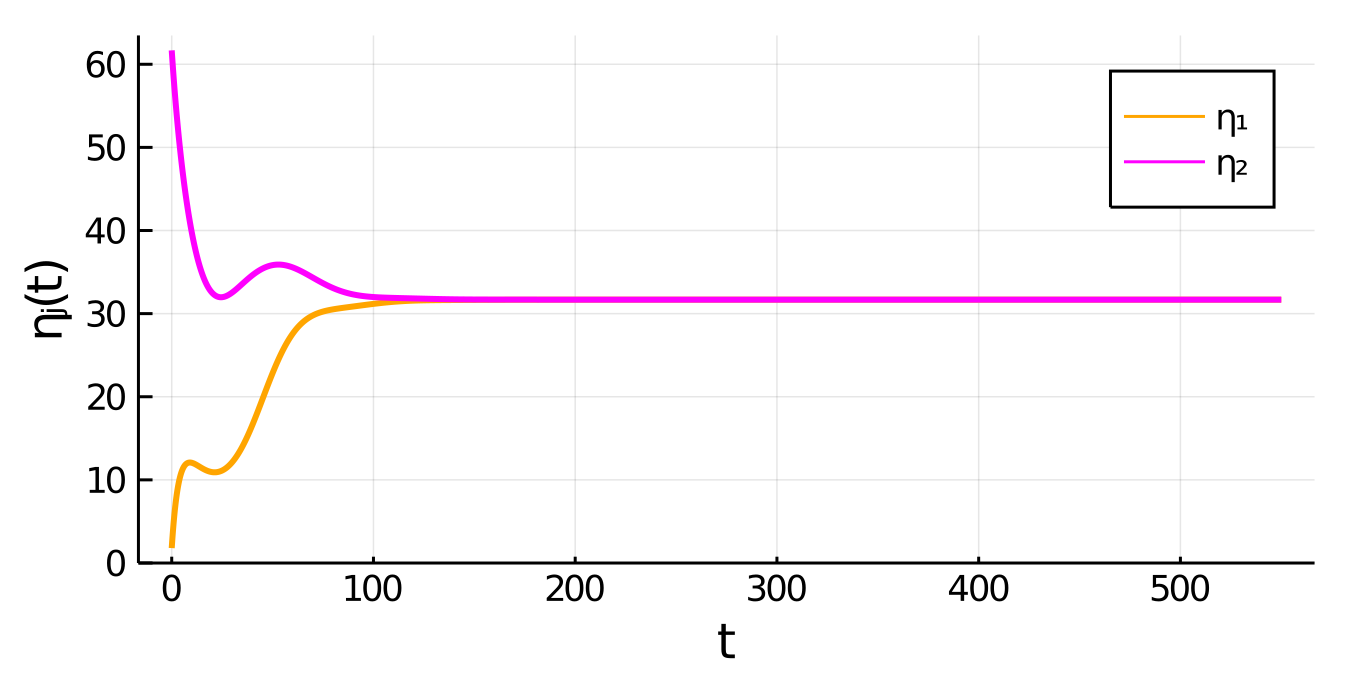}  
		\end{subfigure}
		\begin{subfigure}[b]{1.\textwidth}
			\centering
			\includegraphics[width=1.\linewidth]{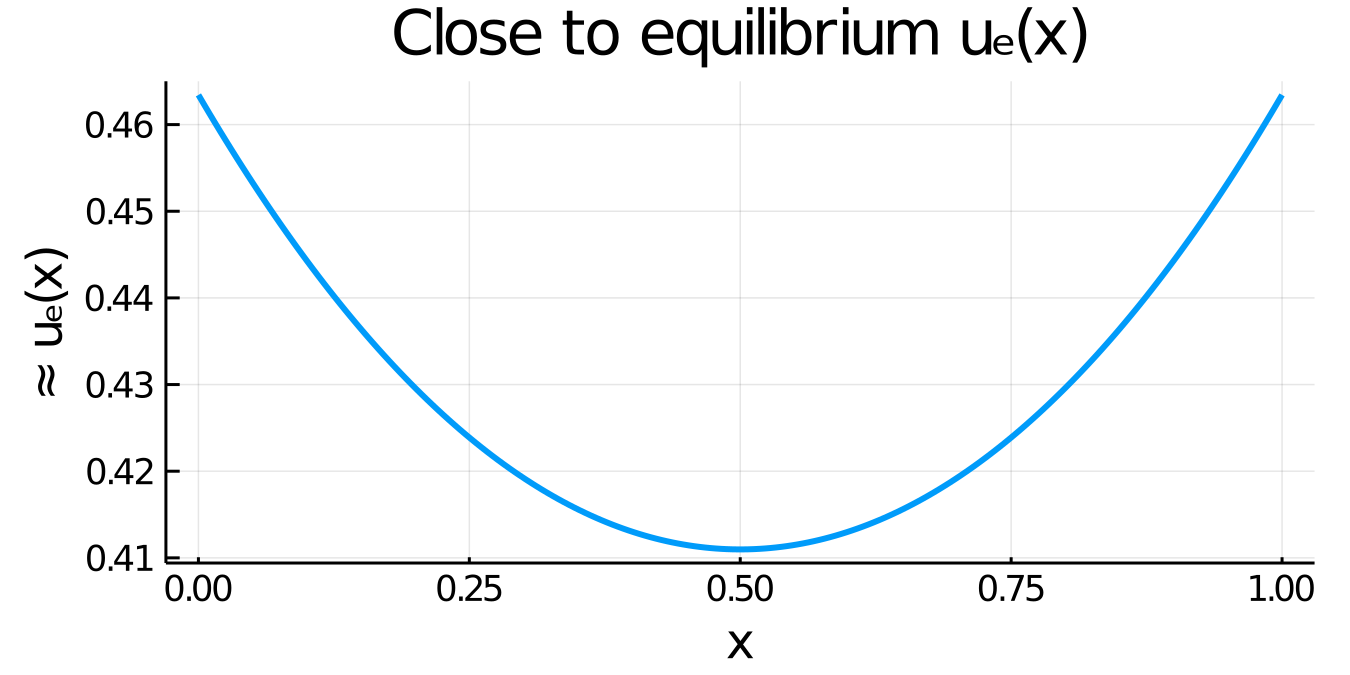}  
		\end{subfigure}
	\end{subfigure}
	\begin{subfigure}[b]{.45\textwidth}
	    \begin{subfigure}[b]{1.\textwidth}
	        \centering
	        \includegraphics[width=1.\linewidth]{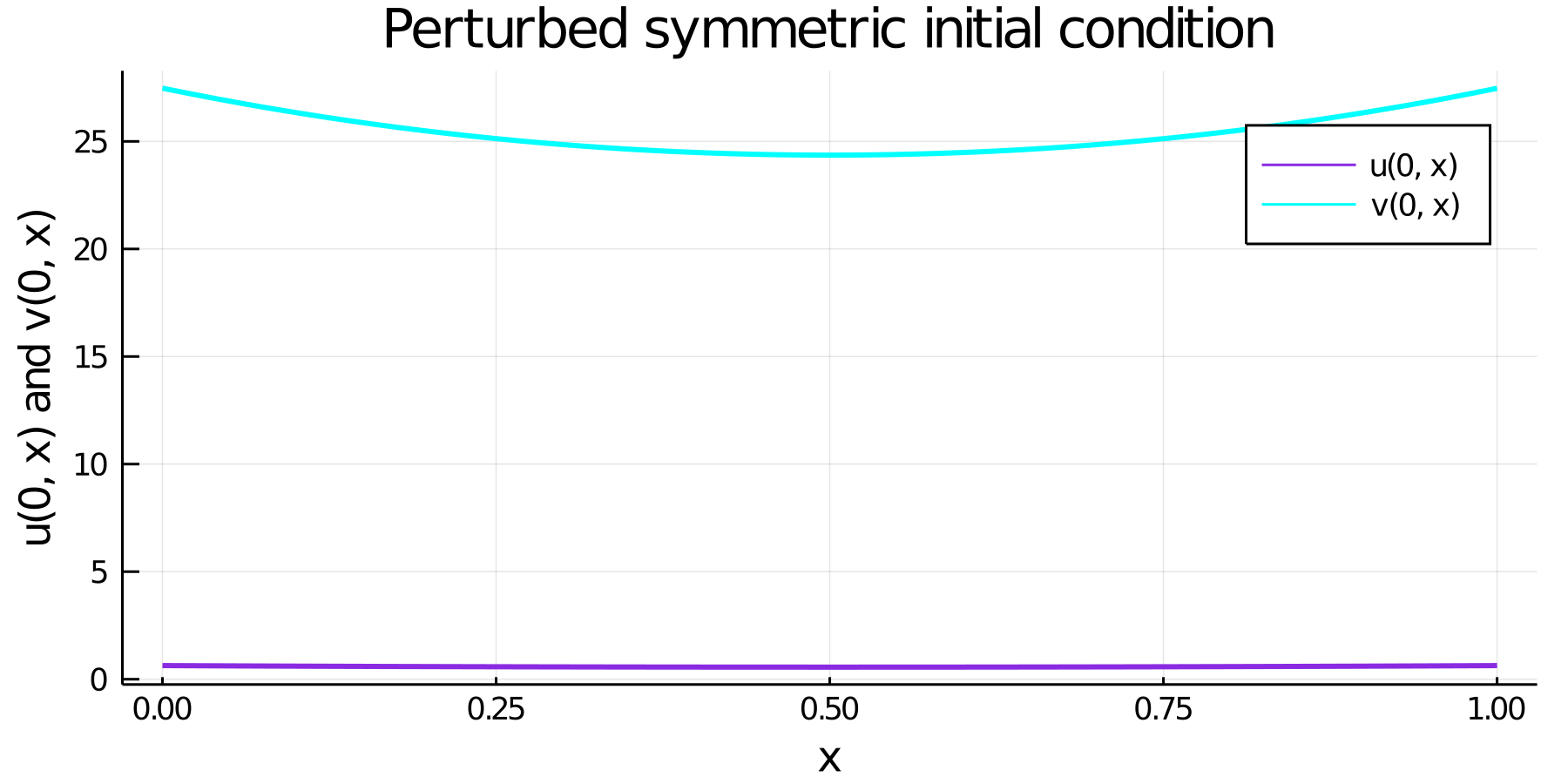}
	    \end{subfigure}
	    \begin{subfigure}[b]{1.\textwidth}
	        \centering
	        \includegraphics[width=1.\linewidth]{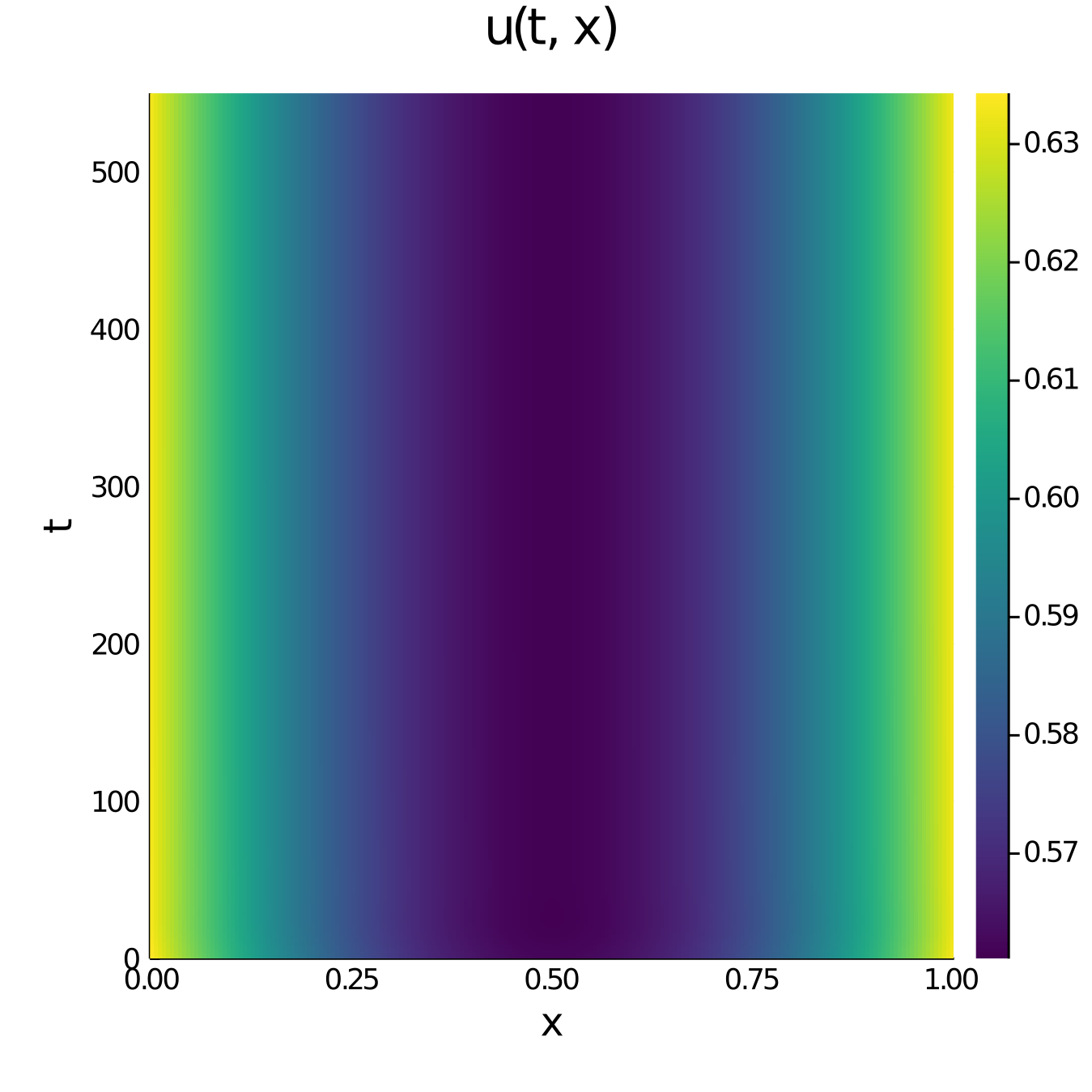}
	    \end{subfigure}
	    \begin{subfigure}[b]{1.\textwidth}
    		\centering
			\includegraphics[width=1.\linewidth]{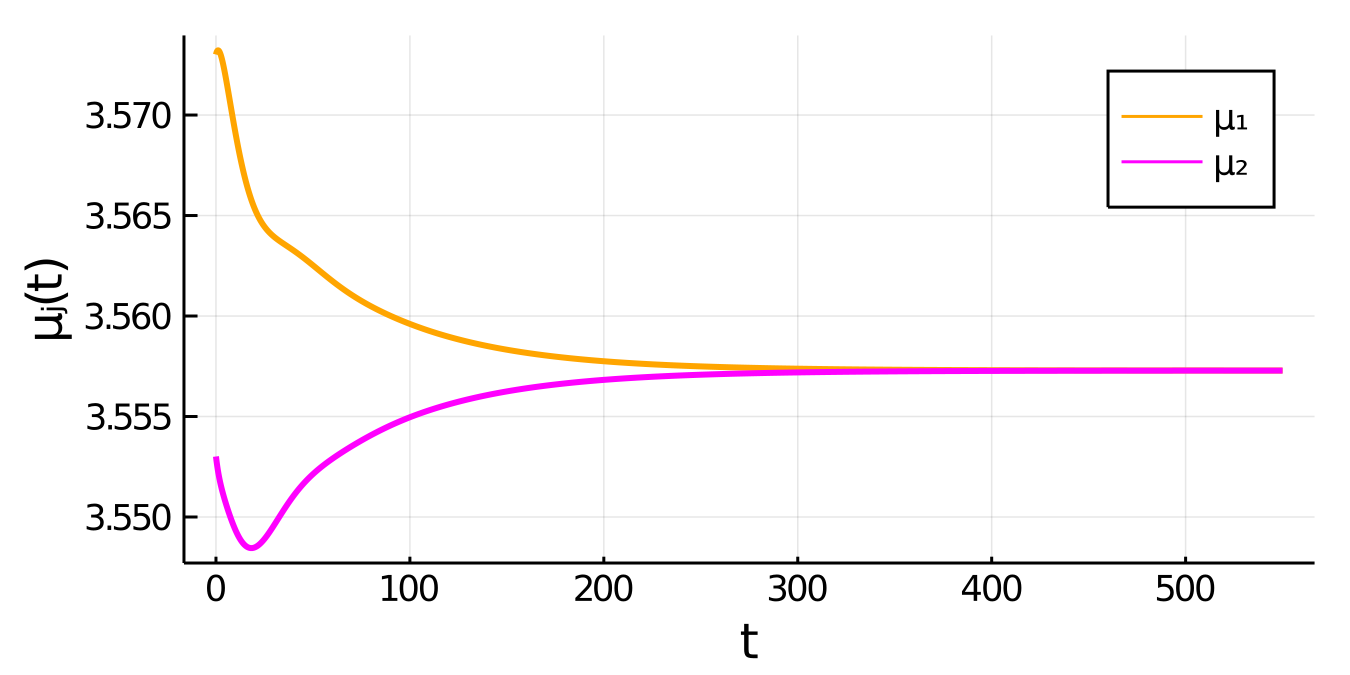}  
		\end{subfigure}
		\begin{subfigure}[b]{1.\textwidth}
			\centering
			\includegraphics[width=1.\linewidth]{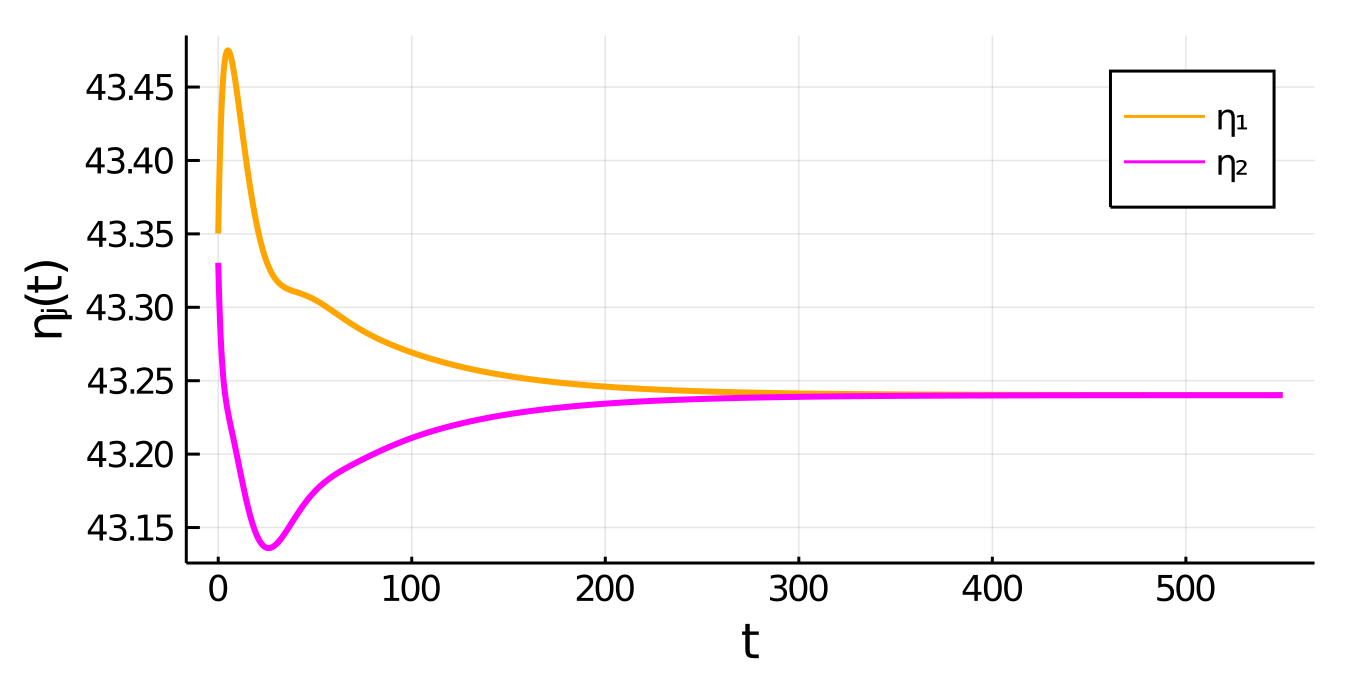}  
		\end{subfigure}
		\begin{subfigure}[b]{1.\textwidth}
			\centering
			\includegraphics[width=1.\linewidth]{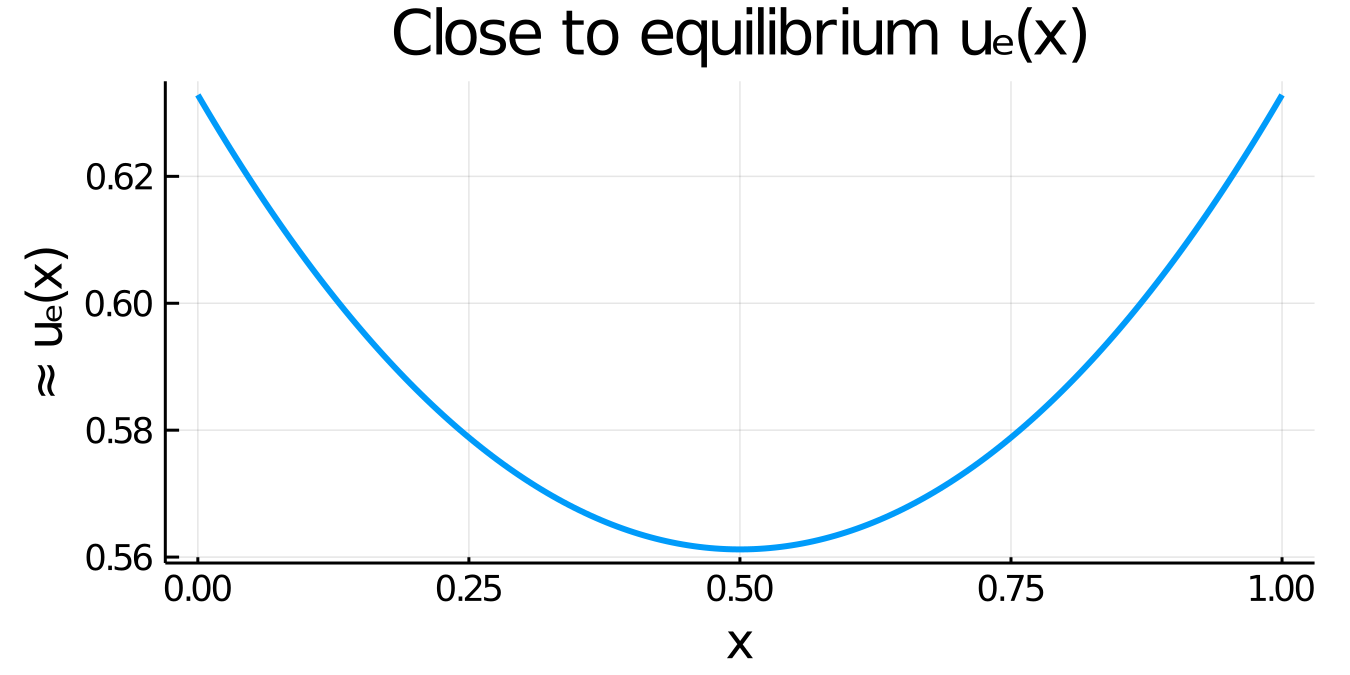}  
		\end{subfigure}
	\end{subfigure}
	\caption{Full numerical computations of (\ref{s:full}) by the
          CN-RK4 IMEX method of Appendix \ref{CN-RK4 IMEX} for GM
          kinetics. Left: convergence to the symmetric branch for
          $\rho=4$ starting close to the symmetric branch. Right:
          convergence to the symmetric branch for $\rho=8$ and
          starting near the symmetric branch. The remaining parameter
          values are as in the caption to
          Fig.~\ref{fig:bifdiagrhoGMiota0p1}.}
	\label{fig:symmhysteresis2cells}
\end{figure}

\subsection{Compartments with Rauch-Millonas kinetics}

We now briefly show that linearly stable asymmetric patterns can occur for
a compartmental-reaction diffusion model using the generic intracellular
reaction kinetics proposed in \cite{rauch} 
\begin{equation}\label{ss:rauch}
    \begin{array}{rclcl}
      \dot{\mu} &=& f(\mu,\eta) := c_u - q_u \mu +
     \frac{\alpha_1^u \mu}{\gamma_1^u+\mu} -
      \frac{\alpha_2^u \mu\eta}{\gamma_2^u+\mu} \\
      \dot{\eta} &=& g(\mu,\eta) := c_v + w_v\mu - q_v\eta \,, 
    \end{array}
  \end{equation}
  where we identify $g_1(\mu)=c_v+w_v\mu$ and $g_2=q_v$. This kinetic model
  was proposed in \cite{rauch} as a simplified universal description
  for signal transduction kinetics in cells.

  Owing to the special form of $g(\mu,\eta)$, we can readily implement
  the steady-state theory of \S \ref{2cellcoupledsystem} when the kinetics
  (\ref{ss:rauch}) are restricted to the domain boundaries. Parameters are
  chosen as in \cite{rauch} to ensure that the uncoupled dynamics
  (\ref{ss:rauch}) has a stable equilibrium point.

\begin{figure}[htbp]
  \centering
  \includegraphics[width=0.48\textwidth]{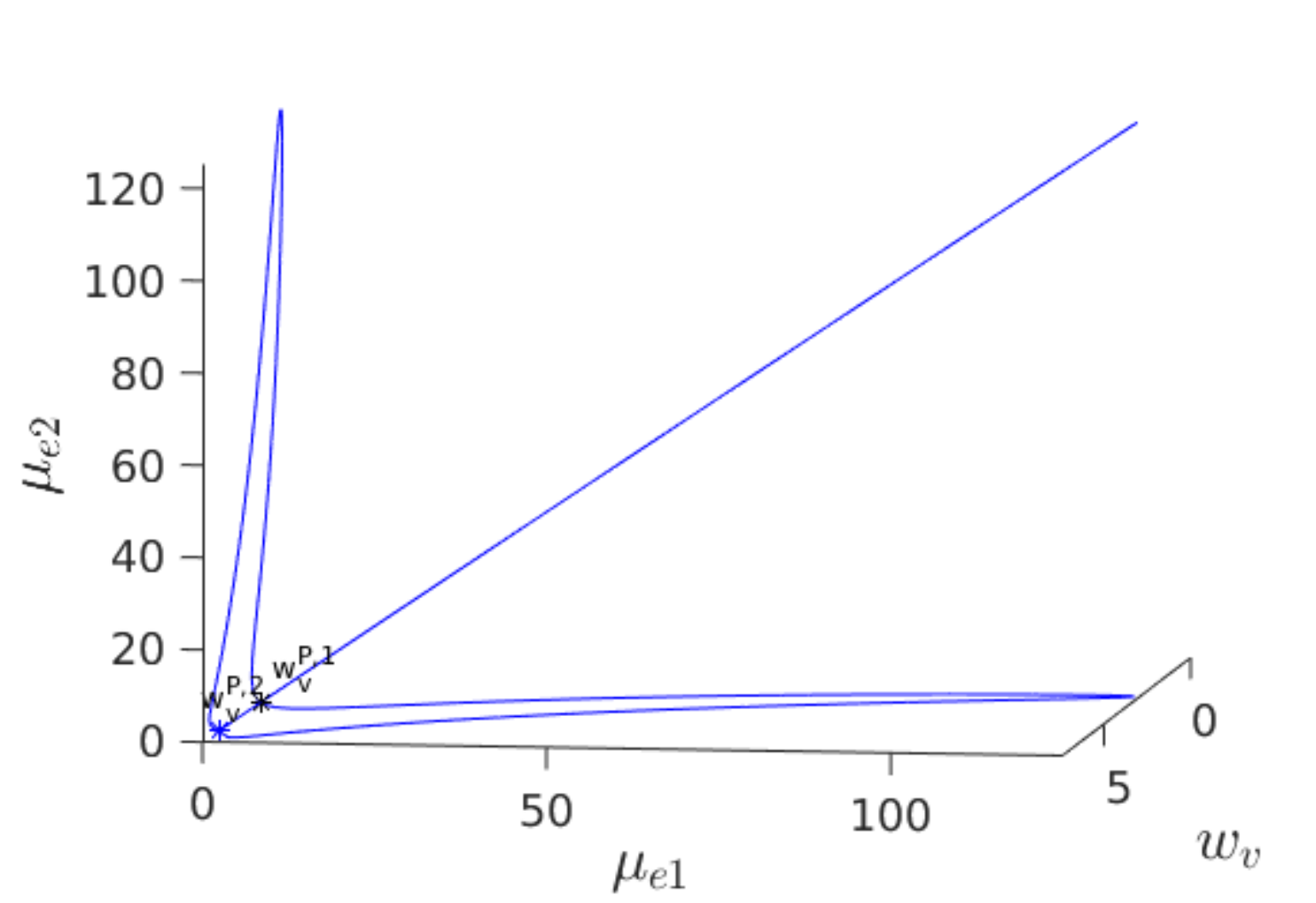}
  \includegraphics[width=0.48\textwidth]{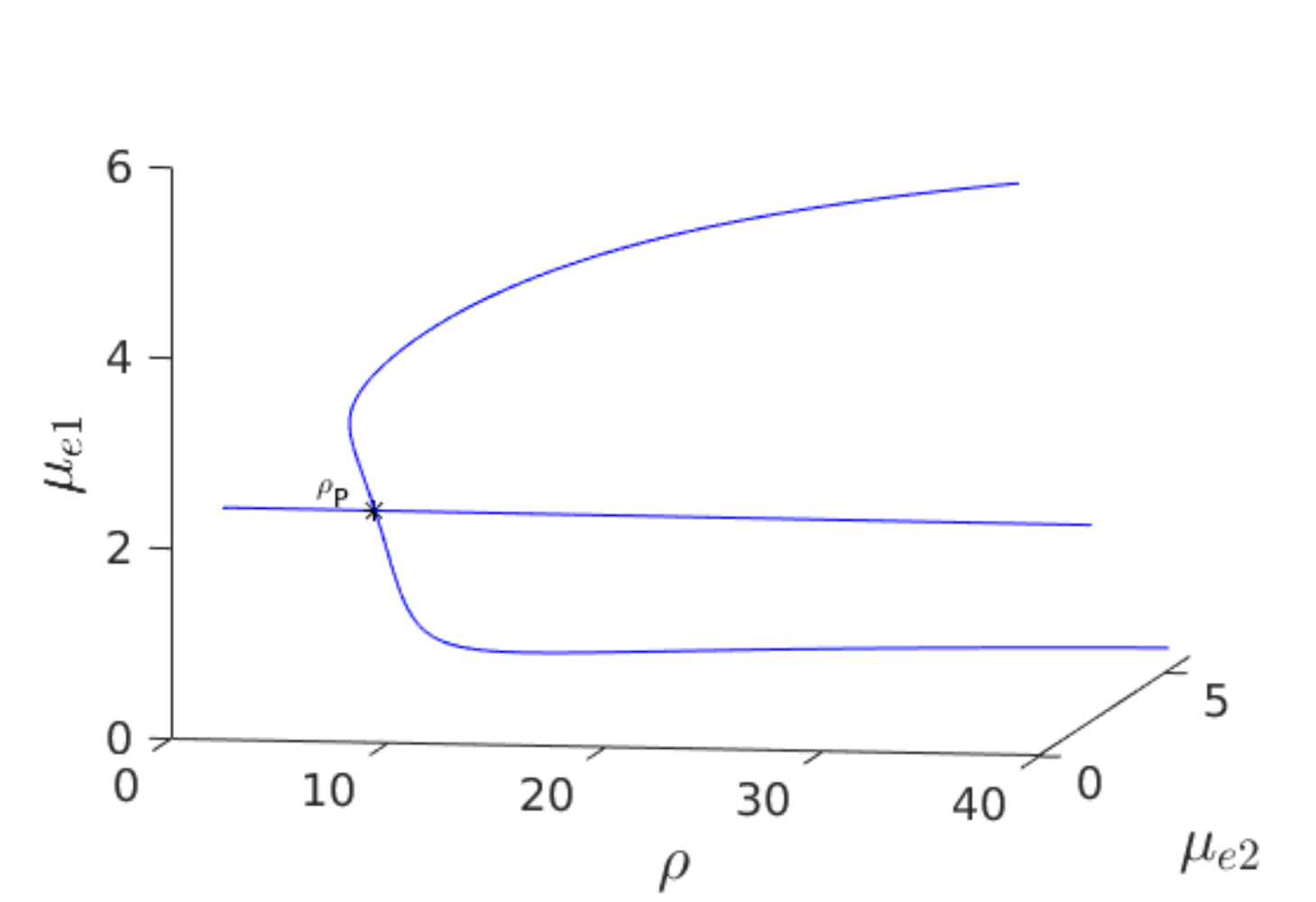}
  \caption{{3-D bifurcation diagrams using MatCont \cite{matcont} for the
    two-cell compartmental-reaction diffusion system with
    Rauch-Millonas kinetics (\ref{ss:rauch}):} Left: Plot of
    $\mu_1^{e}$ showing that asymmetric steady-states occur inside the
    degenerate pitchfork bubble delimited by
    $w_v^{P,1}\approx 6.67610$ and $w_v^{P,2}\approx 7.28516$ when
    $\rho={\beta_v/\beta_u}= 7$. Right: Supercritical pitchfork
    bifurcation from the symmetric branch occurs when
    $w_v^{P,2}\approx 7.28516$.  Linearly stable asymmetric branches
    occur past this threshold in $\rho$.  Parameters:
    $D_u = D_v= 1, \sigma_u=\sigma_v =0.01,L=1,c_u=c_v=1, q_u={1/100},
    q_v=7, \alpha_1^u=600, \alpha_2^u=6, \gamma_1^u=100, \gamma_2^u=
    {1/10}$, and $\beta_u=0.3$.} \label{fig:pbubbleandrhobifRM}
\end{figure}

\begin{figure}[htbp]
	\begin{subfigure}[b]{.49\textwidth}
	    \begin{subfigure}[b]{1.\textwidth}
    		\centering
			\includegraphics[width=1.\linewidth]{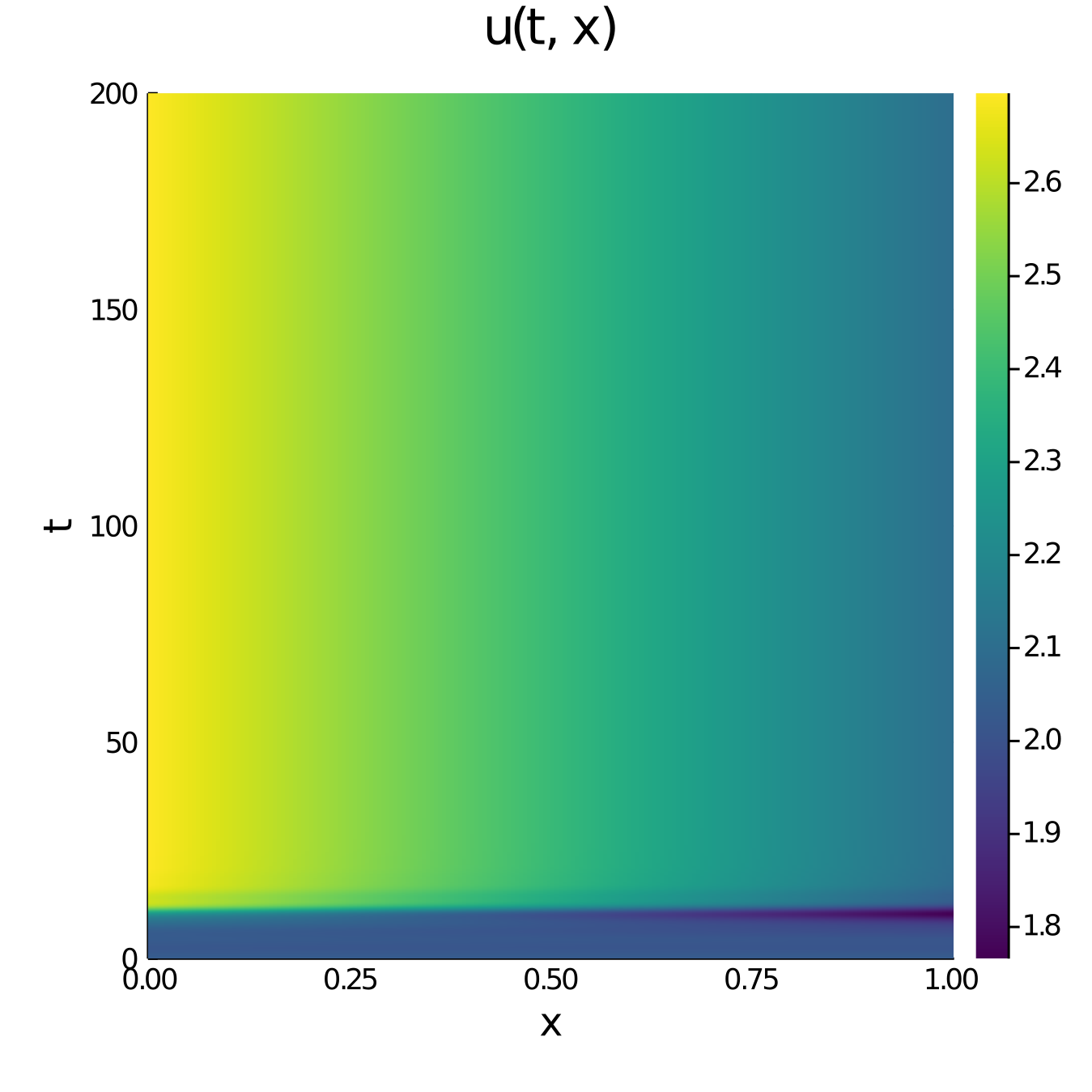}  
		\end{subfigure}
		\begin{subfigure}[b]{1.\textwidth}
    		\centering
			\includegraphics[width=1.\linewidth]{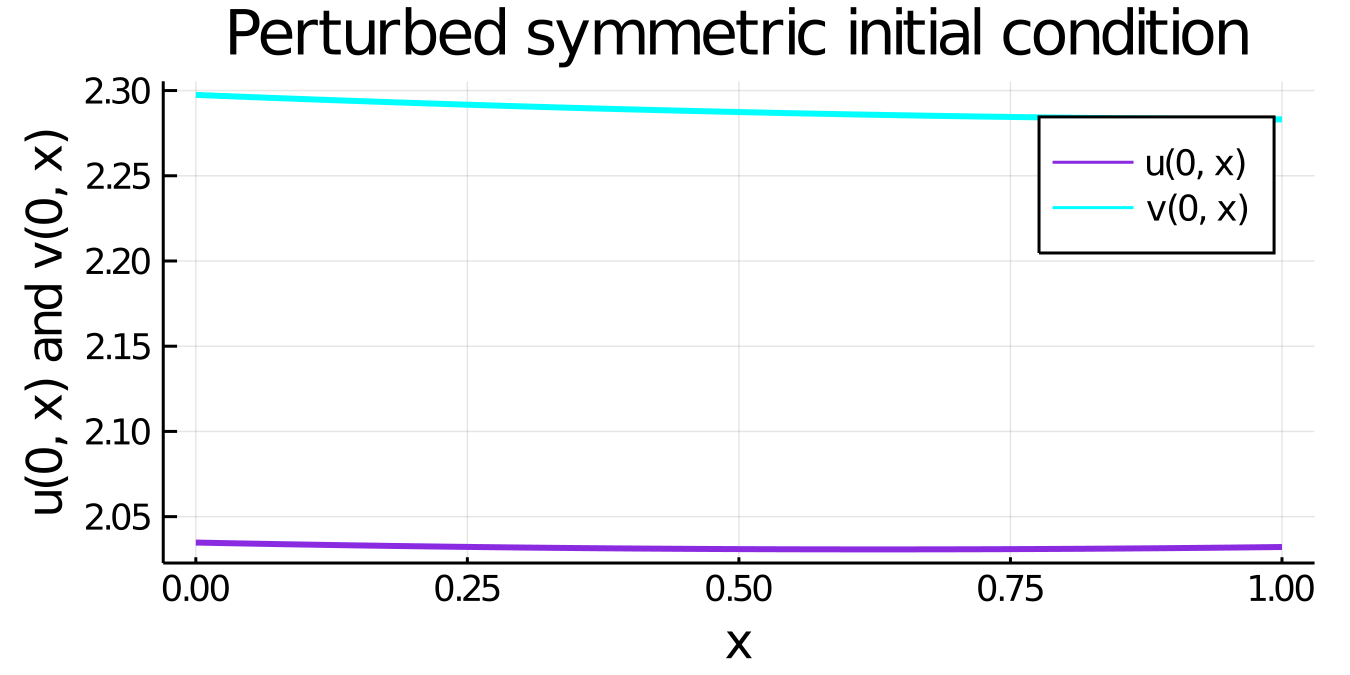}  
		\end{subfigure}
	\end{subfigure}
	\begin{subfigure}[b]{.50\textwidth}
		\begin{subfigure}[b]{1.\textwidth}
    		\centering
			\includegraphics[width=1.\linewidth]{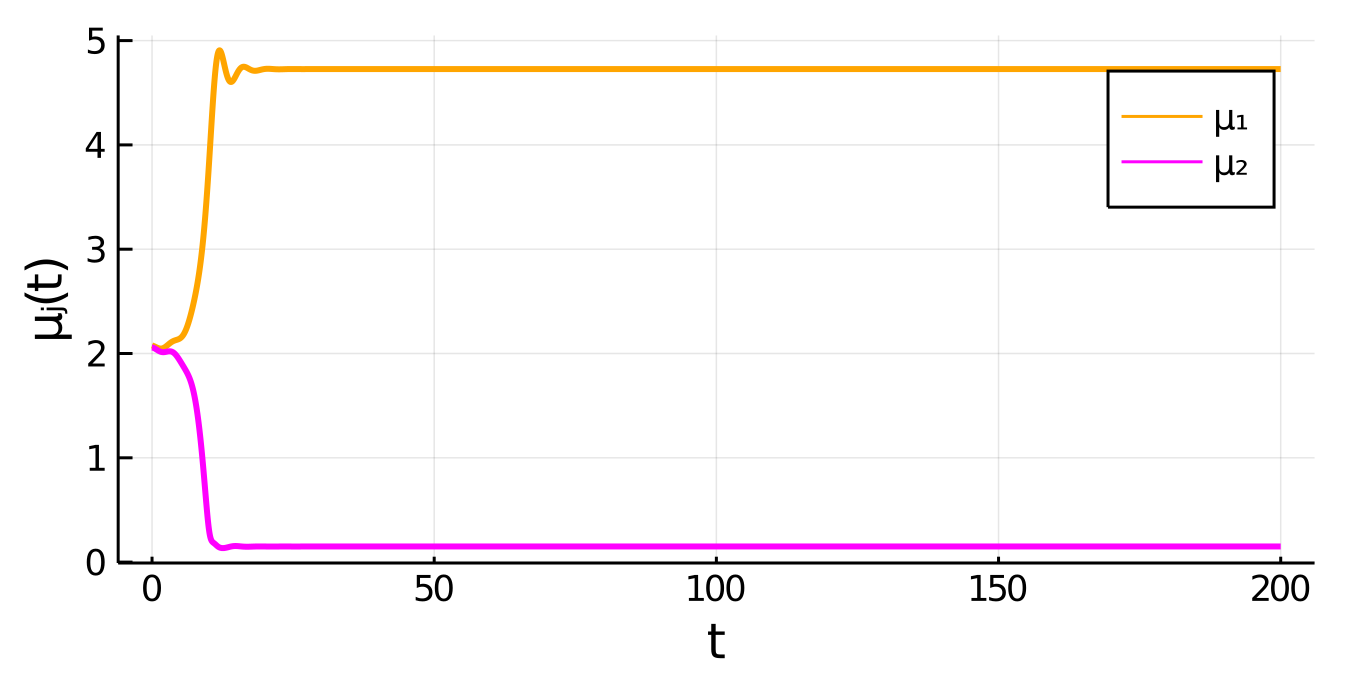}  
		\end{subfigure}
		\begin{subfigure}[b]{1.\textwidth}
			\centering
			\includegraphics[width=1.\linewidth]{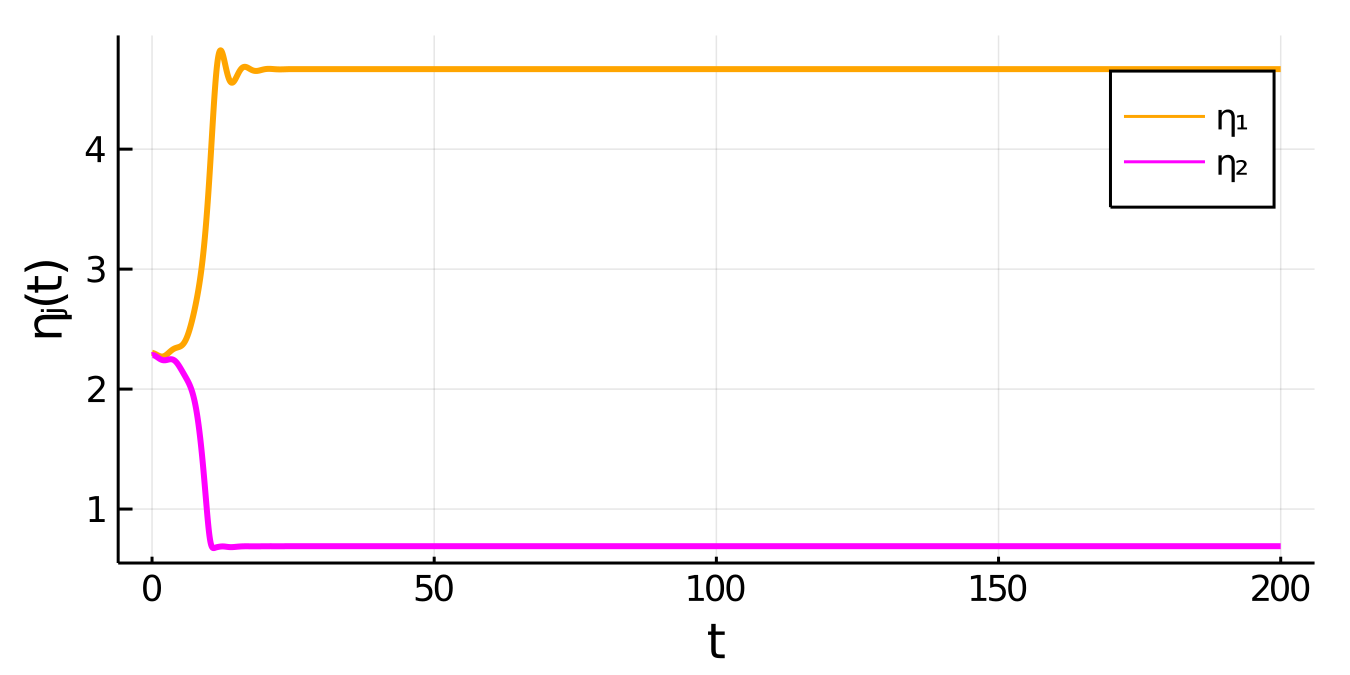}  
		\end{subfigure}
		\begin{subfigure}[b]{1.\textwidth}
			\centering
			\includegraphics[width=1.\linewidth]{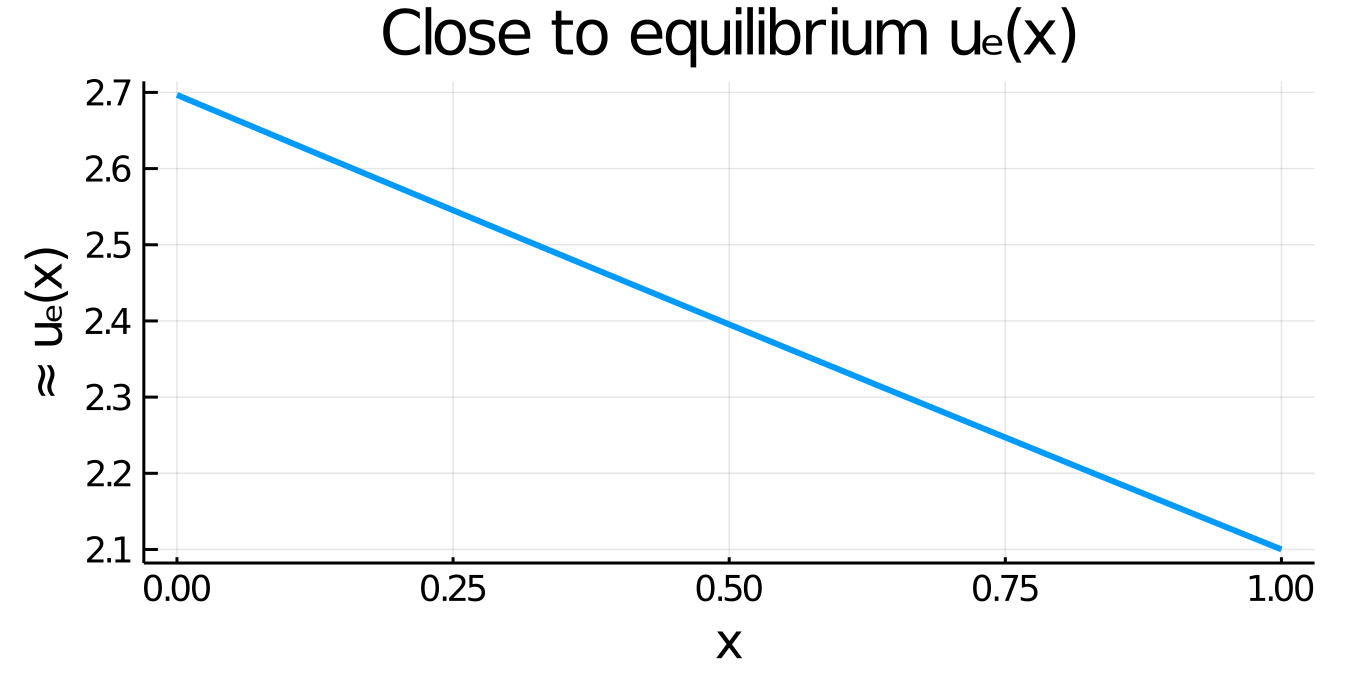}  
		\end{subfigure}
	\end{subfigure}
	\caption{Full numerical computations of (\ref{s:full}) by the
          CN-RK4 IMEX method of Appendix \ref{CN-RK4 IMEX} for
          Rauch-Millonas kinetics (\ref{ss:rauch}). For $\rho=15$ and
          $w_v=w_v^{P,2}\approx 7.28516$, and with an initial
          condition near the unstable symmetric branch (see
          Fig.~\ref{fig:pbubbleandrhobifRM}(right)), the
          time-dependent solution converges to a stable asymmetric
          pattern. Parameters are as in the caption of
          Fig.~\ref{fig:pbubbleandrhobifRM}.}
	\label{fig:RM2cell}
\end{figure}

  With $\omega_v$ or the binding rate ratio
  $\rho={\beta_v/\beta_u}$ as bifurcation parameters, we use MatCont
  \cite{matcont} to numerically compute steady-states of the bulk-cell
  model. With the other parameters similar to those in \cite{rauch},
  in Fig.~\ref{fig:pbubbleandrhobifRM}(right) we observe that
  asymmetric patterns do emerge from the symmetric solution branch as
  $\rho$ exceeds a threshold. Pitchfork bifurcations in $w_v$ also
  occur (see Fig.~\ref{fig:pbubbleandrhobifRM}(left)), and they
  correspond as before to a zero-eigenvalue
  crossing of the anti-phase perturbation of the symmetric
  steady-state. The full-time dependent numerical computations shown
  in Fig.~\ref{fig:RM2cell} confirms the existence of stable
  asymmetric steady-state patterns.

\section{A Periodic Compartmental-Reaction Diffusion Model} \label{ring-cell system}

We now extend the two-compartment model to allow for a periodic array of
identical reaction compartments in 1-D. Our goal is to investigate
symmetry-breaking behavior from a symmetric steady-state for this
extended model. For a periodic chain of compartments there can be
multiple possible modes of instability of the symmetric steady-state.

\begin{figure}[H]
\centering
    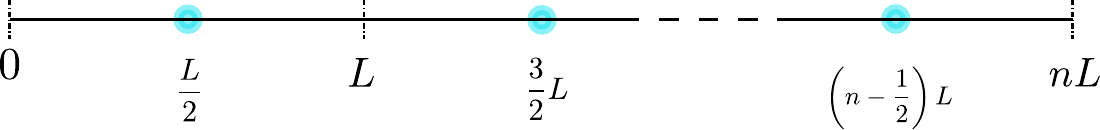
    \caption{The periodic compartmental-reaction
      diffusion model on $[0,nL]$ with compartments centered at the
      blue dots. Two bulk diffusing species provide the
      inter-compartment coupling.}
    \label{fig:schem_per}
\end{figure}

In our formulation, we consider a periodic chain of a {fixed
  length $nL, n\in\bN\backslash\{0,1\}, L>0,$ with compartments
  centered at $x_j = \frac{(2j-1)}{2}L$ for $j\in\{1,...,n\}$,} so
that for $x\in \bR\backslash (nL\cdot \bZ)$ we have the following
bulk-cell system {on the domain $0<x<nL$ (see
  Fig.~\ref{fig:schem_per} for a schematic):}
\begin{subequations}\label{s:pfull}
\begin{eqnarray}
	\text{bulk} &&
	\begin{cases}
          \partial_t u = D_u\; \partial_{xx} u  - \sigma_u u\,, \quad
          &t\in(0,\infty)\,, \;x\in(0,nL)\backslash \bigcup_{j=1}^n\{x_j\}
          \qquad \\
          \partial_t v = D_v\; \partial_{xx} v - \sigma_v v, \quad
          &t\in(0,\infty),
          \;x\in(0,nL)\backslash \bigcup_{j=1}^n\{x_j\}
	\end{cases} \label{s:pfull_1} \\
	\text{bulk boundary} &&
	\begin{cases}
          u(t,0) = u(t,nL)\,, &v(t,0) = v(t,nL) \qquad
          \text{(periodic BC)} \\
          \partial_x u(t,0) = \partial_x u(t,nL)\,, &\partial_x v(t,0) =
          \partial_x v(t,nL)
	\end{cases} \label{s:pfull_2} \\
	\text{reaction fluxes} &&
	\begin{cases}
          [D_u \partial_xu]|_{x=x_j}&=\beta_u\;(u(t,x_j)-\mu_j(t)) \; \;
          \qquad\text{(cell jump conditions)}\\
		[D_v \partial_xv]|_{x=x_j}&=\beta_v\;(v(t,x_j)-\eta_j(t))
	\end{cases} \label{s:pfull_3} \\
	\text{compartments} &&
	\begin{cases}
          \dot{\mu}_j = f(\mu_j,\eta_j) + [D_u \partial_xu]|_{x=x_j}
          \qquad \qquad\text{(reaction kinetics at } x=x_j) \\
		\dot{\eta}_j = g(\mu_j,\eta_j) + [D_v \partial_xv]|_{x=x_j}\,,
	\end{cases} \label{s:pfull_4}
\end{eqnarray}
\end{subequations}
for $j=\lbrace{1,\ldots,n\rbrace}$. Here for any function $\mathcal{F}$ we have
defined $[\mathcal{F}]|_{x=\tilde{x}} := \mathcal{F}(\tilde{x}^+) -
\mathcal{F}(\tilde{x}^-)$.

\subsection{Compartments with general reaction kinetics}

We will first construct a symmetric steady-state solution for (\ref{s:pfull})
and we will analyze the linear stability of this steady-state by using a
Floquet-based approach. Owing to the complexity of (\ref{s:pfull}), we will
not construct asymmetric steady-states of (\ref{s:pfull}).

\subsubsection{Symmetric steady-state and the linear stability
  problem}\label{pf:stab_ana}

To construct a symmetric steady-state $u_e$, $v_e$ for (\ref{s:pfull})
we need only consider the \emph{fundamental domain} $[0,L]$ upon which
we impose periodic boundary conditions
\begin{equation}\label{s:per_fun}
    u_e(0)=u_e(L)\,, \quad v_e(0)=v_e(L) \,, \quad
  \partial_x u_e(0) = \partial_x u_e(L), \quad \partial_x v_e(0)=
  \partial_x v_e(L) \,.
\end{equation}
The symmetric steady-state is obtained by a periodic extension to the
entire chain.

To represent the steady-states in the bulk, we introduce the periodic Green
function $G_\omega(x)$.
\begin{definition}
  The periodic Green function on $[0,L]$ with singularity at
  $x=\frac{L}{2}$ is the solution of
\begin{subequations}\label{p:green}
\begin{align}
  \partial_{xx} G_{\omega} - \omega^2 G_\omega &= -\delta(x-\frac{L}{2})\,,
    \,\,\, x\in[0,L]\,; \quad G_{\omega}(0) = G_{\omega}(L)\,, \quad
   \partial_{x} G_{\omega}(0) = \partial_x G_{\omega}(L)\,, \label{p:green_a}
  \\
  G_\omega(\frac{L}{2}^+) &= G_\omega(\frac{L}{2}^-) \,, \quad
  \partial_{x}G_{\omega}(\frac{L}{2}^+) -\partial_{x}G_{\omega}(\frac{L}{2}^-)
                              = -1\,. \label{p:green_b}
\end{align}                  
\end{subequations}            
\end{definition}
The solution to (\ref{p:green}) is readily calculated as
\begin{equation}\label{p:green_sol}
    G_\omega(x) = 
    \begin{cases}
      \frac{1}{2\omega} \coth\left(\frac{\omega L}{2}\right)
      \cosh(\omega(\frac{L}{2}-x))
      - \frac{1}{2\omega}\sinh(\omega(\frac{L}{2}-x)\,, \quad
      &x\in [0,\frac{L}{2}], \\
      \frac{1}{2\omega} \coth\left(\frac{\omega L}{2}\right)
      \cosh(\omega(\frac{L}{2}-x))
      + \frac{1}{2\omega}\sinh(\omega(\frac{L}{2}-x))\,, \quad
      &x\in [\frac{L}{2},L]\,,
    \end{cases}
\end{equation}
where we identify that
$G_\omega\left({L/2}\right) =\frac{1}{2\omega}\coth\left({\omega L/2}\right)$.

Upon defining $\omega_u:=\sqrt{\sigma_u/D_u}$ and
$\omega_v:=\sqrt{\sigma_v/D_v}$ the bulk steady-state solutions
satisfying (\ref{s:pfull_1}) and (\ref{s:pfull_3}) on the fundamental
domain $[0,L]$ with periodic boundary conditions (\ref{s:per_fun}) are 
\begin{equation}\label{s:p_ueve}
    \begin{array}{rclcl}
      u_e(x) &=& c_1 \frac{ G_{\omega_u}(x)}{ G_{\omega_u}\left({L/2}\right)}
 \qquad \text{with} \qquad c_1 := \frac{\beta_u}
           {\beta_u + 2 D_u\omega_u \tanh\left({\omega_u L/2}\right)} \;\mu_e\,, \\
      v_e(x) &=& c_2 \frac{ G_{\omega_v}(x)}{ G_{\omega_v}\left({L/2}\right)}
  \qquad \text{with} \qquad c_2 := \frac{\beta_v}{\beta_v + 2 D_v\omega_v
 \tanh\left({\omega_v L/2}\right)}\;\eta_e\,.
    \end{array}
\end{equation}
Finally, upon substituting (\ref{s:p_ueve}) into the steady-state of
the intra-compartment dynamics (\ref{s:pfull_4}) we obtain that
$\mu_e$ and $\eta_e$ satisfy the nonlinear algebraic system
\begin{equation}\label{p:non_alg}
 f(\mu_e,\eta_e) = \frac{2 D_u \omega_u \beta_u}
 {\beta_u\coth\left({\omega_u L/2}\right) + 2D_u\omega_u} \;\mu_e\,,
 \qquad
 g(\mu_e,\eta_e) =  \frac{2 D_v \omega_v \beta_v}
        {\beta_v\coth\left({\omega_v L/2}\right) + 2D_v\omega_v} \;
\eta_e \,.
\end{equation}
In this way, for a given reaction-kinetics, we can determine
the symmetric equilibrium $(\mu_e,\eta_e)$ by a numerical root finding on
(\ref{p:non_alg}).

We now formulate and study the linear stability problem for the
symmetric steady-state solution branch. For the linear stability
analysis we perturb the symmetric steady-state by introducing
$u(t,x)=u_e(x)+ \phi(x) e^{\lambda t}, v(t,x)=v_e(x)+\psi(x)
e^{\lambda t}, \mu_j(t)=\mu_e + \xi_j e^{\lambda t}, \eta = \eta_e +
\zeta_j e^{\lambda t}$, where $|\phi|\ll 1, |\psi|\ll 1, |\xi_j|\ll 1$
and $|\zeta_j|\ll 1$ for $j\in\{1,2\}$, into (\ref{s:pfull}) and
linearizing. This yields the linearized eigenvalue problem
\begin{subequations}\label{p:lin_stab}
\begin{eqnarray}
	\text{bulk} &&
	\begin{cases}
          \partial_{xx}\phi - \Omega_u^2 \phi = 0\,, \quad
          &x\in(0,nL)\backslash \bigcup_{j=1}^n\{x_j\} \qquad \\
          \partial_{xx} \psi - \Omega_v^2 \psi = 0\,, \quad
          &x\in(0,nL)\backslash \bigcup_{j=1}^n\{x_j\}
	\end{cases} \\
	\text{bulk boundary} &&
	\begin{cases}
          \phi(0) = \phi(nL)\,, &\psi(0) = \psi(nL) \qquad \qquad 
          \text{(periodic BC)} \\
	    \partial_x\phi(0) = \partial_x \phi(nL)\,,
            &\partial_x\psi(0) = \partial_x\psi (nL)
	\end{cases} \\
	\text{reaction fluxes} &&
	\begin{cases}
          [D_u \partial_x\phi]|_{x=x_j}&=\beta_u\;(\phi(x_j)-\xi_j)
          \quad\qquad \text{(cell jump conditions)}\\
		[D_v \partial_x\psi]|_{x=x_j}&=\beta_v\;(\psi(x_j)-\zeta_j)
	\end{cases} \\
	\text{compartments} &&
	\begin{cases}
          \lambda \xi_j = \partial_\mu f_e \xi_j + \partial_\eta f_e \zeta_j +
          [D_u \partial_x\phi]|_{x=x_j}
          \quad\text{(reaction kinetics at } x=x_j) \\
          \lambda \zeta_j = \partial_\mu g_e \xi_j + \partial_\eta
          g_e \zeta_j + [D_v \partial_x]|_{x=x_j}
	\end{cases}\,,
\end{eqnarray}
\end{subequations}
for $j\in\{1,...,n\}$. Here we have defined
$\Omega_u:=\sqrt{(\lambda+\sigma_u)/D_u}$ and
$\Omega_v:=\sqrt{(\lambda+\sigma_v)/D_v}$.

To study this eigenvalue problem, it is convenient to use a Floquet-based
approach where we consider the \emph{fundamental cell problem} on $[0,L]$
and impose a Floquet-type boundary condition:
\begin{subequations}\label{p:flin_stab}
\begin{eqnarray}
	\text{bulk} &&
	\begin{cases}
          \partial_{xx}\phi - \Omega_u^2 \phi = 0\,, \quad
          &x\in(0,L)\backslash \{\frac{L}{2}\} \qquad \\
          \partial_{xx}\psi - \Omega_v^2 \psi = 0\,, \quad
          &x\in(0,L)\backslash \{\frac{L}{2}\}
	\end{cases} \label{p:flin_stab_1}\\
	\text{bulk boundary} &&
	\begin{cases}
          \phi(0) = Z\phi(L), &\psi(0) = Z\psi(L) \qquad\qquad\qquad
          \text{(Floquet BC)} \\
          \partial_x\phi(0) = Z\partial_x\phi(L)\,, &\partial_x\psi(0) = Z
          \partial_x\psi(L)
	\end{cases} \label{p:flin_stab_2}\\
	\text{reaction fluxes} &&
	\begin{cases}
          [D_u \partial_x\phi]|_{x=\frac{L}{2}}&=\beta_u\;
          (\phi(\frac{L}{2})-\xi) \;\;\quad \qquad \quad\text{(cell
            jump conditions)}\\
          [D_v\partial_x\psi]|_{x=\frac{L}{2}}&=\beta_v\;(\psi(\frac{L}{2})
          -\zeta)
	\end{cases} \label{p:flin_stab_3} \\ 
	\text{intracellular} &&
	\begin{cases}
          \lambda \xi = \partial_\mu f_e \xi + \partial_\eta f_e \zeta +
          [D_u \partial_x\phi]|_{x=\frac{L}{2}} \quad \;\;\quad
          \text{(reaction kinetics at } x=\frac{L}{2}) \\
          \lambda \zeta = \partial_\mu g_e \xi + \partial_\eta g_e \zeta +
          [D_v \partial_x\psi]|_{x=\frac{L}{2}} \,,
	\end{cases} \label{p:flin_stab_4}
\end{eqnarray}
\end{subequations}
for $Z\in \bC$. Upon using translational invariance, we get the chain
\begin{equation*}
    \phi(0)=Z\phi(L)=Z^2\phi(2L)=...=Z^n\phi(nL)\,,
\end{equation*}
so that in order to satisfy $\phi(0)=\phi(nL)$, we must have that  
\begin{equation}\label{p:unity}
  Z^n=1 \quad \Leftrightarrow \quad Z_k=e^{2\pi i k/n} \,,
  \quad \text{for}\,\,\, k\in\{0,...,n-1\} \,.
\end{equation}

{Such a Floquet-inspired approach to more readily derive the
  eigenvalue equation in this periodic setting has been used
  previously in other contexts. In \S 3 of \cite{kww} a similar
  approach was used to analyze the stability of a periodic pattern of
  localized hot-spots or ``spikes'' for a reaction-diffusion model of
  urban crime. It was also used in \S 5.2 of \cite{gou2017} to analyze
  the linear stability of a compartment-reaction diffusion model with
  a periodic array of compartments, but with only one-bulk diffusing
  species.}

\begin{remark}
  Since $Z=1$ when $k=0$, the mode $k=0$ represents in-phase perturbations of
  the symmetric steady-state. There are $n-1$ other possible modes of
  instability, which we shall refer to as ``asymmetric modes''. For
  the special case $n=2$, we have $Z_1=-1$ and $\phi(0)=-\phi(L)$, and
  we refer to this mode as the anti-phase mode.
\end{remark}

In order to derive an explicit formula for the eigenvalues $\lambda$ of
(\ref{p:flin_stab}) for a given $Z$, we must first introduce the
the \emph{quasi-periodic} Green function $G_{\Omega,Z}$.
\begin{definition}
  The \emph{quasi-periodic Green function} on the fundamental domain
  $[0,L]$ with singularity at $x=\frac{L}{2}$ is the solution of
\begin{subequations}\label{pf:green}
\begin{align}
  \partial_{xx} G_{\Omega,Z} &- \Omega^2 G_{\Omega,Z} = -\delta(x-\frac{L}{2})\,,
              \quad x\in[0,L]\,, \label{pf:green_a}\\
  G_{\Omega,Z}(0) &= Z G_{\Omega,Z}(L)\,,  \quad
  \partial_{x} G_{\Omega,Z}(0) = Z \partial_x G_{\Omega,Z}(L)\,, \label{pf:green_b}
  \\
  G_{\Omega,Z}(\frac{L}{2}^+) &= G_{\Omega,Z}(\frac{L}{2}^-) \,, \quad
  \partial_{x}G_{\Omega,Z}(\frac{L}{2}^+) -\partial_{x}G_{\Omega,Z}(\frac{L}{2}^-)
                              = -1\,. \label{pf:green_c}
\end{align}                  
\end{subequations}            
\end{definition}

Upon satisfying the continuity and jump conditions in (\ref{pf:green_c}), the
explicit solution to (\ref{pf:green_a}) can be written in terms of
two constants $A_{\Omega}$ and $B_{\Omega}$ as
\begin{equation}\label{p:G_omeg}
    G_{\Omega,Z}(x) =
    \begin{cases}
      A_\Omega\cosh\left(\Omega(\frac{L}{2}-x)\right)+B_\Omega
    \sinh\left(\Omega(\frac{L}{2}-x)\right)\,, \quad &x\in[0,\frac{L}{2}]\,, \\
    A_\Omega\cosh\left(\Omega(\frac{L}{2}-x)\right)+
    \left(\frac{1}{\Omega}+B_\Omega\right)\sinh\left
      (\Omega(\frac{L}{2}-x)\right)\,,
    \quad &x\in[\frac{L}{2},L]\,.
    \end{cases}
\end{equation}
Then, by satisfying the quasi-periodic boundary conditions in
(\ref{pf:green_a}), we obtain a matrix problem 
\begin{equation*}
    E 
    \begin{pmatrix}
        A_{\Omega} \\
        B_{\Omega}
    \end{pmatrix}
    =  \frac{Z}{\Omega}
    \begin{pmatrix}
        -\sinh\left(\frac{\Omega L}{2}\right) \\
        \cosh\left(\frac{\Omega L}{2}\right)
      \end{pmatrix}\,,
      \quad 
 \mbox{with} \quad  E:=
    \begin{pmatrix}
      (1-Z)\cosh\left(\frac{\Omega L}{2}\right) &
      (1+Z)\sinh\left(\frac{\Omega   L}{2}\right) \\
      (1+Z)\sinh\left(\frac{\Omega L}{2}\right) &
      (1-Z)\cosh\left(\frac{\Omega L}{2}\right)
    \end{pmatrix}\,.
\end{equation*}
Next, we invert the matrix $E$ by using
$\det(E) =1-2Z\cosh(\Omega L) + Z^2$ and $Z+{1/Z}=2\Re(Z)$ when $|Z|^2=1$.
This determines $A_{\Omega}$ and $B_{\Omega}$ as
\begin{equation}\label{p:AB}
        \begin{pmatrix}
            A_{\Omega} \\
            B_{\Omega}
        \end{pmatrix}
        = \frac{1}{2\Omega\left(\Re(Z)-\cosh(\Omega L)\right)}
        \begin{pmatrix}
            -\sinh(\Omega L) \\
            -Z+\cosh(\Omega L)
        \end{pmatrix} \,.
\end{equation}
The explicit formula for $G_{\Omega,Z}$ immediately follows from
(\ref{p:AB}) and (\ref{p:G_omeg}). By setting $x={L/2}$ in
(\ref{p:G_omeg}) and using (\ref{p:AB}), we get
\begin{equation}\label{p:gf_mid}
  G_{\Omega,Z}\left(\frac{L}{2}\right) = \frac{\sinh(\Omega L)}
  {2\Omega \left[\cosh(\Omega L) - \Re(Z)\right]} \,.
\end{equation}
As a remark, for $Z=1$ in (\ref{p:G_omeg}) and (\ref{p:AB}) we
recover the periodic Green function in (\ref{p:green_sol}).

In terms of this Green function, the solution to
(\ref{p:flin_stab_1})--(\ref{p:flin_stab_3}) is
\begin{equation*}
    \begin{array}{rclccl}
   \phi(x) &=& d_1 \frac{G_{\Omega_u,Z}(x)}{G_{\Omega_u,Z}\left({L/2}\right)}
                  \quad &\text{with} \quad d_1 &=&
           \frac{\beta_u G_{\Omega_u,Z}\left({L/2}\right)}
             {\beta_u G_{\Omega_u,Z}\left({L/2}\right) + D_u}\xi\,, \\
   \psi(x) &=& d_2 \frac{G_{\Omega_v,Z}(x)}{G_{\Omega_v,Z}\left({L/2}\right)}
               \quad &\text{with} \quad d_2 &=& \frac{\beta_v G_{\Omega_v,Z}
  \left({L/2}\right)}{\beta_v G_{\Omega_v,Z}\left({L/2}\right) +
                                                D_v}\zeta\,.
    \end{array}
\end{equation*}
Finally, from (\ref{p:flin_stab_4}) we obtain an explicit nonlinear
matrix eigenvalue problem for determining $\lambda$
\begin{subequations}\label{f:f_mat}
  \begin{align}
    & \qquad \qquad \qquad \qquad  M_Z(\lambda) 
    \begin{pmatrix}
        \xi \\
        \zeta
    \end{pmatrix} =      \begin{pmatrix}
        0 \\
        0
    \end{pmatrix} \,, \\
\mbox{where} \quad M_{Z}(\lambda) &:=
    \begin{pmatrix}
      \lambda - \partial_\mu f_e + \frac{\beta_u D_u}
      {\beta_u G_{\Omega_u,Z}\left({L/2}\right) + D_u} & - \partial_\eta f_e \\
      - \partial_\mu g_e & \lambda - \partial_\eta g_e +
      \frac{\beta_v D_v}{\beta_v G_{\Omega_v,Z}\left({L/2}\right) + D_v} 
    \end{pmatrix}\,.
  \end{align}
\end{subequations}
In calculating $G_{\Omega_v,Z}\left({L/2}\right)$ and
$G_{\Omega_u,Z}\left({L/2}\right)$ from (\ref{p:gf_mid}) we recall
that $\Omega_u=\sqrt{(\lambda+\sigma_u)/D_u}$ and
$\Omega_v=\sqrt{(\lambda+\sigma_v)/D_v}$.

It follows that $\lambda$ is an eigenvalue of the linearization
(\ref{p:lin_stab}) if and only if for a given $Z=Z_k$ with
$k\in \lbrace{0,\ldots, n-1\rbrace}$, we have
$\det(M_Z(\lambda))=0$. To determine zero-eigenvalue crossings for a
given mode $Z_k$ when a parameter, such as the binding rate
ratio $\rho={\beta_v/\beta_u}$, is varied, we must simultaneously
solve $\det(M_{Z_k}(0))=0$ together with the nonlinear algebraic
system (\ref{p:non_alg}) for the steady-state to identify the
bifurcation point. Such zero-eigenvalue crossings for mode $Z_k$
correspond to symmetry-breaking bifurcations from the symmetric
steady-state.

We now clarify how the Floquet modes $Z_k$ are related to the form of
the perturbation near each cell centered at
$x=x_j=\left(j-{1/2}\right)L$ for $j\in\lbrace{1,\ldots,n\rbrace}$,
whenever we have a zero-eigenvalue crossing $\lambda=0$ obtained from
setting $\det(M_{Z_k}(0))=0$ for any
$k\in \lbrace{0,\ldots,n-1\rbrace}$. Labeling
$\phi_{\mbox{cell} j,k}:=\phi\left( (j-{1/2})L\right)$, we have by
translation invariance that
$\phi_{\mbox{cell} j,k}= Z_k^{j-1}\phi\left({L/2}\right)$ when
$Z=Z_k$. In this way, we can define a vector characterizing the
perturbation of the symmetric steady-state $u_e$ at the cell locations
by
\begin{equation}\label{fl:cell}
  \mathbf{\phi}_{\mbox{cell},k} :=
  \begin{pmatrix}
    \phi_{\mbox{cell} 1,k} \\
    \phi_{\mbox{cell} 2,k} \\
    \vdots  \\
      \phi_{\mbox{cell} n,k} 
    \end{pmatrix} =    \begin{pmatrix}
    1 \\
    Z_k \\
    \vdots  \\
    Z_{k}^{n-1} 
    \end{pmatrix} \phi\left({L/2}\right) \,.
\end{equation}
In a similar way, we can identify for (\ref{p:lin_stab}) in terms of
the fundamental cell variables that $\psi_{\mbox{cell} j,k}:=
\psi\left( (j-{1/2})L\right)=Z_k^{j-1}\psi\left({L/2}\right)$,
$\xi_{jk}=Z_k^{j-1}\xi$, and $\zeta_{jk}=Z_k^{j-1}\zeta$
for $j\in\lbrace{1,\ldots,n\rbrace}$. 

Since $G_{\Omega,Z_k}\left({L/2}\right)$ from (\ref{p:gf_mid}) and,
consequently, $M_{Z_k}(\lambda)$ in (\ref{f:f_mat}) are real-valued
when $\lambda$ is real, we conclude that $\phi\left({L/2}\right)$ is
real-valued when $\lambda$ is real. By taking the real and imaginary
parts of (\ref{fl:cell}) we conclude that
\begin{equation}\label{fl:cell_real}
  \mathbf{\phi}_{\mbox{cell},k} \in
  \mbox{span}\lbrace{ \mathbf{v}_{ck}, \mathbf{v}_{sk}\rbrace}
  \,, \quad \mathbf{v}_{ck} :=
  \begin{pmatrix}
    1 \\
   \vdots \\
   \cos\left(\frac{2\pi(j-1) k}{n}\right) \\
       \vdots\\
       \cos\left(\frac{2\pi(n-1) k}{n}\right)
         \end{pmatrix} \,, \quad 
         \mathbf{v}_{sk} : =
         \begin{pmatrix}
    0 \\
   \vdots \\
   \sin\left(\frac{2\pi(j-1) k}{n}\right) \\
       \vdots\\
       \sin\left(\frac{2\pi(n-1) k}{n}\right)
         \end{pmatrix} \,.
\end{equation}

When $n=2$, we have either $Z_0=1$ or $Z_0=-1$. For the in-phase mode
$Z_0=1$, we have $\mathbf{v}_{sk}={\mathbf 0}$, and so
(\ref{fl:cell_real}) yields
$\mathbf{\phi}_{\mbox{cell},0}\in \mbox{span}\lbrace{
  \mathbf{v}_{c0}\rbrace}= \mbox{span}\lbrace{ (1,1)^T\rbrace}$. Likewise,
for the anti-phase mode $Z_0=-1$ we have
$\mathbf{\phi}_{\mbox{cell},1} \in \mbox{span}\lbrace{
  \mathbf{v}_{c1}\rbrace}= \mbox{span}\lbrace{ (1,-1)^T\rbrace}$.
For any $n\geq 2$, we always have the in-phase mode $Z_0=1$ for which
$\mathbf{\phi}_{\mbox{cell},0} \in \mbox{span}\lbrace{
  \mathbf{v}_{c0}\rbrace}= \mbox{span}\lbrace{ (1,\ldots,1)^T\rbrace}$. 

Now suppose that $n>2$ is even. Then, since $Z_{n/2}=e^{i\pi}=-1$, we
will have a zero-eigenvalue crossing for an anti-phase perturbation
$\mathbf{\phi}_{\mbox{cell},{k/2}} \in
\mbox{span}\lbrace{(1,-1,\ldots,1,-1)^T \rbrace}$, whenever there is a
bifurcation parameter for which $\det(M_{Z_{n/2}}(0))=0$. Furthermore,
since $\Re Z_k=\Re Z_{n-k}$ and, consequently,
$M_{Z_k}(\lambda)=M_{Z_{n-k}}(\lambda)$ in (\ref{f:f_mat}) for
$k\in\lbrace{1,\ldots,\frac{n}{2}-1\rbrace}$, we conclude that the
bifurcation parameter thresholds for which $\det(M_{Z_k})(0)=0$ and
$\det(M_{Z_n-k})(0)=0$ are identical. As a result, we have ${n/2}-1$
``degenerate'' asymmetric modes when $n>2$ is even. Since
$\mathbf{v}_{ck}=\mathbf{v}_{c(n-k)}$ and
$\mathbf{v}_{sk}=-\mathbf{v}_{s(n-k)}$ for
$k\in\lbrace{1,\ldots,\frac{n}{2}-1 \rbrace}$, we conclude from
(\ref{fl:cell_real}) that
$\mathbf{\phi}_{\mbox{cell},k} \in\mbox{span}\lbrace{ \mathbf{v}_{ck},
  \mathbf{v}_{sk}\rbrace}$ for $k\in\lbrace{1,\ldots,\frac{n}{2}-1\rbrace}$.

In contrast, for $n\geq 3$ with $n$ odd, there is no anti-phase mode
and there are exactly $\frac{n+1}{2}-1$ possible degenerate asymmetric
modes for which the bifurcation parameter thresholds are determined
from setting $\det(M_{Z_k})(0)=0$ for
$k\in\lbrace{1,\ldots,\frac{n+1}{2}-1 \rbrace}$.  We again conclude
from (\ref{fl:cell_real}) that
$\mathbf{\phi}_{\mbox{cell},k} \in \mbox{span}\lbrace{
  \mathbf{v}_{ck}, \mathbf{v}_{sk}\rbrace}$ for
$k\in\lbrace{1,\ldots,\frac{n+1}{2}-1\rbrace}$. In particular, when
$n=3$, we conclude that at the zero-eigenvalue crossing for either
$Z=Z_1$ or $Z=Z_2$ there is a two-dimensional eigenspace for the vector 
$\mathbf{\phi}_{\mbox{cell},1}$ of cell perturbations given by the plane
$(1,1,1)^T\mathbf{\phi}_{\mbox{cell},1}=0$.

\subsection{Two Compartments with $g(\mu,\eta)
  =g_1(\mu)-g_2\eta$}\label{per:two}

The analysis so far for the periodic problem has focused on
constructing the symmetric steady-state and formulating the linear
stability problem so as to readily detect symmetry-breaking
bifurcation points for any $n\geq 2$ and arbitrary intra-compartment
reactions $f$ and $g$.

In this subsection, we briefly focus on the two-compartment case $n=2$ when
$g(\mu,\eta)=g_1(\mu)-g_2\eta$ with an aim of constructing both
symmetric and asymmetric branches of equilibria. Since the analysis is
similar to that in \S \ref{2cellcoupledsystem} we only briefly highlight it
here.

For $n=2$ compartments, the steady-state for $u_e$ has the form
\begin{eqnarray*}
        u_e(x) = 
        \begin{cases}
          A_1\sinh\left(\omega_u(\frac{L}{2}-x)\right) +
          A_2\sinh(\omega_u x)\,, \quad &x\in[0,\frac{L}{2}]\,, \\
          A_3\sinh\left(\omega_u(\frac{3L}{2}-x)\right) +
          A_4\sinh\left(\omega_u(x-\frac{L}{2})\right)\,,
          &x\in[\frac{L}{2},\frac{3L}{2}]\,, \\
          A_5\sinh\left(\omega_u(2L-x)\right) +
          A_6\sinh\left(\omega_u(x-\frac{3L}{2})\right)\,, &x\in[\frac{3L}{2},2L]\,,
        \end{cases}
\end{eqnarray*}
where $\omega_u:=\sqrt{\sigma_u/D_u}$. Upon imposing that
$u_e(0)=u_e(2L)$, $\partial_x u_e(0)=\partial_x u_e(2L)$, and
$u_e$ is continuous at $x={L/2}$ and $x={3L/2}$, we obtain that
\begin{equation*}
  A_1=A_6=A_3+A_4\,, \qquad A_2=A_3\frac{\sinh(\omega_u L)}{\sinh\left(\omega_u
      {L/2}\right)}\,, \qquad  A_5 =
  A_4\frac{\sinh(\omega_u L)}{\sinh\left(\omega_u {L/2}\right)}\,.
\end{equation*}
By enforcing the jump conditions for $u_e$ at $x={L/2}$ and $x={3L/2}$,
we determine $A_3$ and $A_4$ as
 \begin{equation*}
        \begin{pmatrix}
            A_3 \\
            A_4
        \end{pmatrix}
        =\frac{\beta_u}{D_u\omega_u(\gamma_u^2-4)}
        \begin{pmatrix}
            \gamma_u & 2 \\
            2        & \gamma_u
        \end{pmatrix}
        \begin{pmatrix}
            \mu_1^e \\
            \mu_2^e
          \end{pmatrix} \,, \quad \mbox{with} \quad \gamma_u :=
          2\cosh(\omega_u L)
          +\frac{\beta_u}{D_u\omega_u}\sinh(\omega_u L)\,.
\end{equation*}
Likewise, we can calculate $v_e(x)$. Then, from the steady-state
of the intra-compartment dynamics, we proceed as in
\S \ref{2cellcoupledsystem} to obtain a non-linear
algebraic equation for $\mu_1^e$ and $\mu_2^e$ characterizing all steady-states
of the two-compartment system. We obtain as in \S
\ref{2cellcoupledsystem} that
\begin{equation}\label{per2:eq}
    \begin{pmatrix}
        f(\mu_1^e, (1,0)A^{-1}(g_1(\mu_1^e),g_1(\mu_2^e))^T) \\
        f(\mu_2^e, (0,1)A^{-1}(g_1(\mu_1^e),g_1(\mu_2^e))^T)
    \end{pmatrix}
    -\frac{\beta_u}{\gamma_u^2-4} B
    \begin{pmatrix}
        \mu_1^e \\
        \mu_2^e
    \end{pmatrix}
    = 0\,,
\end{equation}
where $A$ and $B$ are now defined by
\begin{subequations}
  \begin{align}
    A &:= \frac{\beta_v}{\gamma_v^2-4} \tilde{A} + g_2 I\,, \qquad
        \tilde{A} :=  \begin{pmatrix}
                2\gamma_v \cosh(\omega_v L) -4
                & 4\cosh(\omega_v L) - 2\gamma_v \\
                4\cosh(\omega_v L) - 2\gamma_v
                & 2\gamma_v \cosh(\omega_v L) -4
              \end{pmatrix} \,,\\
    B & := \begin{pmatrix}
                2 \gamma_u \cosh(\omega_u L) -4
                & 4\cosh(\omega_u L) - 2\gamma_u \\
                4\cosh(\omega_u L) - 2\gamma_u
                & 2 \gamma_u \cosh(\omega_u L) -4
              \end{pmatrix} \,.
  \end{align}
\end{subequations}
Here
$\gamma_v:=2\cosh(\omega_v L) +\frac{\beta_v}{D_v\omega_v}
\sinh(\omega_v L)$.

In place of (\ref{ss:eig}), the eigenvalues $a_1, b_1$ and $a_2, b_2$
of the matrices $A$ and $B$ for the in-phase $q_1=(1,1)^T$ and
anti-phase $q_2=(1,-1)^T$ modes are
\begin{subequations}\label{p2:new_eig}
\begin{align}
  a_1 &= \frac{\beta_v}{\gamma_v^2-4} \left[ 2\gamma_v \left(
        \cosh(\omega_v L)-1\right) + 4 \left(\cosh(\omega_v L)-1\right)\right]
         +g_2 \\
    a_2 &= \frac{\beta_v}{\gamma_v^2-4} \left[ 2\gamma_v \left(
          \cosh(\omega_v L)+1\right) - 4 \left(\cosh(\omega_v L)+1\right)
          \right] +g_2 \,, \\
  b_1 & = 2\gamma_u \left(\cosh(\omega_u L)-1\right) + 4
        \left(\cosh(\omega_u L)-1\right) \,,\\
  b_2 & = 2\gamma_u \left(\cosh(\omega_u L)+1\right) - 4
        \left(\cosh(\omega_u L)+1\right) \,.
\end{align}
\end{subequations}
In terms of $a_1$ and $b_1$, and in analogy with (\ref{ss:nonl_symm}),
the symmetric steady-state $(\mu_1^e,\mu_2^e)=\mu_e q_1^{T}$, is obtained
from the scalar algebraic equation
\begin{equation}\label{p2:symm}
  f(\mu_e, \frac{g_1(\mu_e)}{a_1})-\frac{\beta_u}{\gamma_u^2-4} b_1\mu_e = 0
  \,.
\end{equation}
As a remark, for the special case where $g=g_1(\mu)-g_2\eta$, we
can verify after a direct but tedious calculation that the steady-state
condition (\ref{p:non_alg}) is equivalent to (\ref{p2:symm}).

\subsection{FitzHugh-Nagumo kinetics} \label{ring-cell FN system}

To illustrate the theory, we first choose the FN kinetics as given in
(\ref{s:FN}). For $n=2$, for which (\ref{per2:eq}), (\ref{p2:symm}) and
(\ref{p2:new_eig}) apply, the pitchfork points $z_{P,1}$ and
$z_{P,2}$ delimiting the pitchfork bubble in $z$ are recovered from
the parameter constraint (\ref{ssFN:q}) in which $a_1$, $a_2$, $b_1$,
and $b_2$ are now given as in (\ref{p2:new_eig}). For $n=2$, we can
use MatCont \cite{matcont} on (\ref{per2:eq}) to calculate both
symmetric and asymmetric steady-states using either
$z$ or the binding rate ratio $\rho={\beta_v/\beta_u}$ as parameters.
The bifurcation diagrams are qualitatively similar to those in
Fig.~\ref{fig:FN_bubble} (not shown).

\begin{figure}[]
	\begin{subfigure}[b]{.50\textwidth}
	    \begin{subfigure}[b]{1.\textwidth}
    		\centering
			\includegraphics[width=1.\linewidth]{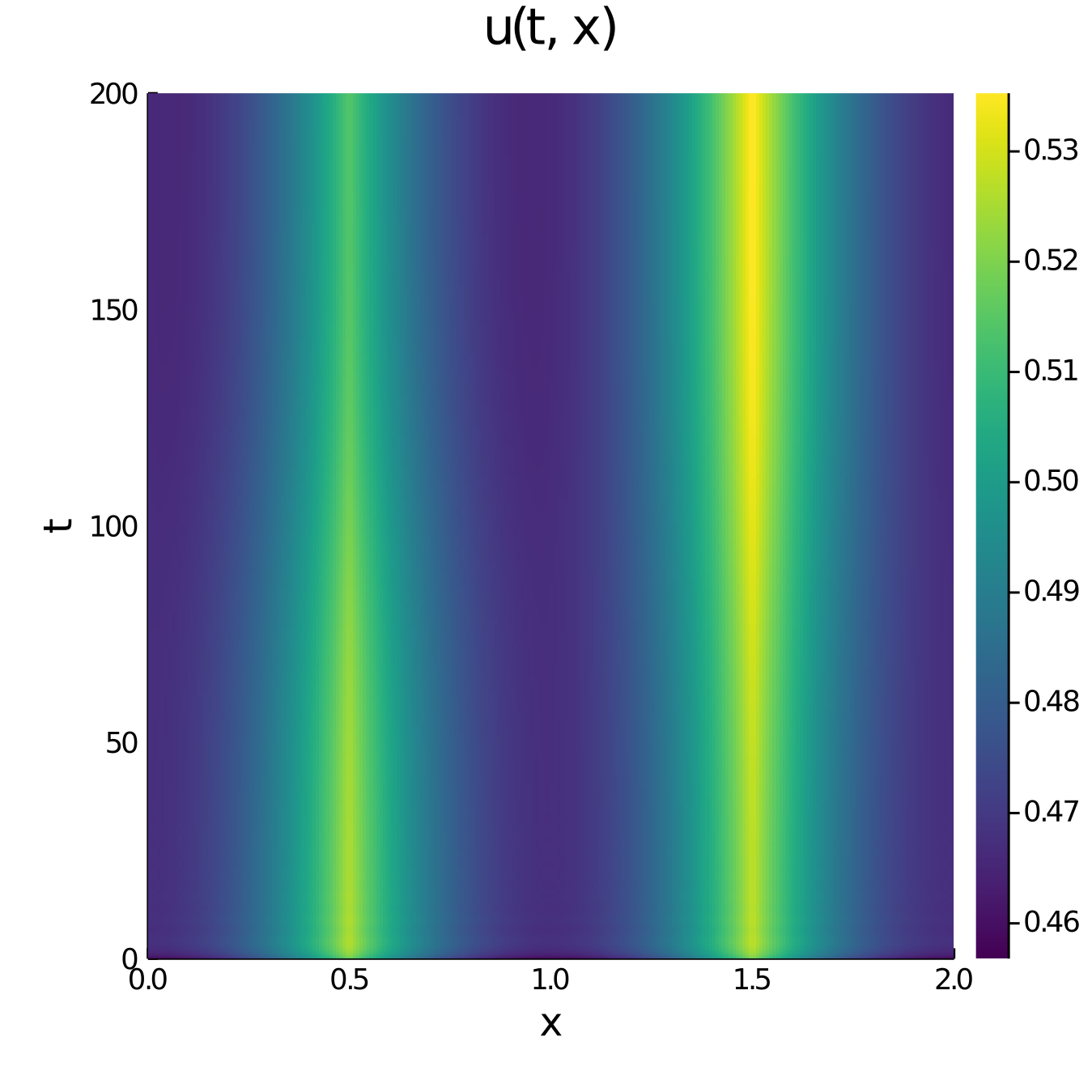}  
		\end{subfigure}
		\begin{subfigure}[b]{1.\textwidth}
    		\centering
			\includegraphics[width=1.\linewidth]{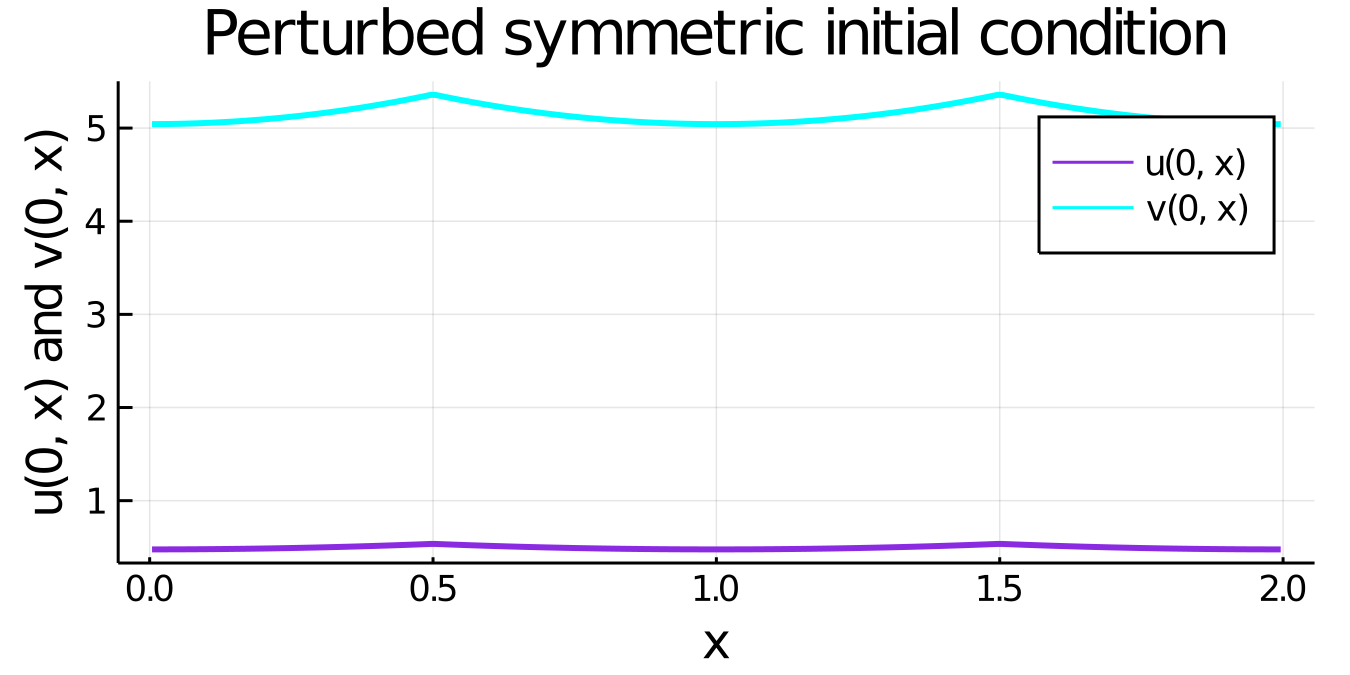}  
		\end{subfigure}
	\end{subfigure}
	\begin{subfigure}[b]{.49\textwidth}
		\begin{subfigure}[b]{1.\textwidth}
    		\centering
			\includegraphics[width=1.\linewidth]{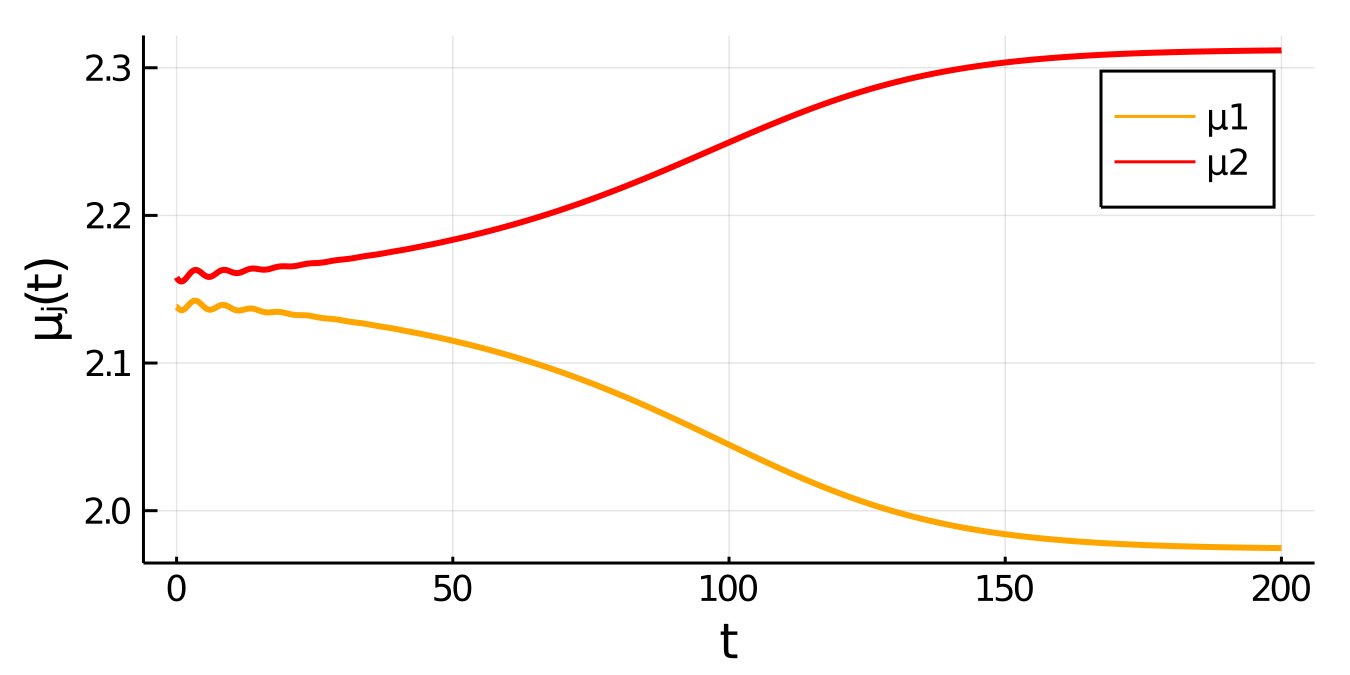}  
		\end{subfigure}
		\begin{subfigure}[b]{1.\textwidth}
			\centering
			\includegraphics[width=1.\linewidth]{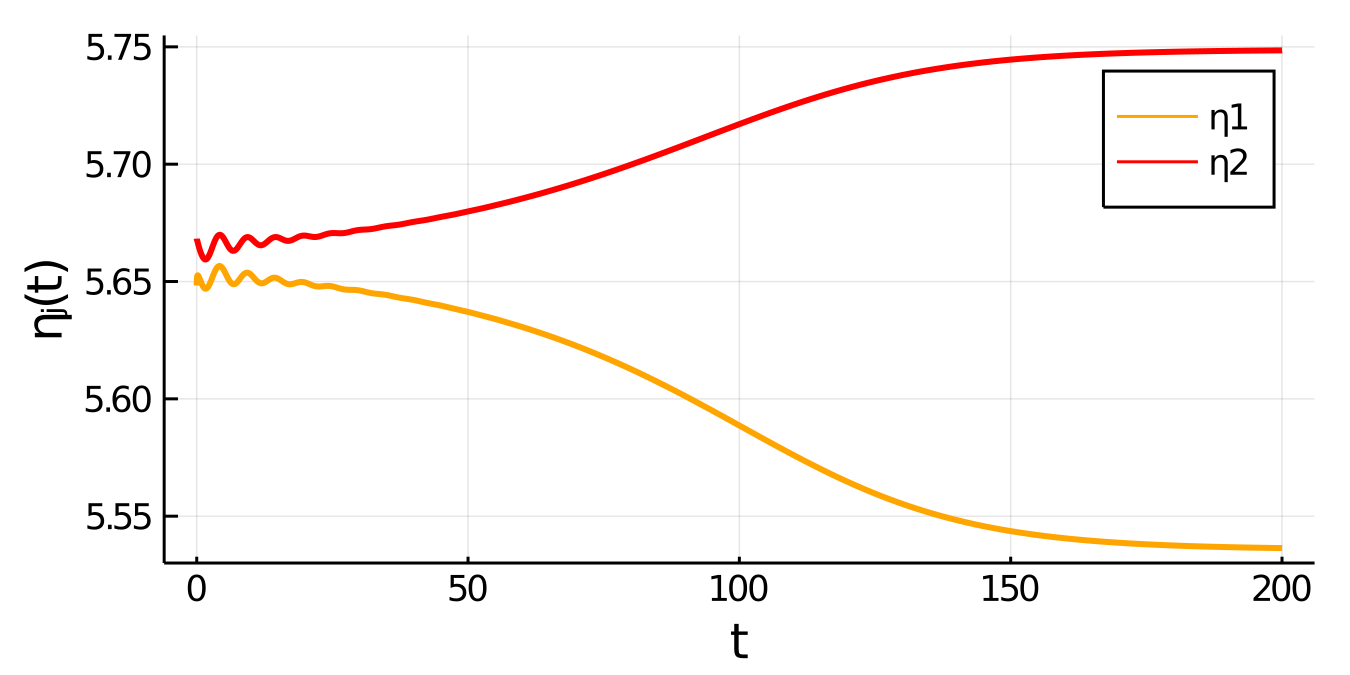}  
		\end{subfigure}
		\begin{subfigure}[b]{1.\textwidth}
			\centering
			\includegraphics[width=1.\linewidth]{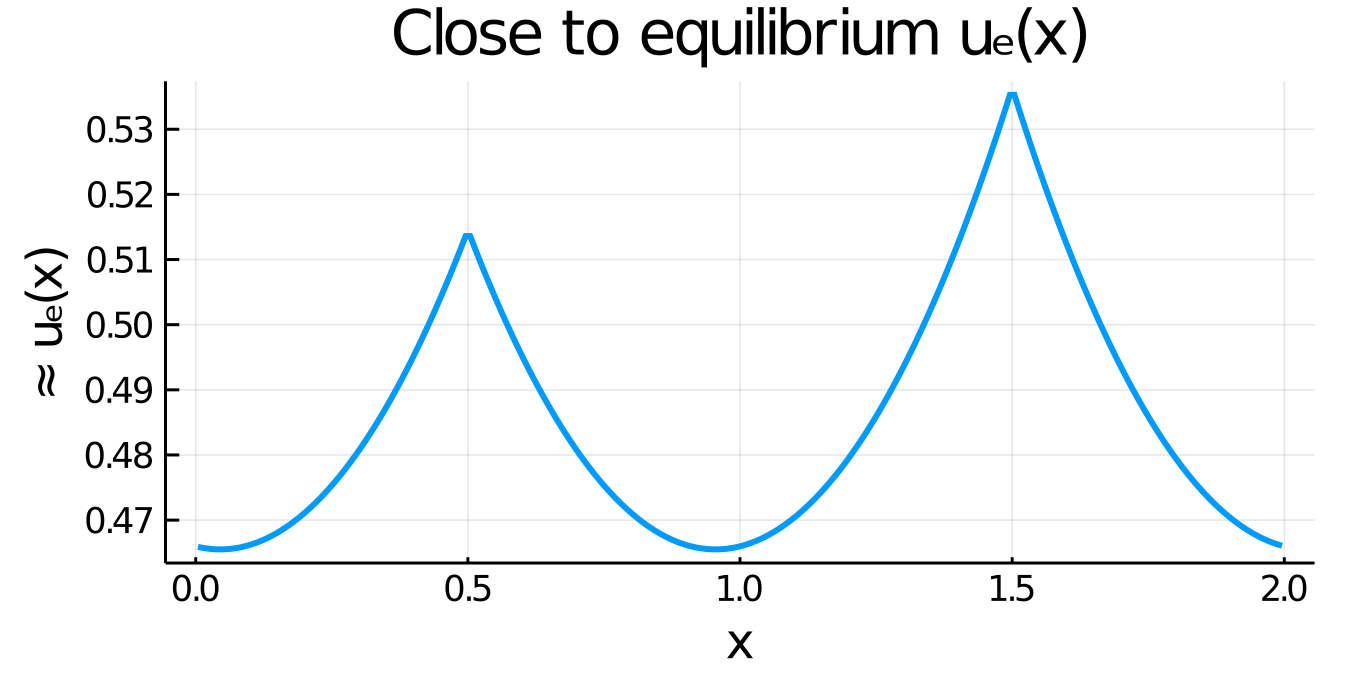}  
		\end{subfigure}
	\end{subfigure}
	\caption{FN kinetics with $n=2$: For an initial condition near
          the unstable symmetric branch, and for $z=z_{P,2}\approx 6.63675$ and
          $\rho={\beta_v/\beta_u}=60$, the full time-dependent
          solution computed using the BE-RK4 IMEX scheme of Appendix
          \ref{num:ring} converges to a stable asymmetric
          steady-state. For $z=6.63675$, the pitchfork bifurcation
          point in $\rho$ is $\rho_p=50$. Parameters:
          $D_u=1,D_v=2,\sigma_u=\sigma_v=1,\varepsilon=0.6,q=1,L=1$,
          and $\beta_u=0.3$.}\label{fig:per:FN2}
\end{figure}

\begin{figure}[]
    \centering
	\begin{subfigure}[b]{.47\textwidth}
	    \begin{subfigure}[b]{1.\textwidth}
	        \centering
	        \includegraphics[width=1.\linewidth]{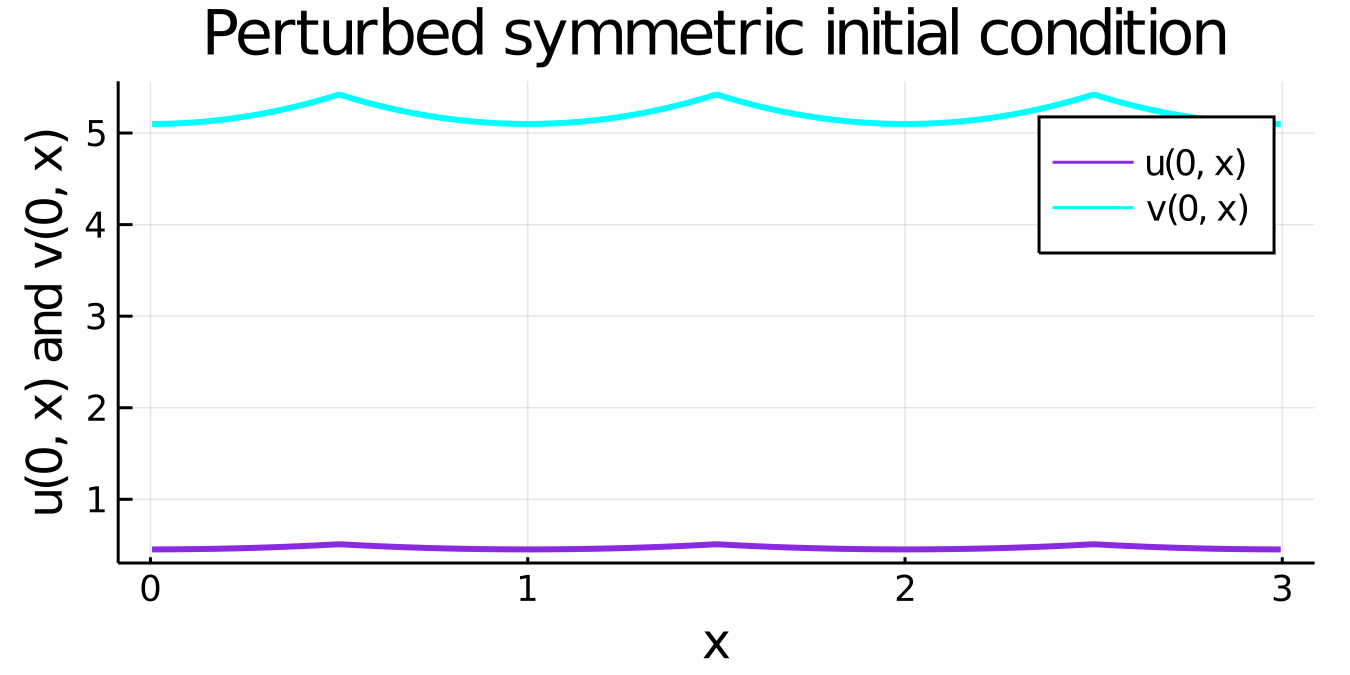}
	    \end{subfigure}
	    \begin{subfigure}[b]{1.\textwidth}
	        \centering
	        \includegraphics[width=1.\linewidth]{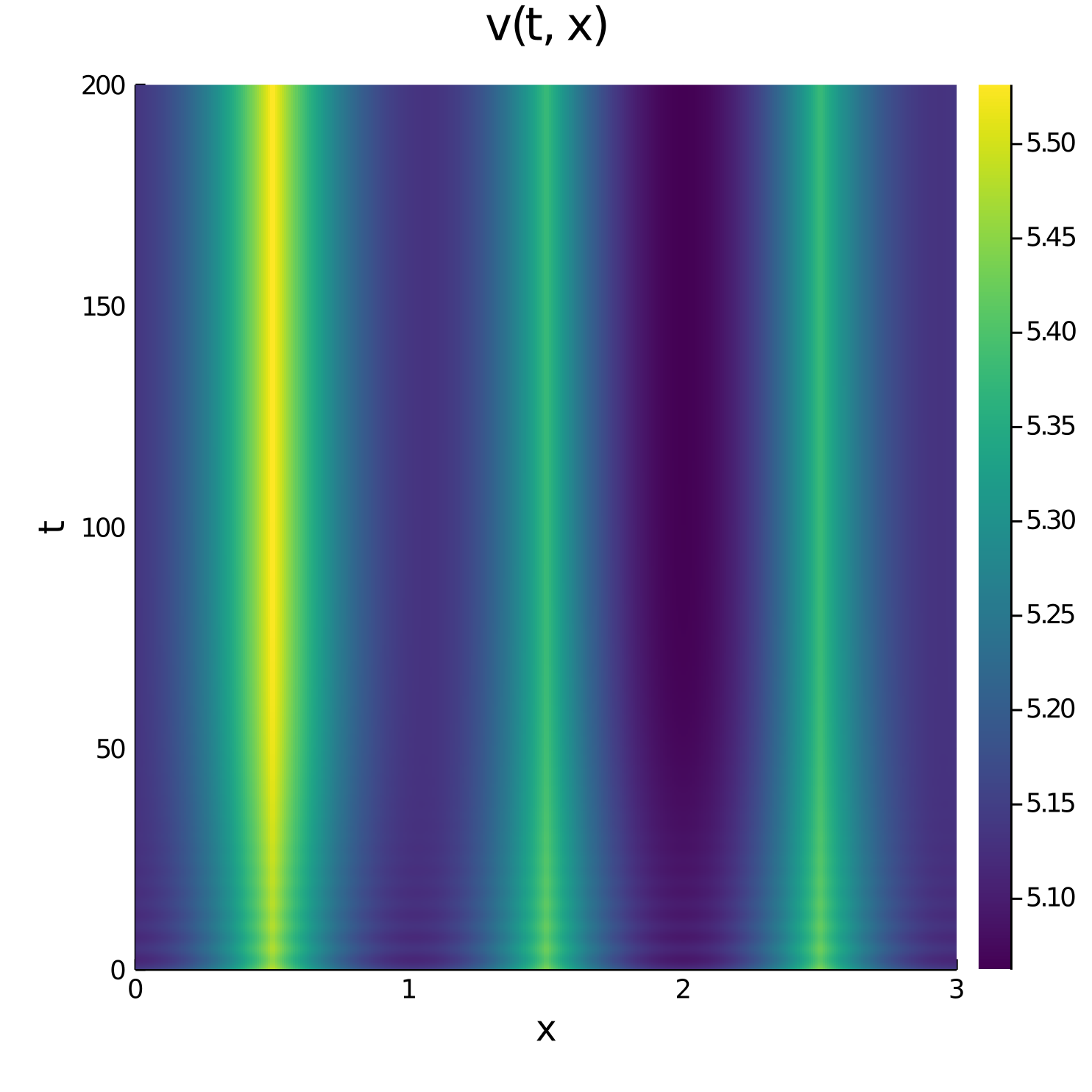}
	    \end{subfigure}
	    \begin{subfigure}[b]{1.\textwidth}
    		\centering
			\includegraphics[width=1.\linewidth]{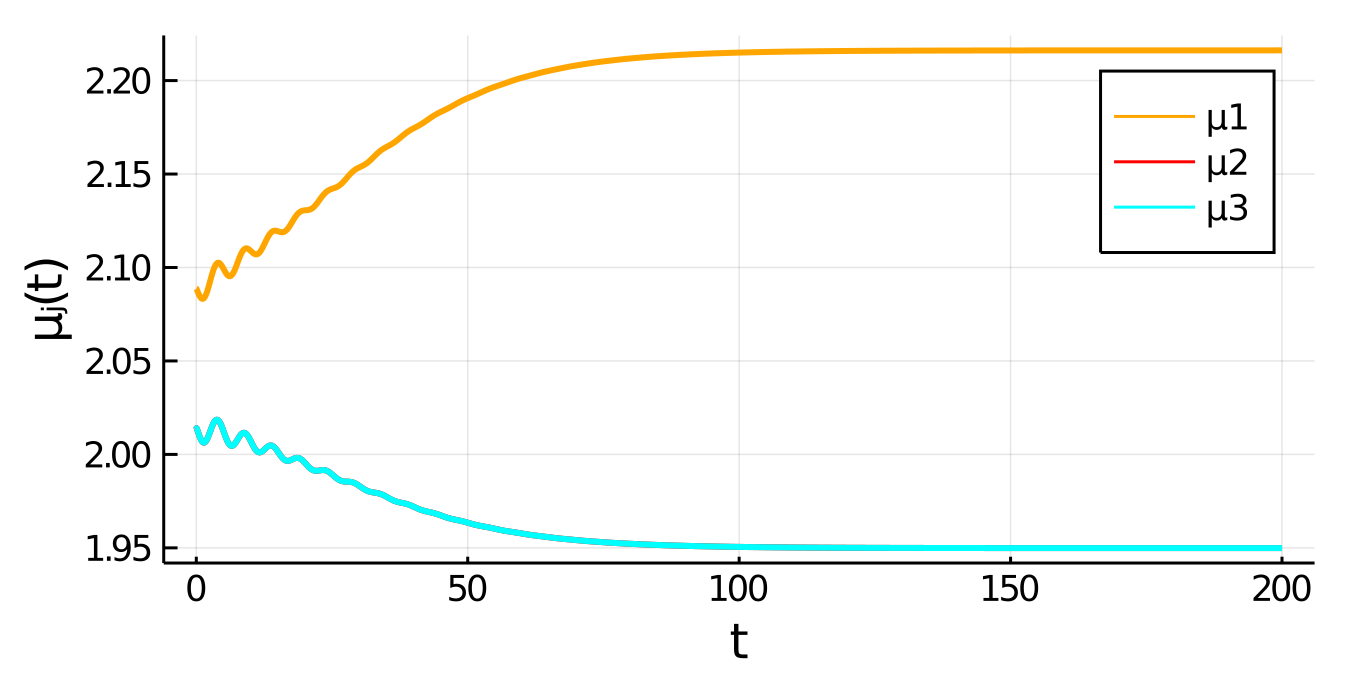}  
		\end{subfigure}
		\begin{subfigure}[b]{1.\textwidth}
			\centering
			\includegraphics[width=1.\linewidth]{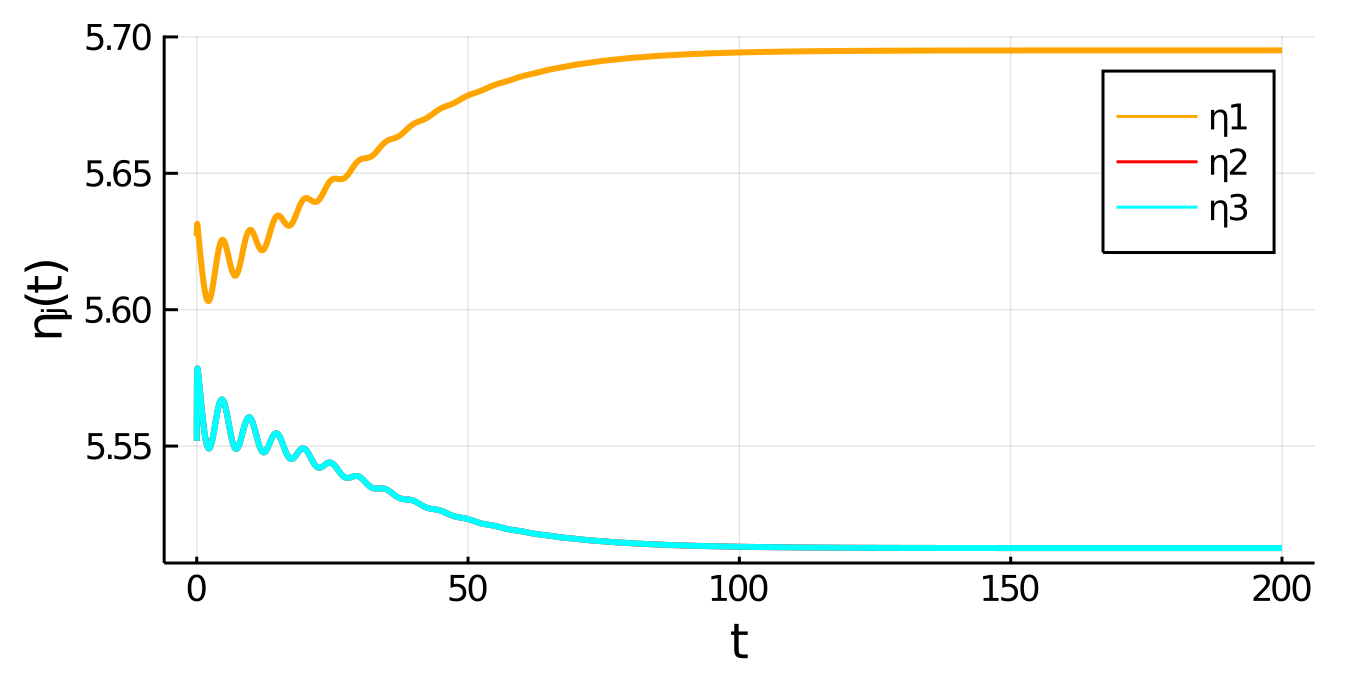}  
		\end{subfigure}
		\begin{subfigure}[b]{1.\textwidth}
			\centering
			\includegraphics[width=1.\linewidth]{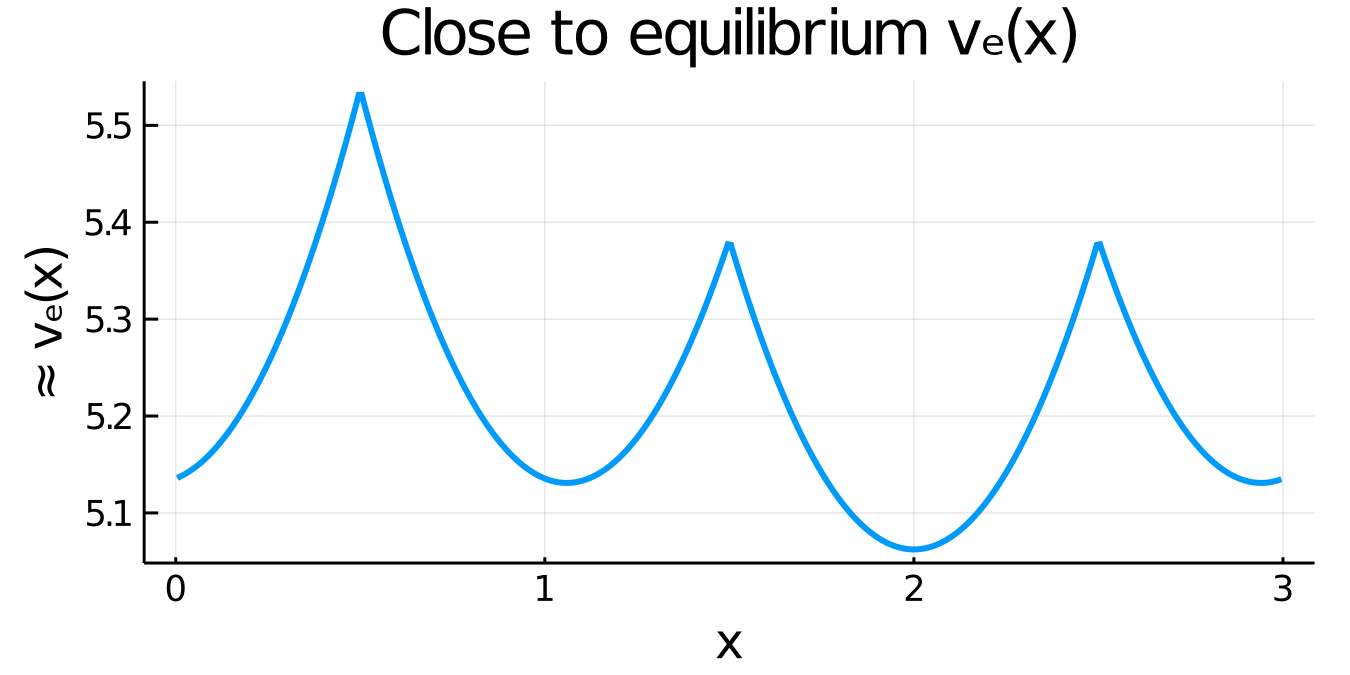}  
		\end{subfigure}
	\end{subfigure}
	\begin{subfigure}[b]{.47\textwidth}
	    \begin{subfigure}[b]{1.\textwidth}
	        \centering
	        \includegraphics[width=1.\linewidth]{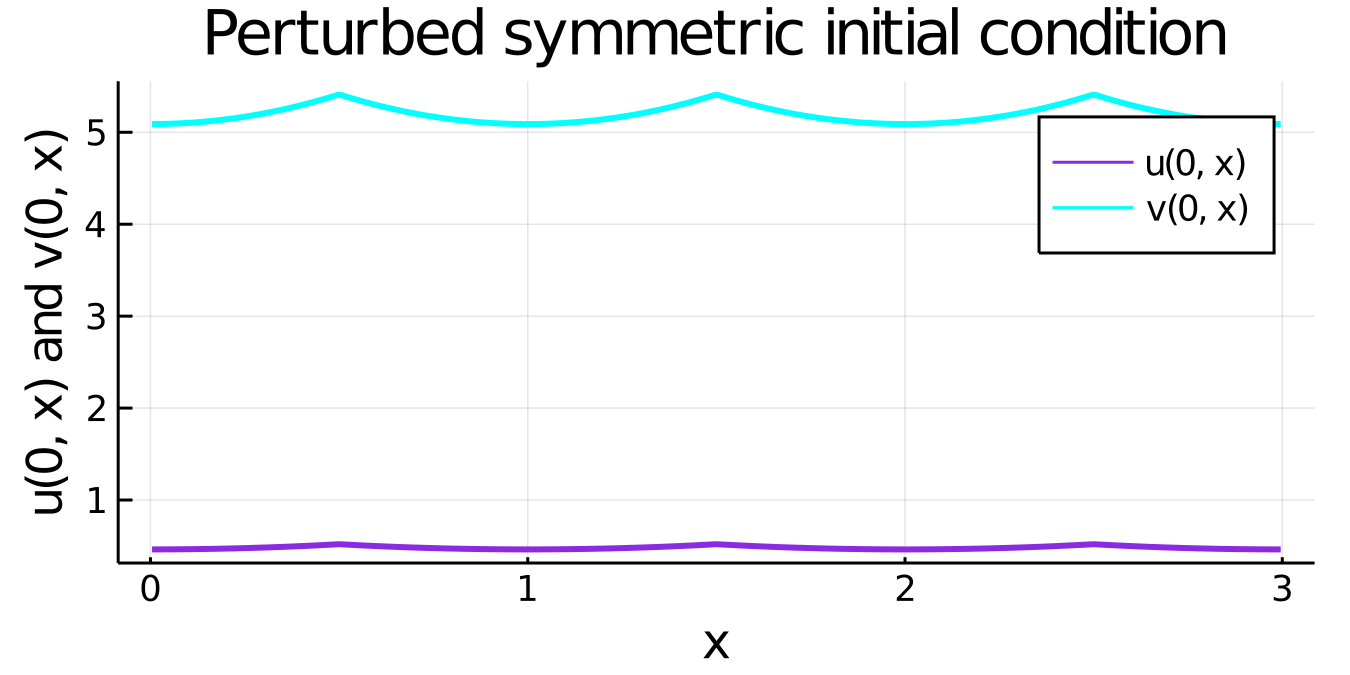}
	    \end{subfigure}
	    \begin{subfigure}[b]{1.\textwidth}
	        \centering
	        \includegraphics[width=1.\linewidth]{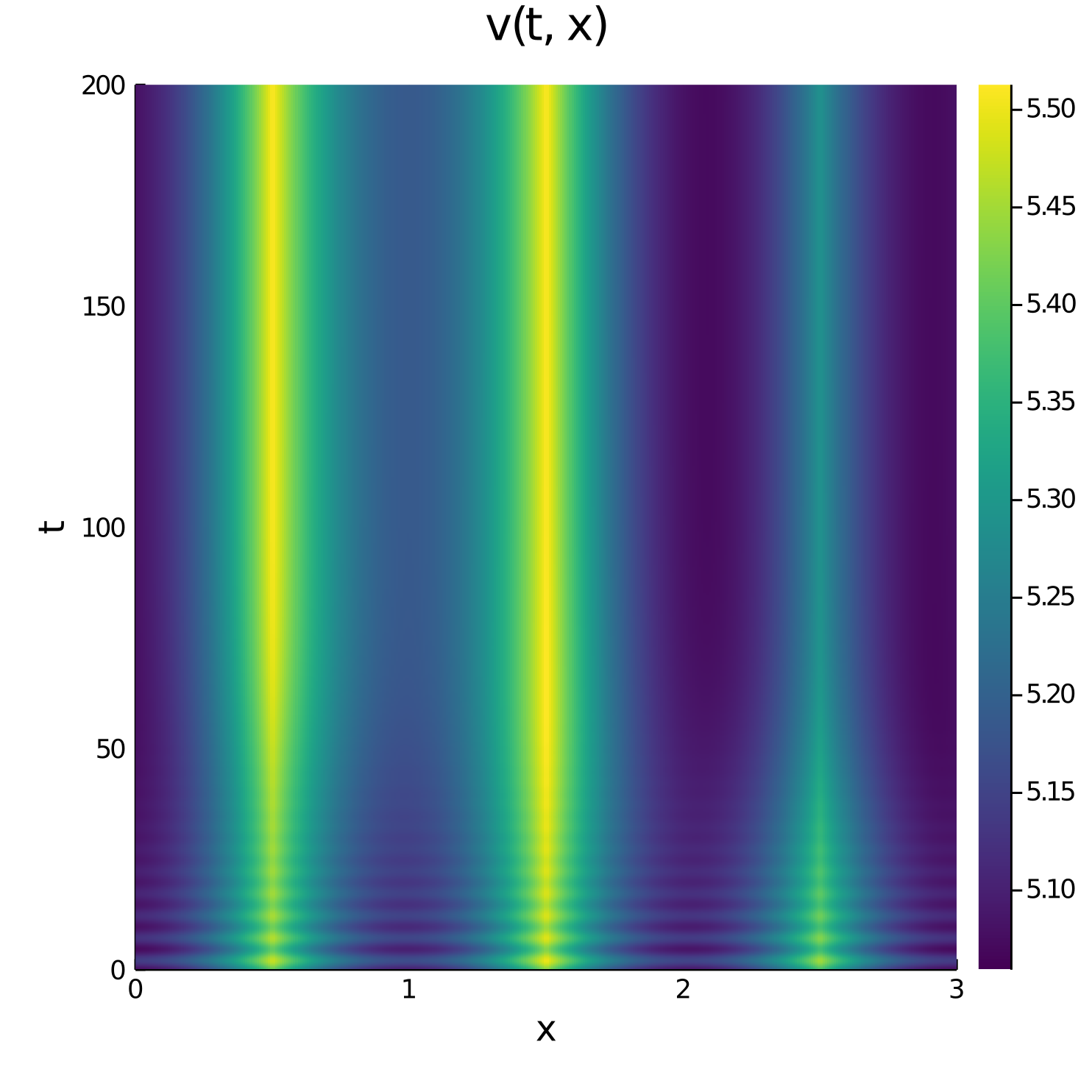}
	    \end{subfigure}
	    \begin{subfigure}[b]{1.\textwidth}
    		\centering
			\includegraphics[width=1.\linewidth]{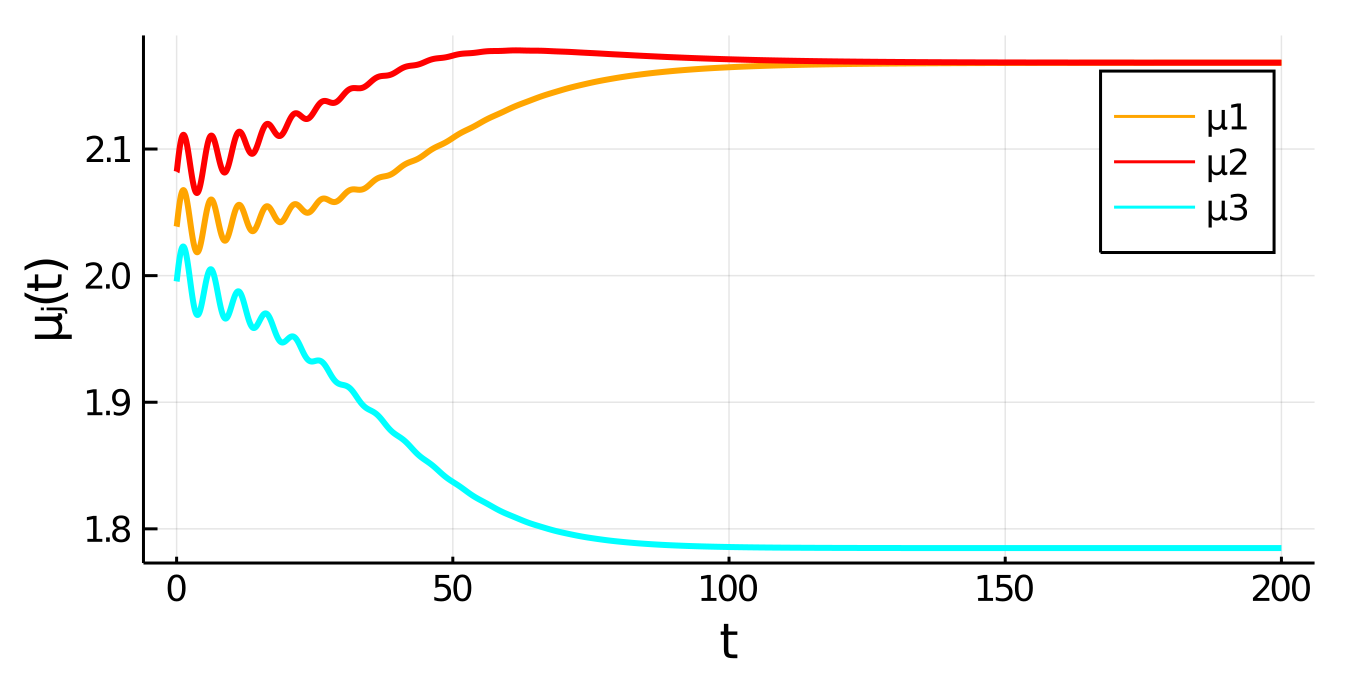}  
		\end{subfigure}
		\begin{subfigure}[b]{1.\textwidth}
			\centering
			\includegraphics[width=1.\linewidth]{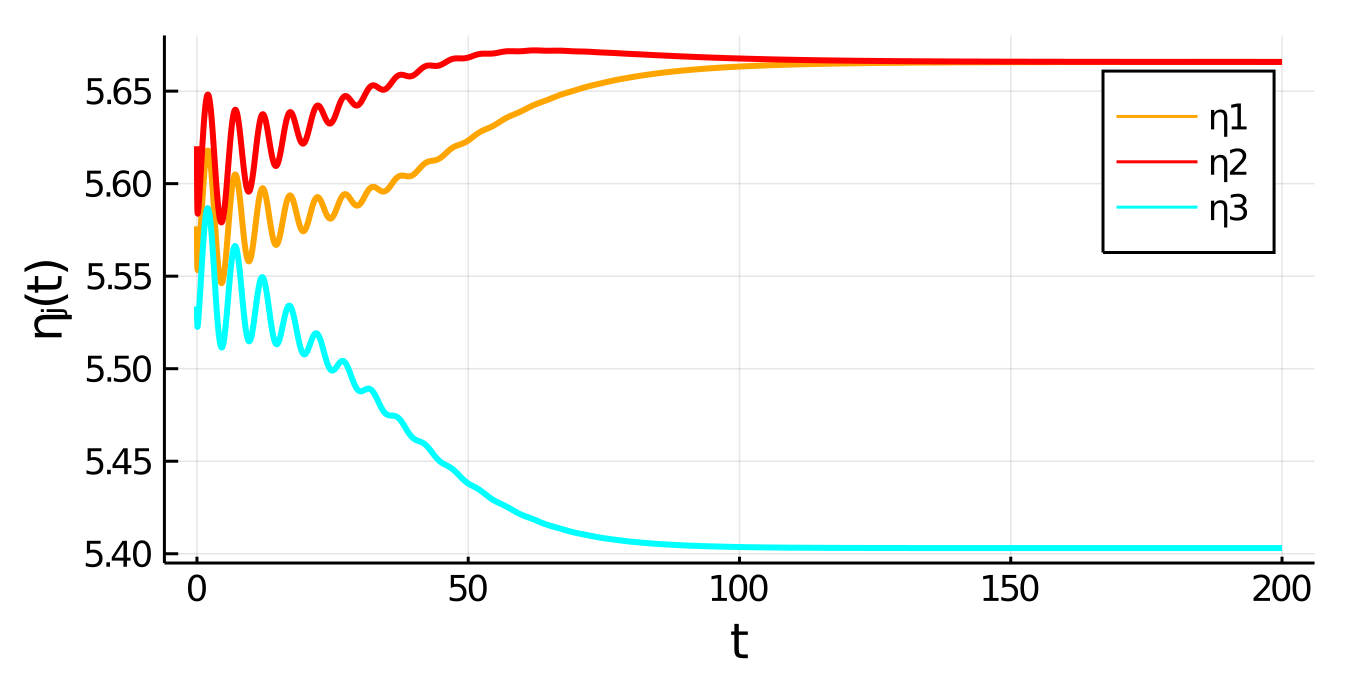}  
		\end{subfigure}
		\begin{subfigure}[b]{1.\textwidth}
			\centering
			\includegraphics[width=1.\linewidth]{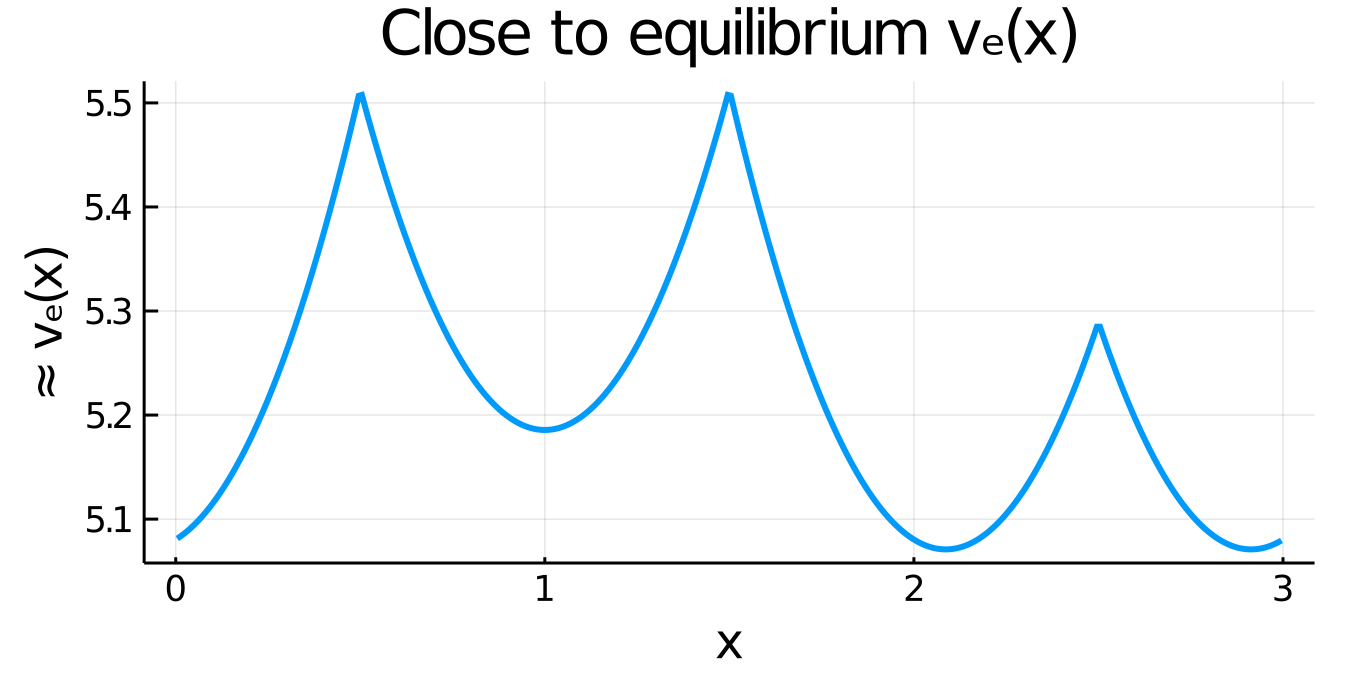}  
		\end{subfigure}
	\end{subfigure}
	\caption{FN kinetics with $n=3$: Full-numerical time-dependent
          results from the BE-RK4 IMEX scheme of Appendix
          \ref{num:ring}. Left: convergence to a stable asymmetric
          pattern due to an initial cell perturbation with
          $\mathbf{\phi}_{\mbox{cell},1}=
          \mathbf{v}_{c1}=\left(1,{-1/2},{-1/2}\right)^T$. Right:
          convergence to a different stable asymmetric branch due to
          an initial cell perturbation with
          $\mathbf{\phi}_{\mbox{cell},1}=
          \left(0,{\sqrt{3}/2},-{\sqrt{3}/2}\right)^T$.  Parameters:
          $D_u=1,D_v=2,\sigma_u=\sigma_v=1,\varepsilon=0.6,q=1,
          L=1,\beta_u=0.3,\rho=120$, and $z=7$.}\label{fig:per:FN3}
\end{figure}

For $n=2$, in Fig.~\ref{fig:per:FN2} we show, for an initial condition
near the unstable symmetric branch, that the time-dependent numerical
solution computed using the BE-RK4 IMEX scheme of Appendix \ref{num:ring}
tends to a stable asymmetric pattern. For $n=3$, and for a parameter set
where the symmetric steady-state is unstable, in
Fig.~\ref{fig:per:FN3} we show the time-dependent convergence to two
different stable asymmetric patterns depending on the specific form
of the initial perturbation of the symmetric state. For $n=3$, we
conclude from (\ref{fl:cell_real}) and the discussion in
\S \ref{pf:stab_ana} that at the zero-eigenvalue crossing for an
asymmetric mode there is a two-dimensional eigenspace for the
vector cell perturbation $\mathbf{\phi}_{\mbox{cell},1}$, given by the plane
$(1,1,1)^T\mathbf{\phi}_{\mbox{cell},1}=0$.

\subsection{Gierer-Meinhardt kinetics}

\begin{figure}[htbp]
    \centering
    \includegraphics[width=1.\textwidth]{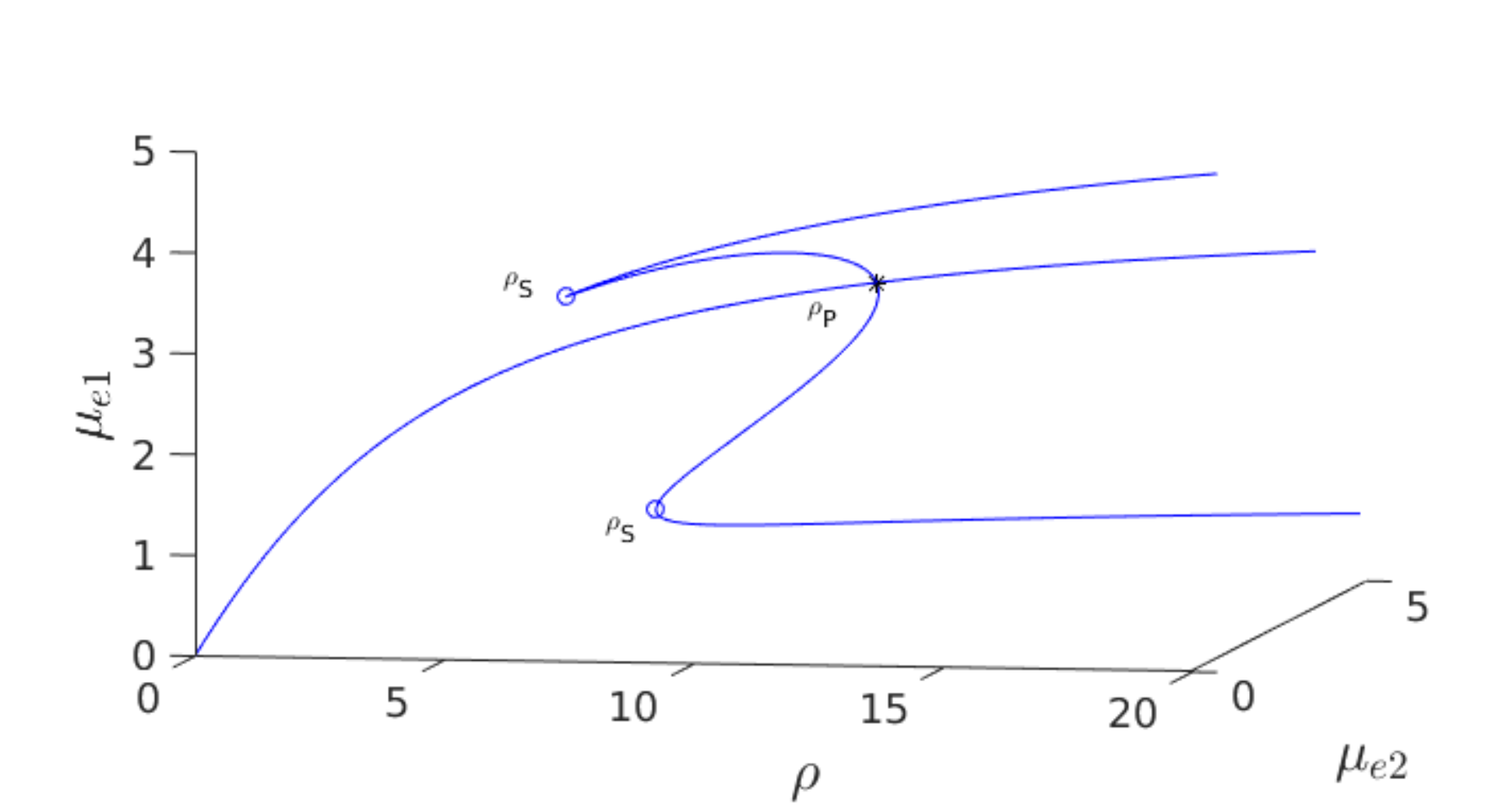}
    \caption{{3-D Bifurcation diagram, exhibiting hysteresis, with
      bifurcation parameter $\rho={\beta_v/\beta_u}$ for steady-states
      with GM kinetics and $n=2$ computed from (\ref{per2:eq}) using
      MatCont \cite{matcont}.} The subcritical pitchfork bifurcation
      point from the symmetric steady-state is at
      $\rho=\rho_p \approx 11.42015$ and the fold points on the
      asymmetric branches are at $\rho_f \approx 6.82631$. The
      symmetric steady-state is linearly stable only for
      $1<\rho<\rho_p$, while the asymmetric branches become stable
      only past the fold points.  Parameters:
      $D_u=D_v=1, \sigma_u=\sigma_v=1, L=1$, and $\beta_u=0.3$.  Note
      that the geometry of the asymmetric branches near the fold
      points is just mirrored and it looks different in the figure
      because of the 3-D plot.}
    \label{fig:per:gm_bif}
\end{figure}

Next, we consider the periodic problem with GM intra-compartmental
kinetics (\ref{cell:GM}). By solving (\ref{per2:eq}) and
(\ref{p2:symm}) using MatCont \cite{matcont}, in
Fig.~\ref{fig:per:gm_bif} we show that the bifurcation diagram of
steady-states for $n=2$ compartments and with $D_u=D_v=1$ exhibits a
hysteresis structure between the symmetric and asymmetric steady-state
solution branches.

\begin{figure}[]
    \centering
	\begin{subfigure}[b]{.47\textwidth}
	    \begin{subfigure}[b]{1.\textwidth}
	        \centering
	        \includegraphics[width=1.\linewidth]{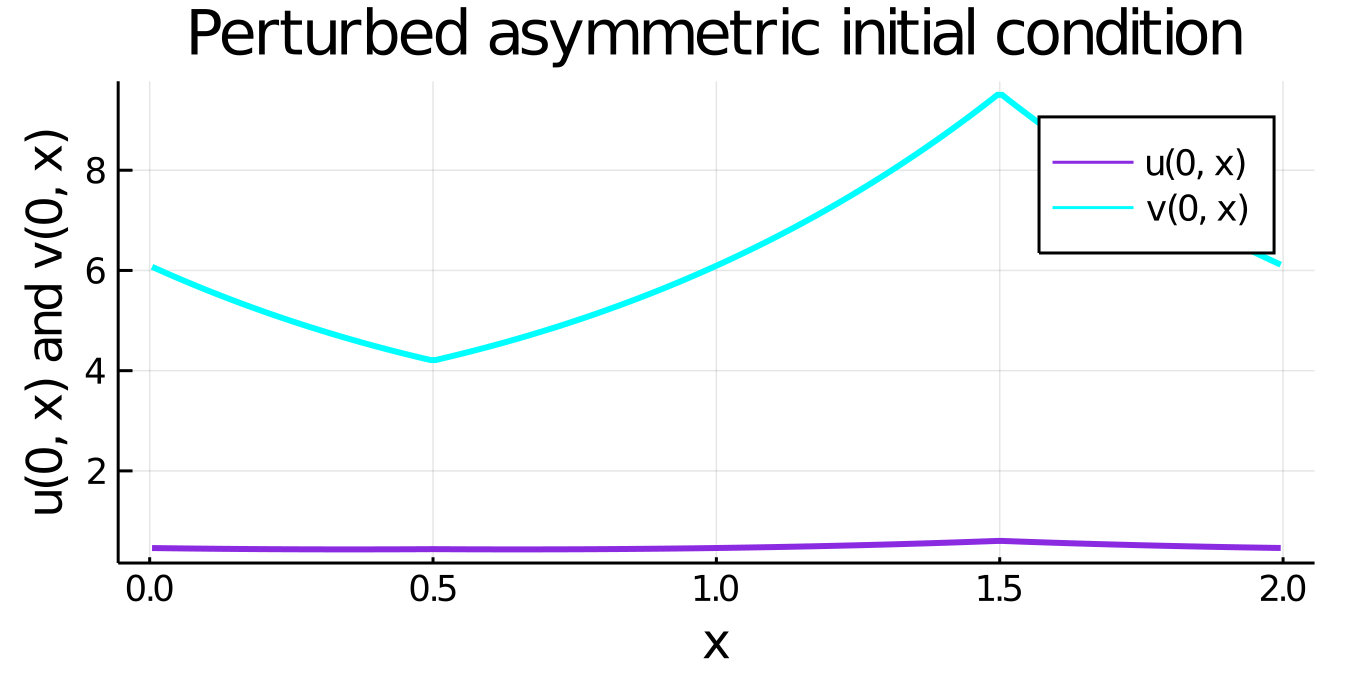}
	    \end{subfigure}
	    \begin{subfigure}[b]{1.\textwidth}
	        \centering
	        \includegraphics[width=1.\linewidth]{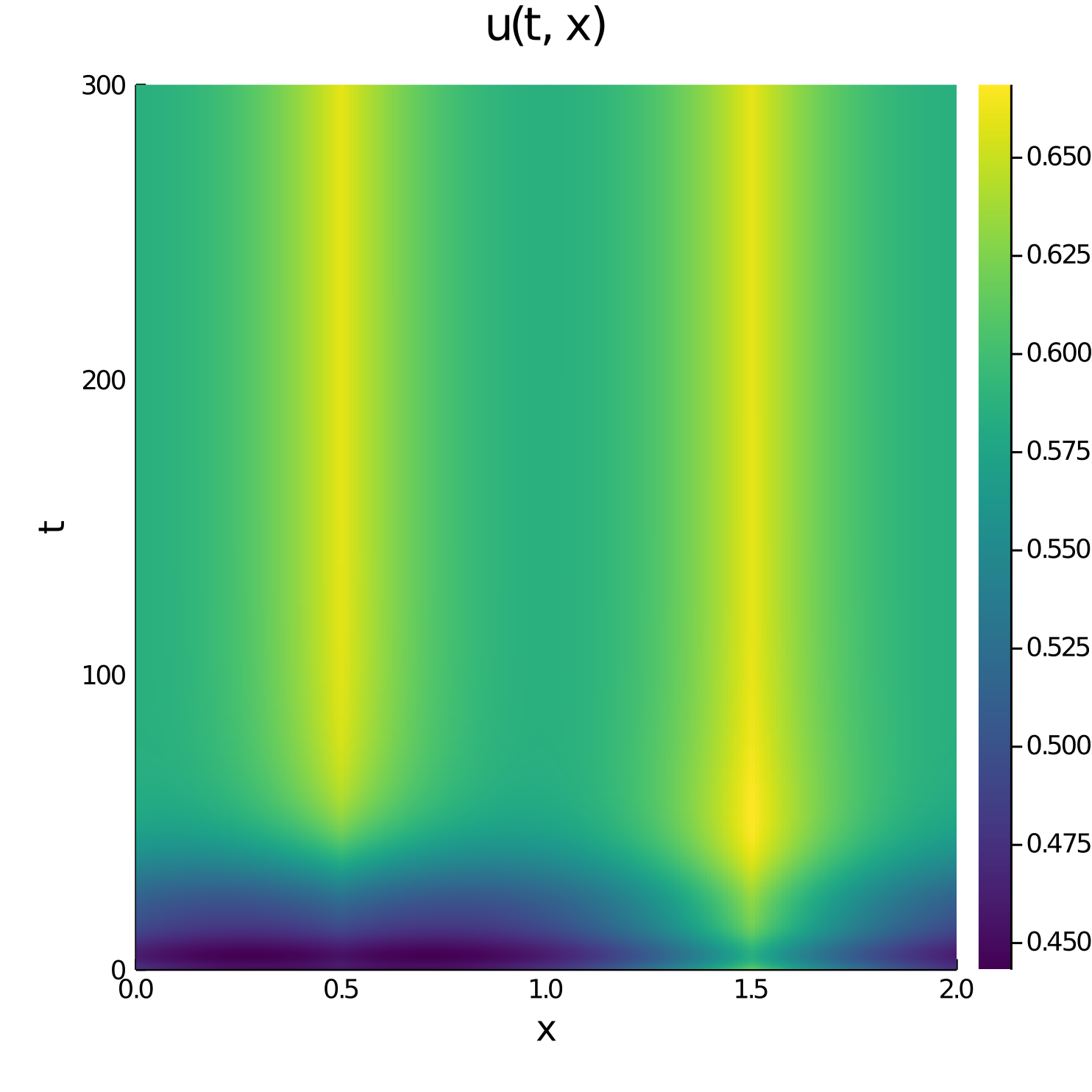}
	    \end{subfigure}
	    \begin{subfigure}[b]{1.\textwidth}
    		\centering
			\includegraphics[width=1.\linewidth]{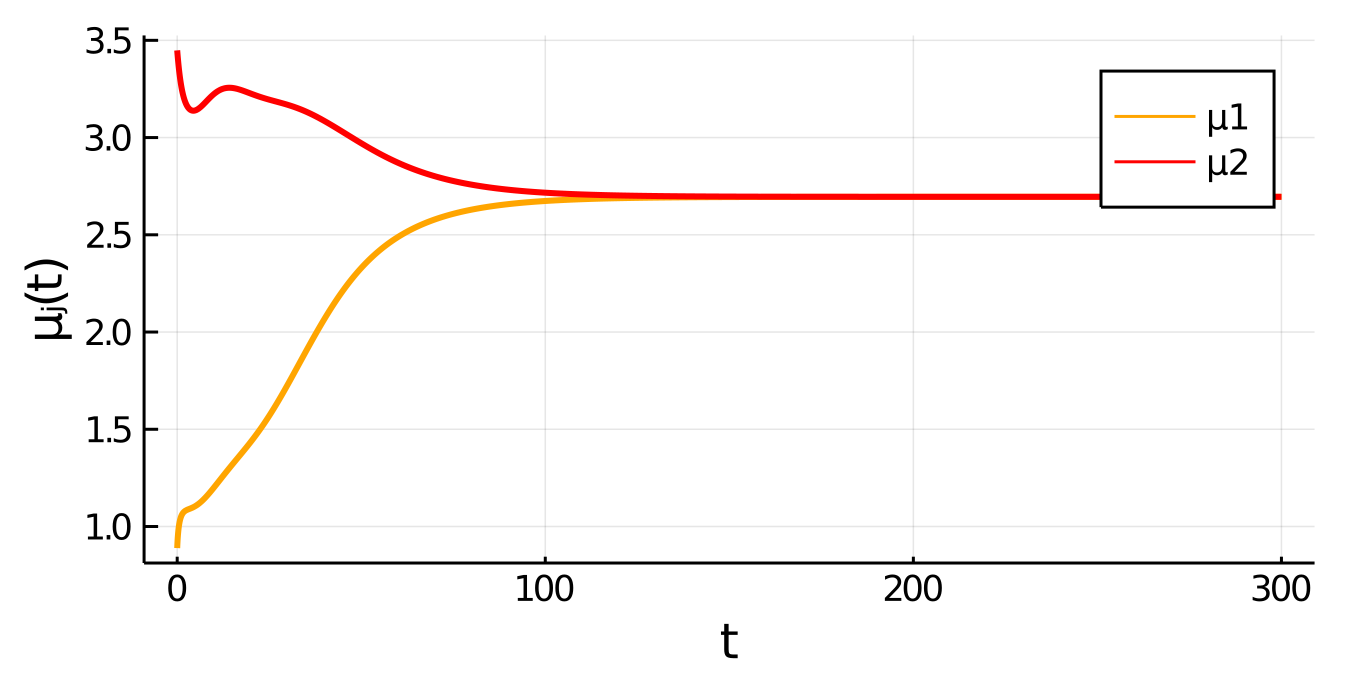}  
		\end{subfigure}
		\begin{subfigure}[b]{1.\textwidth}
			\centering
			\includegraphics[width=1.\linewidth]{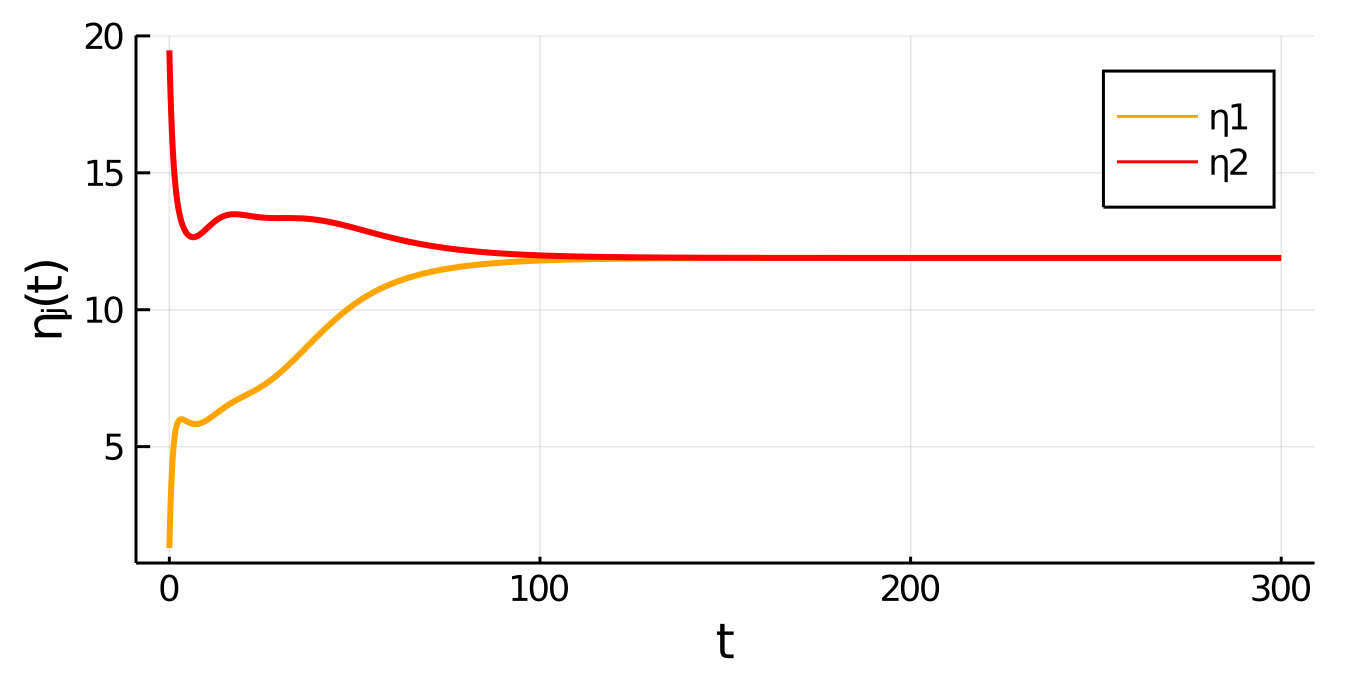}  
		\end{subfigure}
		\begin{subfigure}[b]{1.\textwidth}
			\centering
			\includegraphics[width=1.\linewidth]{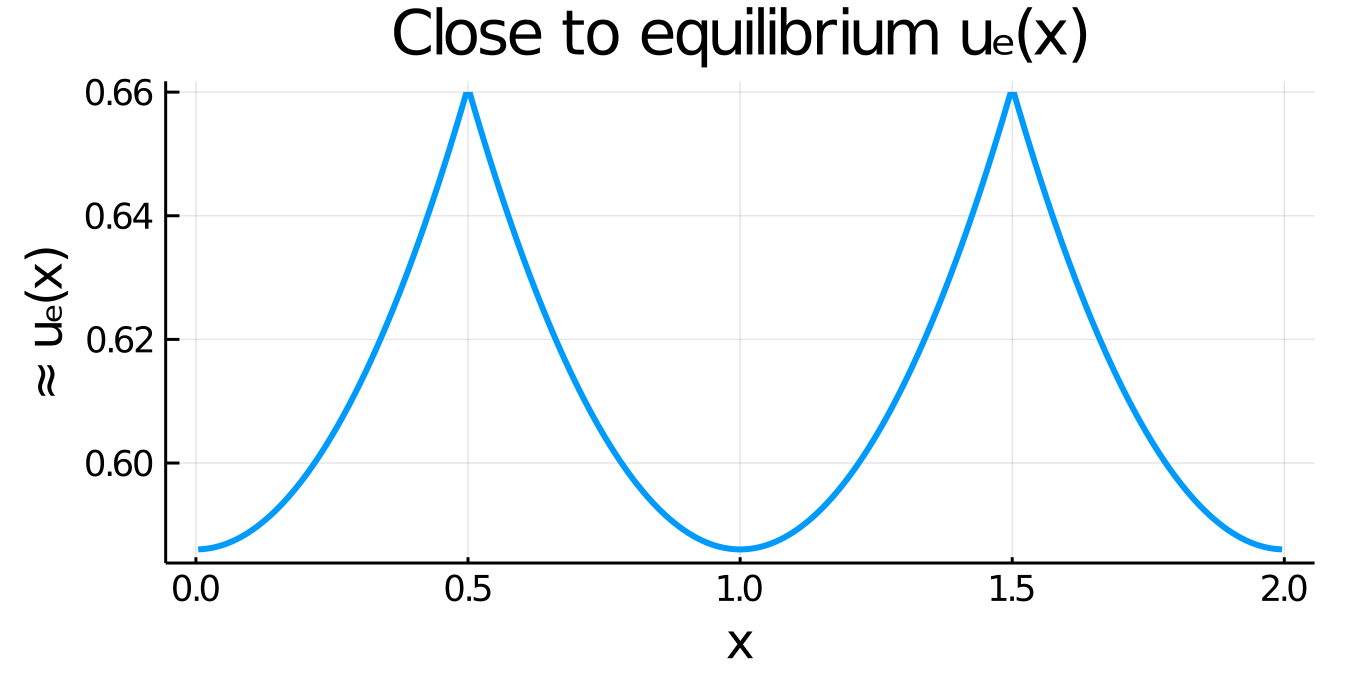}  
		\end{subfigure}
	\end{subfigure}
	\begin{subfigure}[b]{.47\textwidth}
	    \begin{subfigure}[b]{1.\textwidth}
	        \centering
	        \includegraphics[width=1.\linewidth]{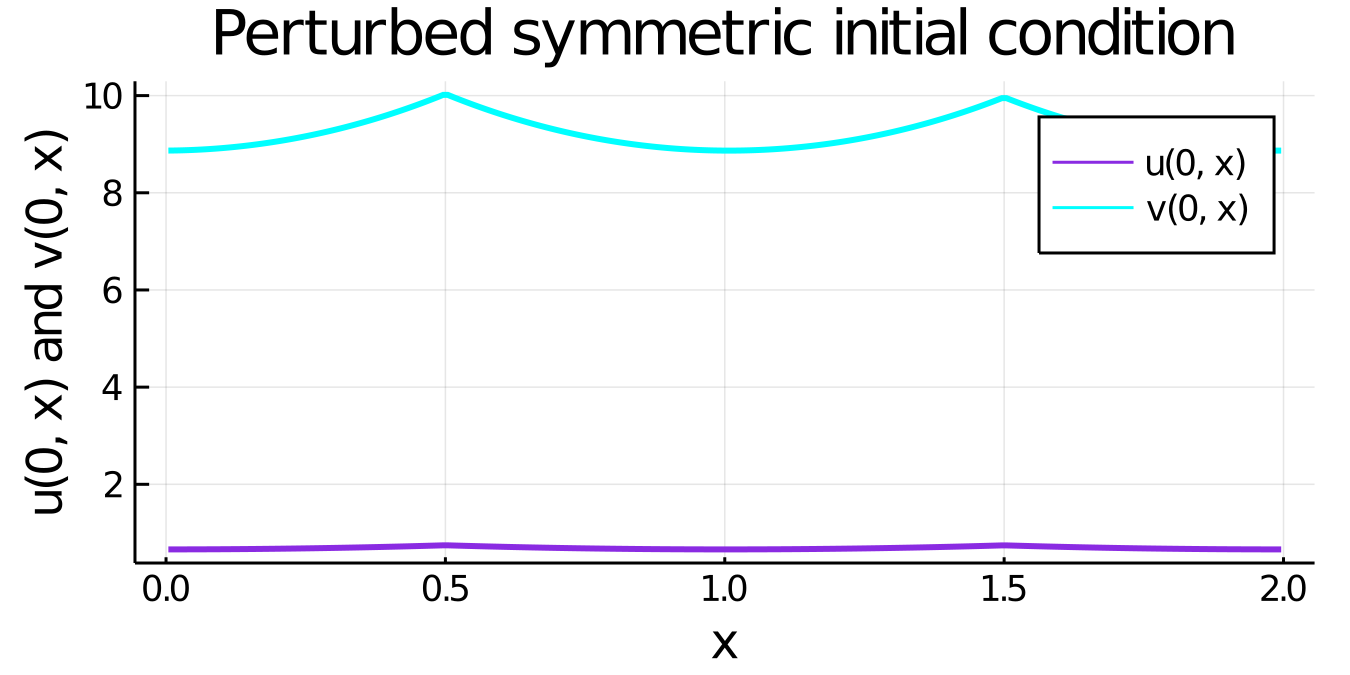}
	    \end{subfigure}
	    \begin{subfigure}[b]{1.\textwidth}
	        \centering
	        \includegraphics[width=1.\linewidth]{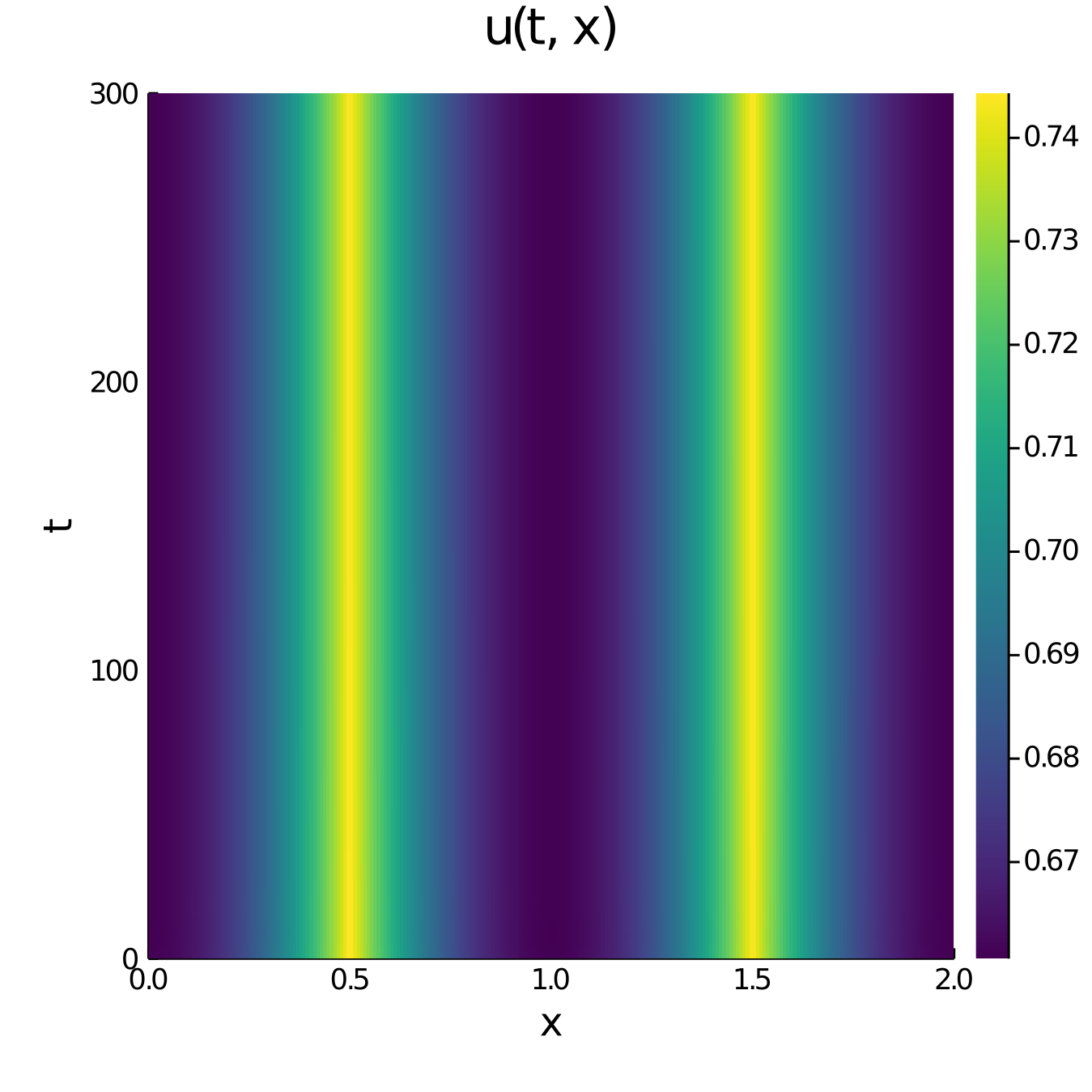}
	    \end{subfigure}
	    \begin{subfigure}[b]{1.\textwidth}
    		\centering
			\includegraphics[width=1.\linewidth]{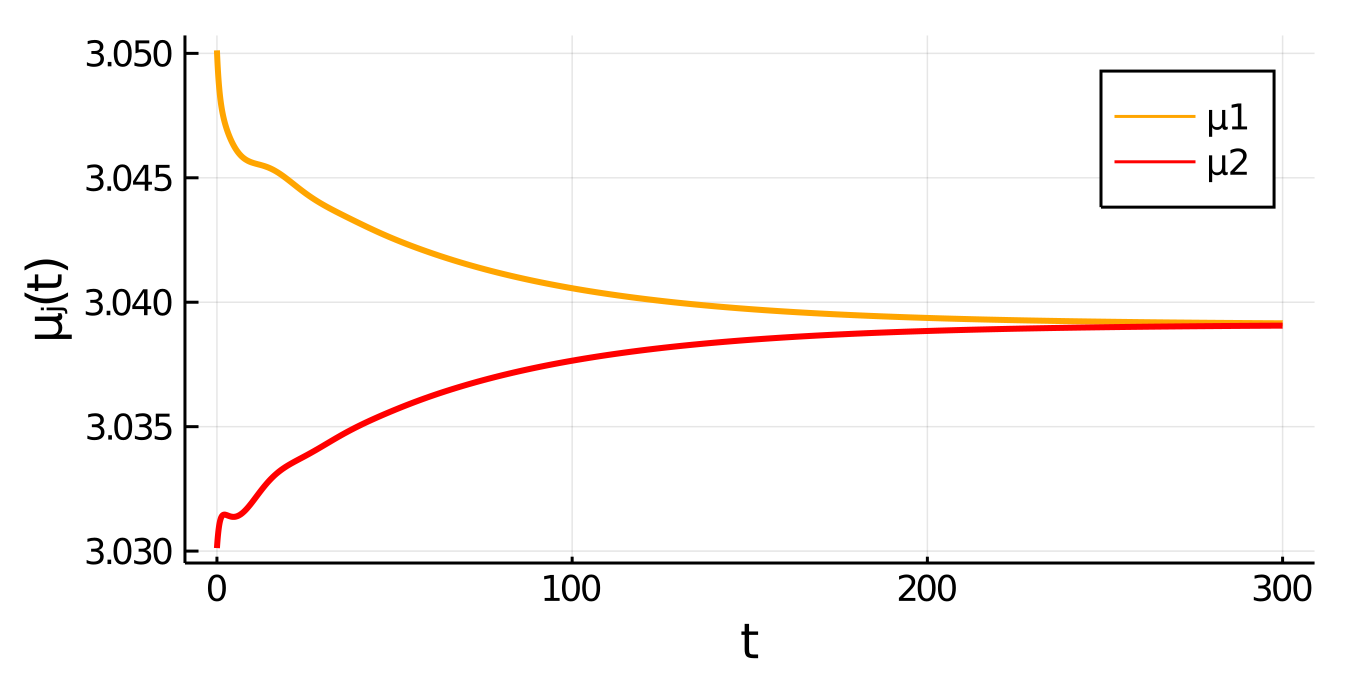}  
		\end{subfigure}
		\begin{subfigure}[b]{1.\textwidth}
			\centering
			\includegraphics[width=1.\linewidth]{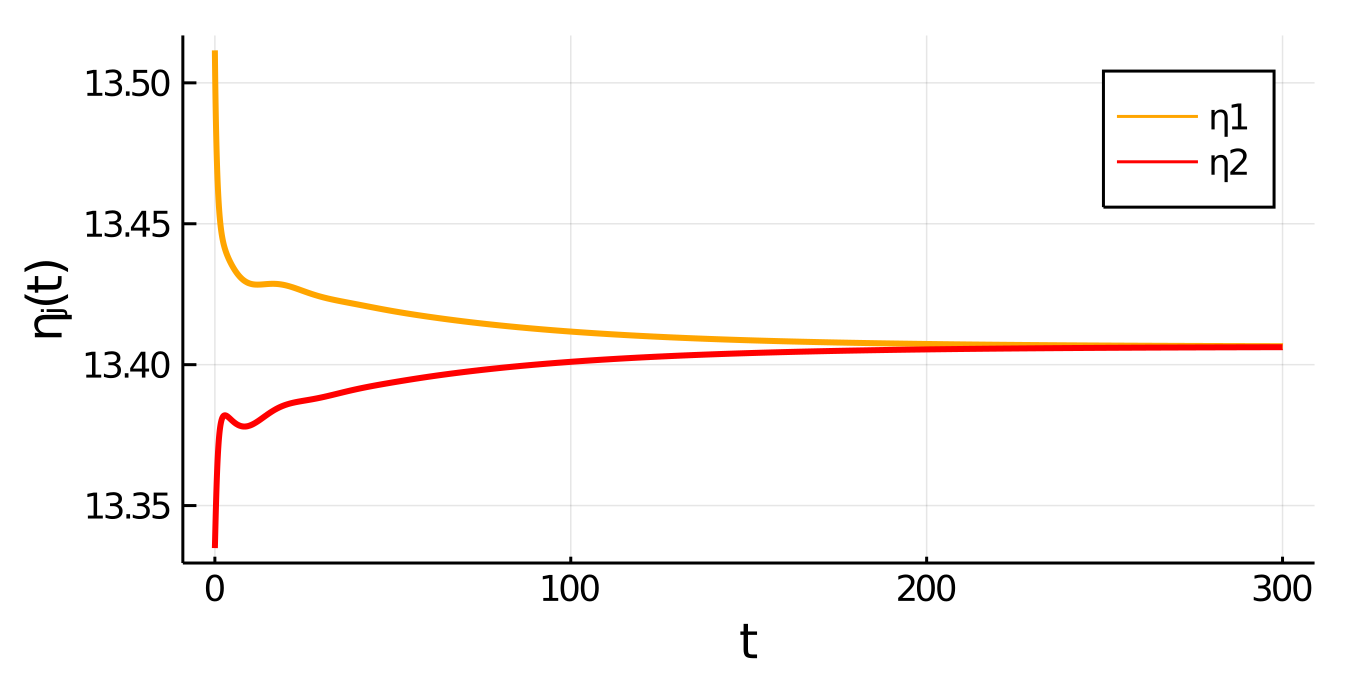}  
		\end{subfigure}
		\begin{subfigure}[b]{1.\textwidth}
			\centering
			\includegraphics[width=1.\linewidth]{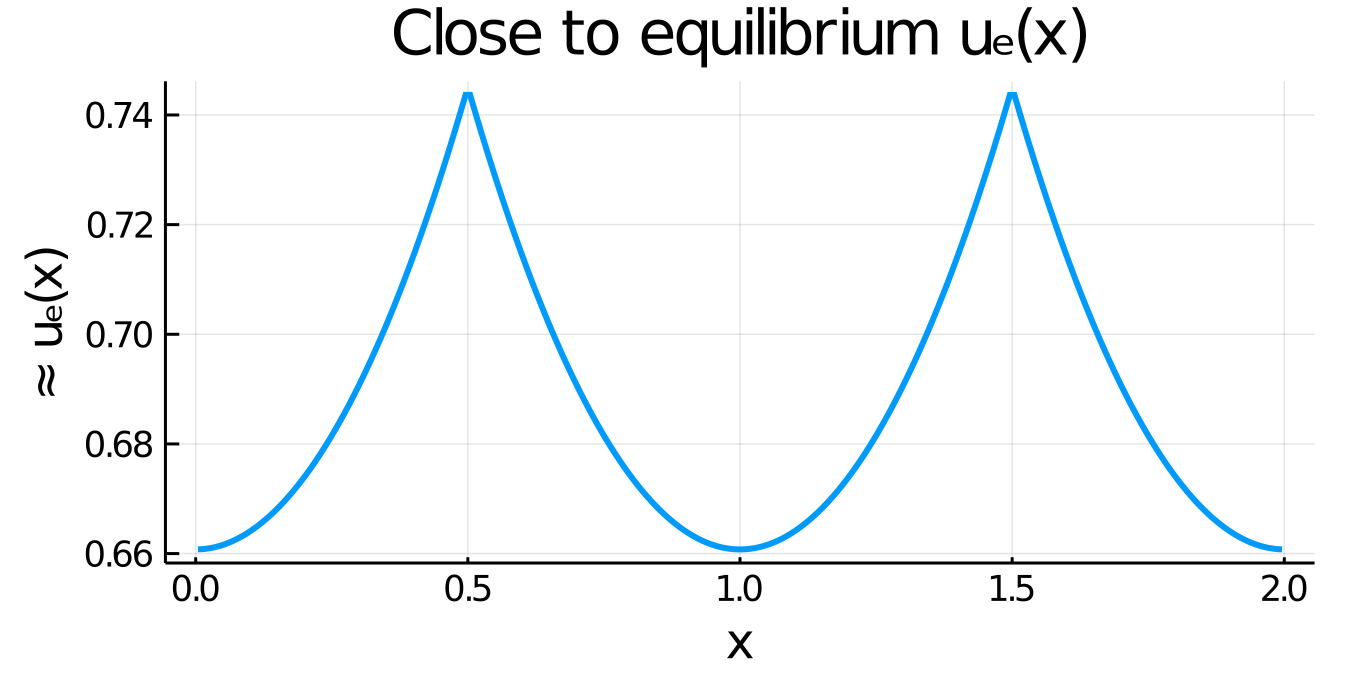}  
		\end{subfigure}
	\end{subfigure}
	\caption{GM kinetics with $n=2$. Full numerical time-dependent results
          from the BE-RK4 IMEX scheme. Left: convergence to the stable
          symmetric branch when $\rho=6$ starting just prior to the
          fold point where a stable asymmetric steady-state is born.
          Right: convergence to the stable symmetric branch for
          $\rho=9$ with an initial condition near the symmetric branch.
          Remaining parameter values are as in the caption to
          Fig.~\ref{fig:per:gm_bif}.}
	\label{fig:symmhysteresisring}
\end{figure}
\begin{figure}[] 
    \centering
	\begin{subfigure}[b]{.47\textwidth}
	    \begin{subfigure}[b]{1.\textwidth}
	        \centering
	        \includegraphics[width=1.\linewidth]{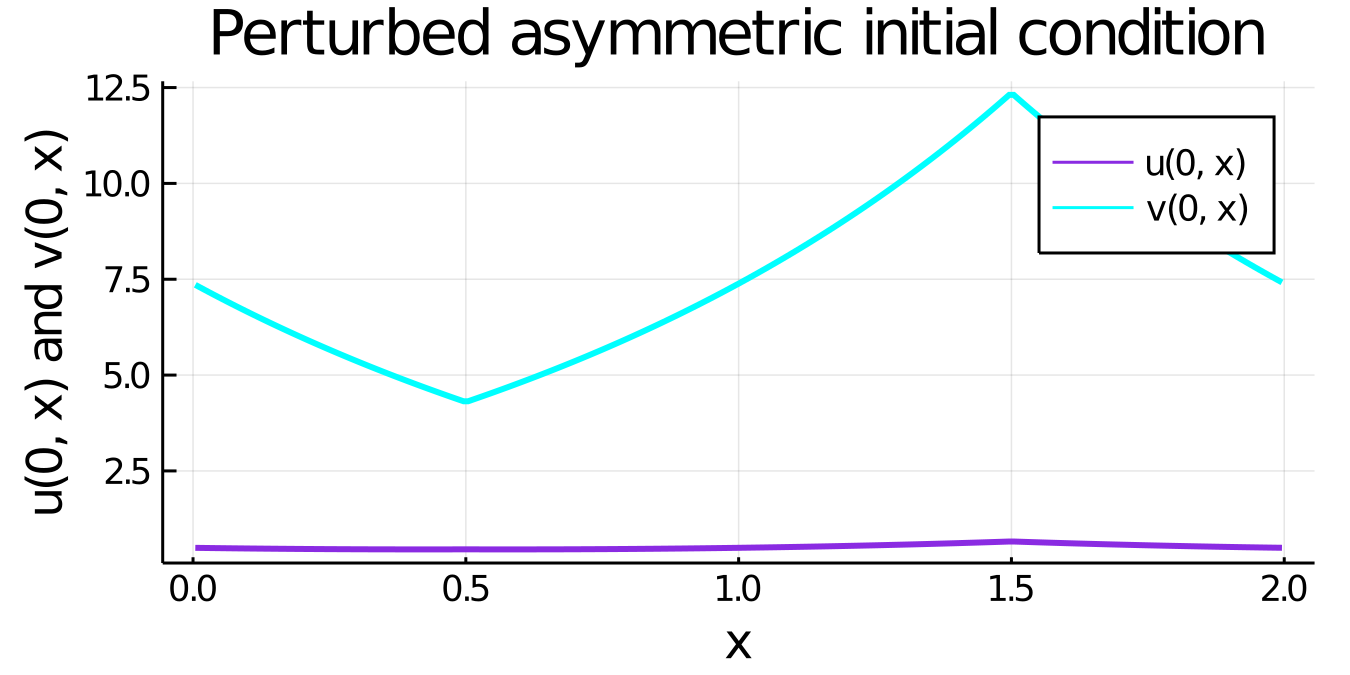}
	    \end{subfigure}
	    \begin{subfigure}[b]{1.\textwidth}
	        \centering
	        \includegraphics[width=1.\linewidth]{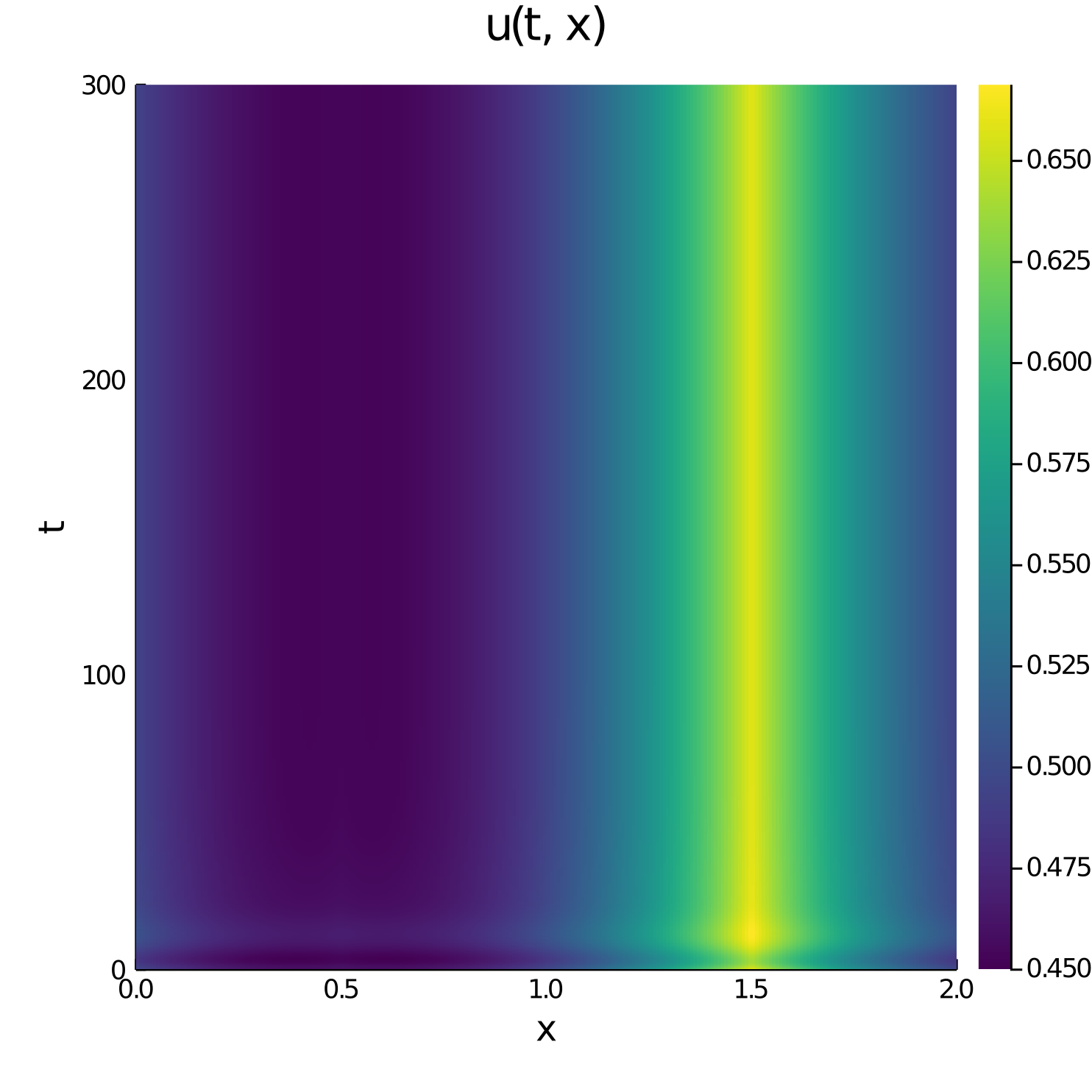}
	    \end{subfigure}
	    \begin{subfigure}[b]{1.\textwidth}
    		\centering
			\includegraphics[width=1.\linewidth]{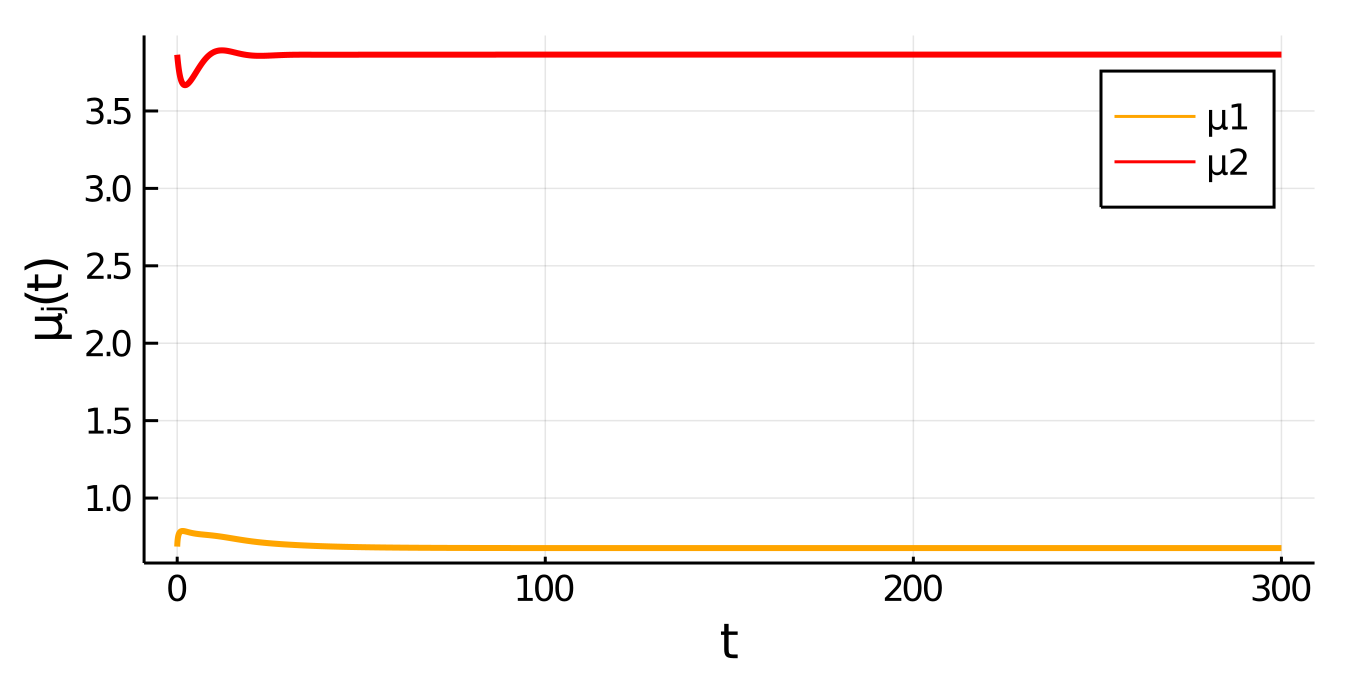}  
		\end{subfigure}
		\begin{subfigure}[b]{1.\textwidth}
			\centering
			\includegraphics[width=1.\linewidth]{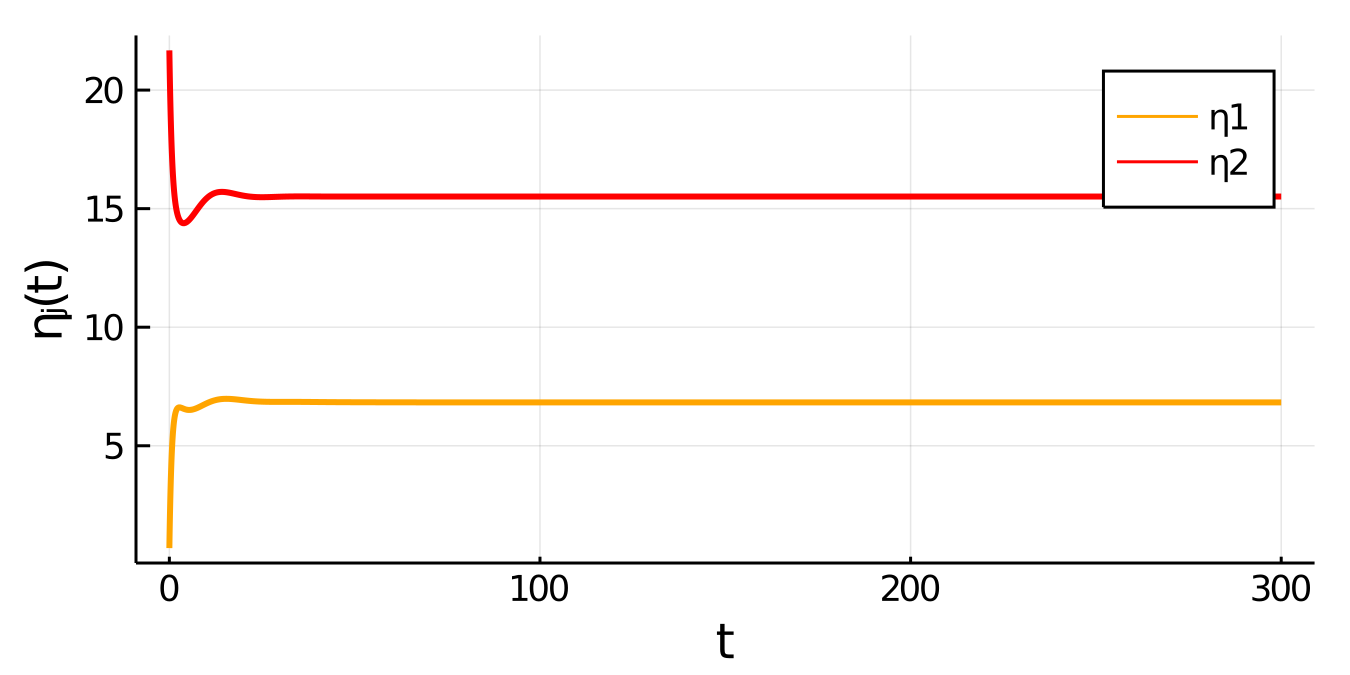}  
		\end{subfigure}
		\begin{subfigure}[b]{1.\textwidth}
			\centering
			\includegraphics[width=1.\linewidth]{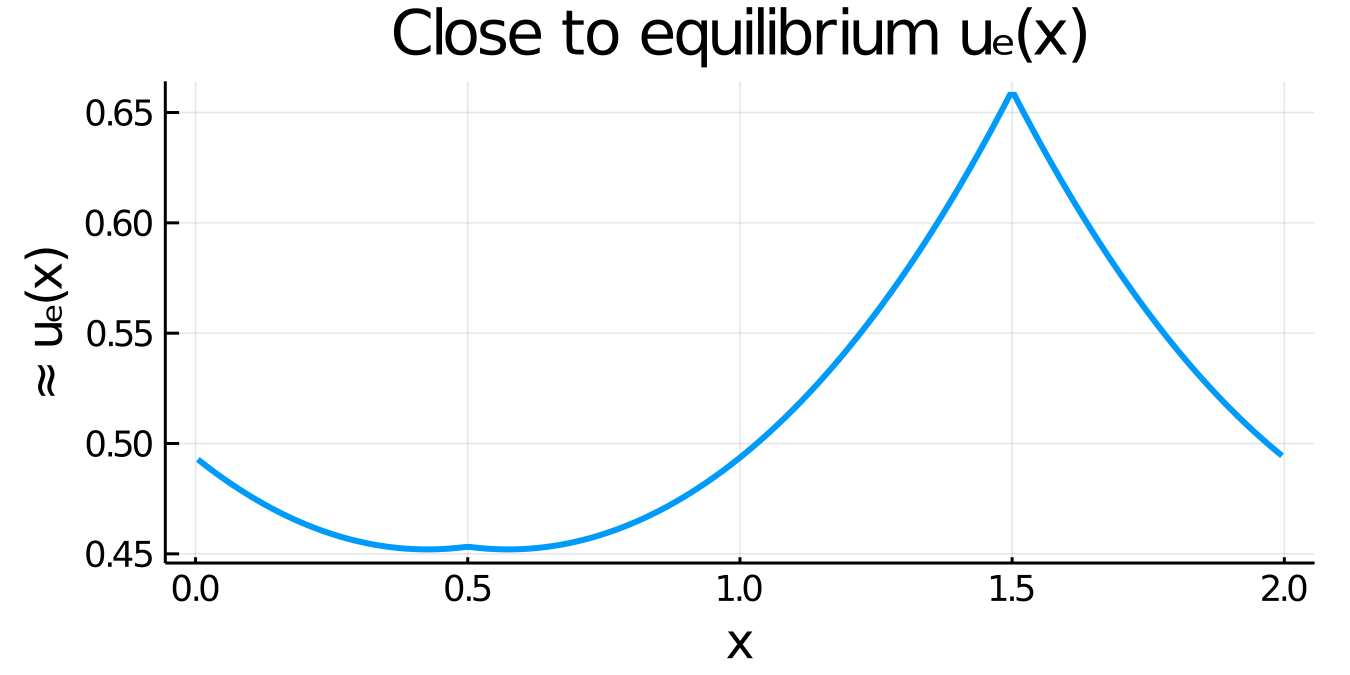}  
		\end{subfigure}
	\end{subfigure}
	\begin{subfigure}[b]{.47\textwidth}
	    \begin{subfigure}[b]{1.\textwidth}
	        \centering
	        \includegraphics[width=1.\linewidth]{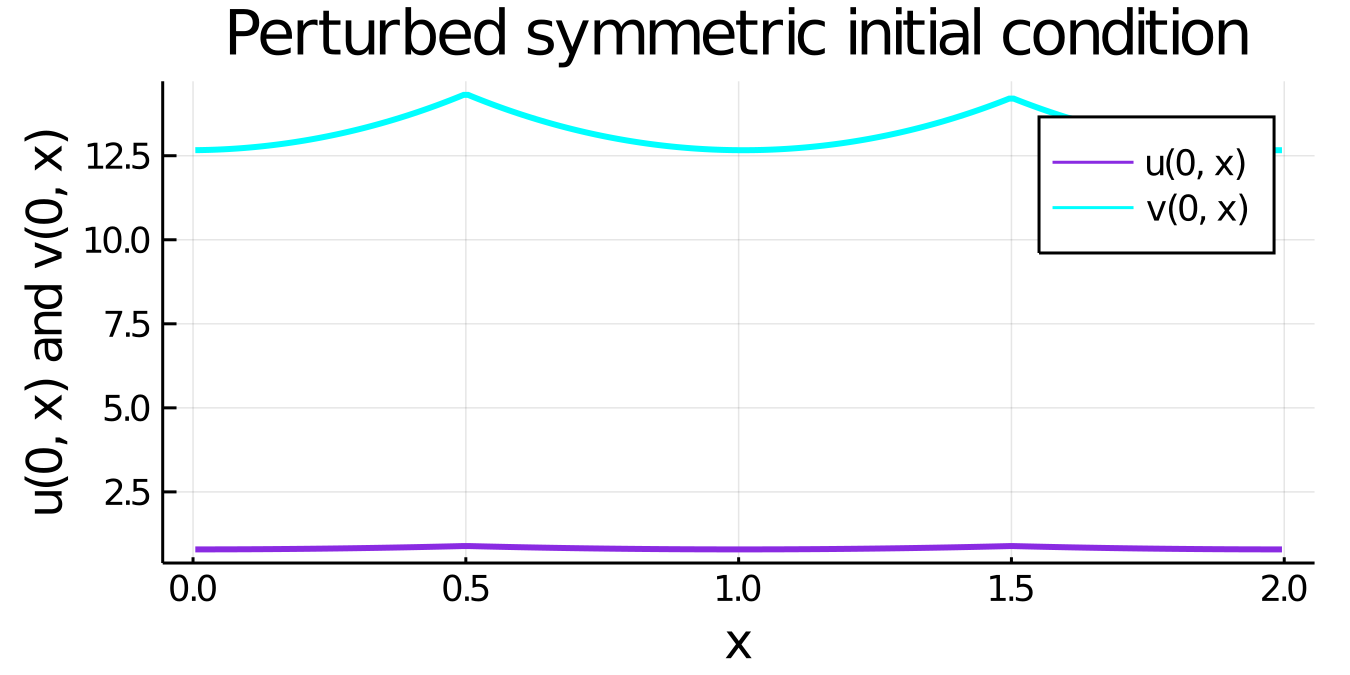}
	    \end{subfigure}
	    \begin{subfigure}[b]{1.\textwidth}
	        \centering
	        \includegraphics[width=1.\linewidth]{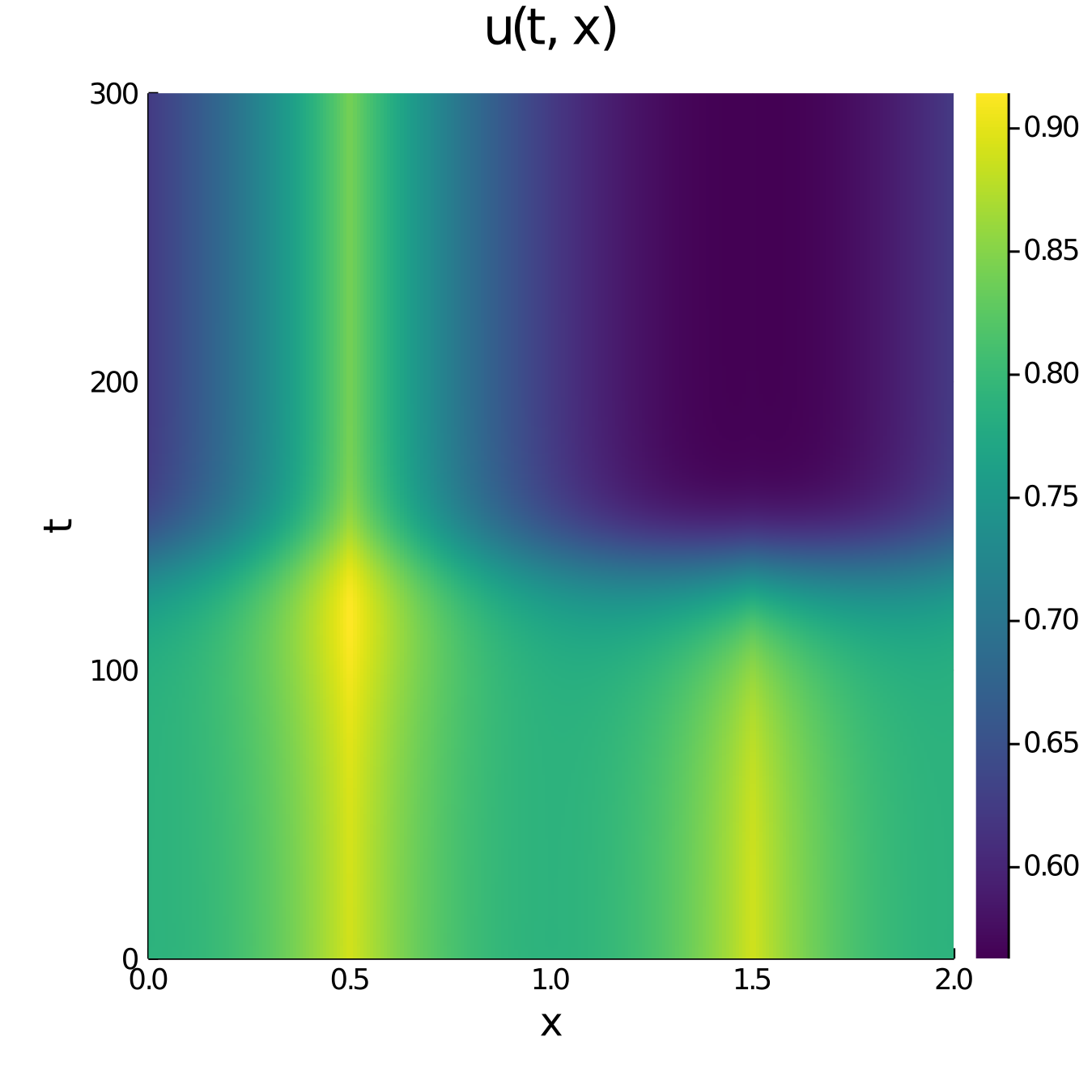}
	    \end{subfigure}
	    \begin{subfigure}[b]{1.\textwidth}
    		\centering
			\includegraphics[width=1.\linewidth]{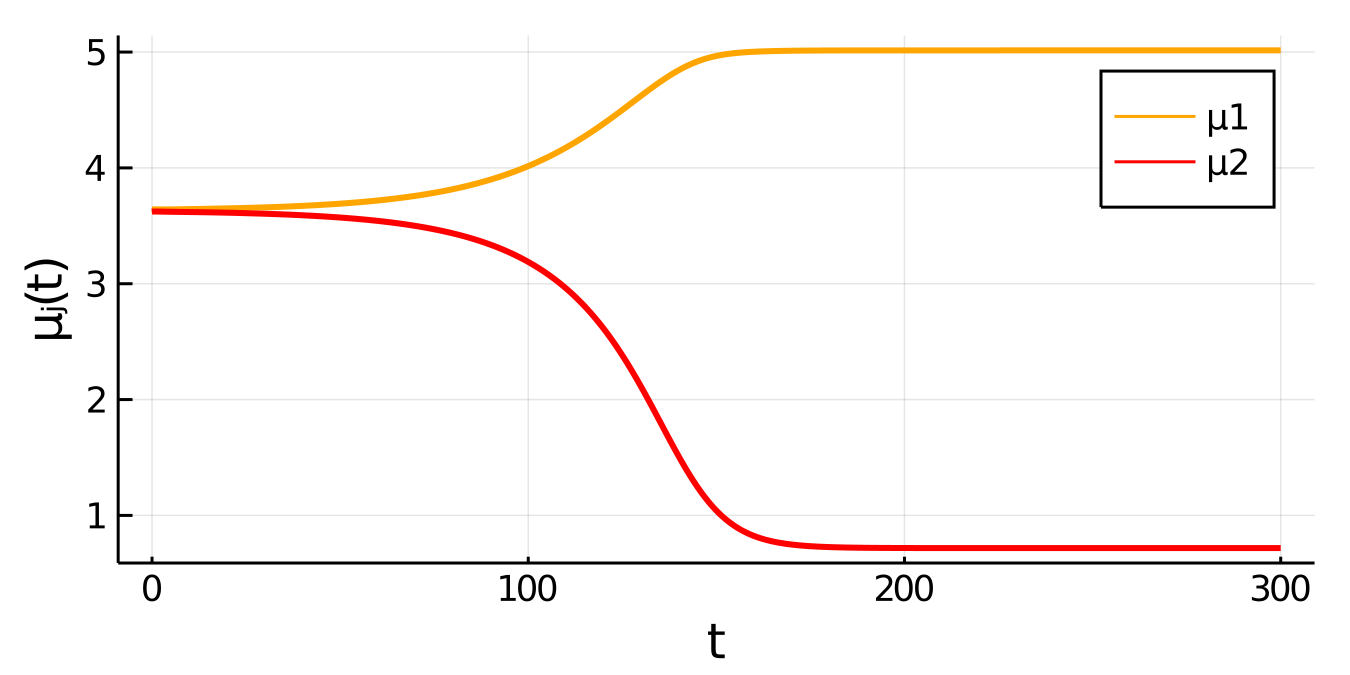}  
		\end{subfigure}
		\begin{subfigure}[b]{1.\textwidth}
			\centering
			\includegraphics[width=1.\linewidth]{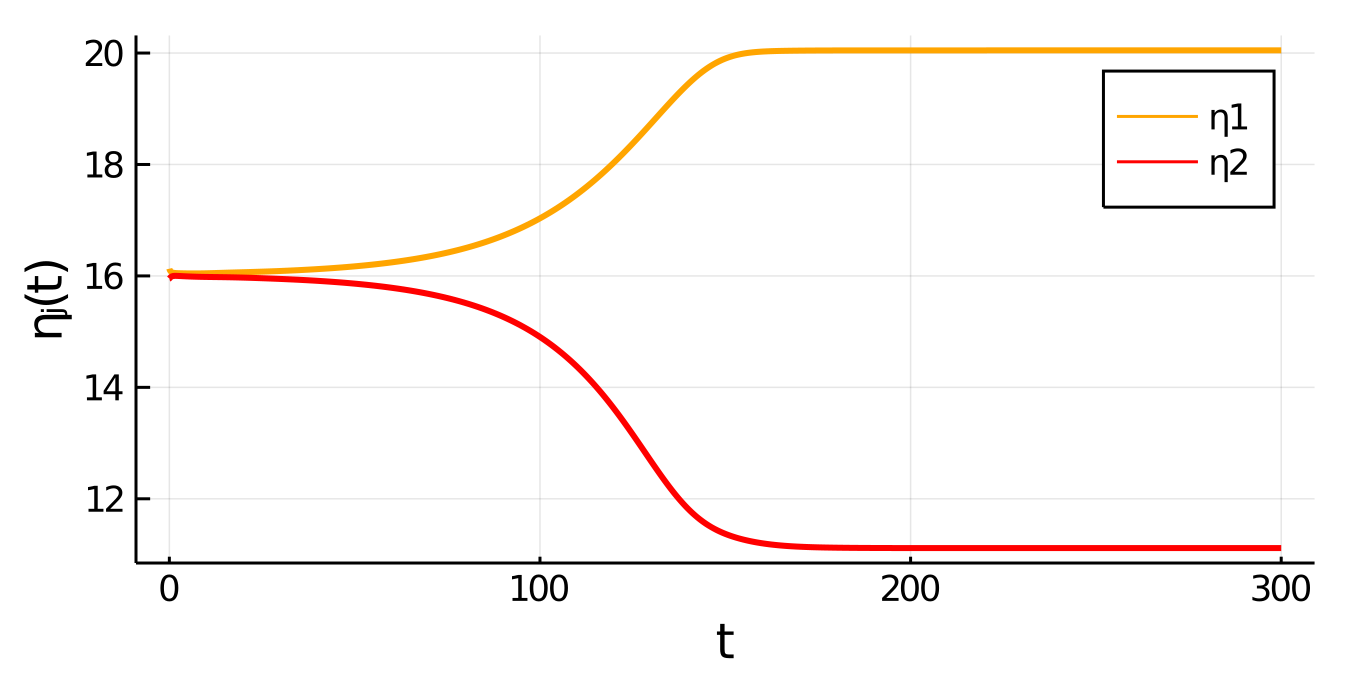}  
		\end{subfigure}
		\begin{subfigure}[b]{1.\textwidth}
			\centering
			\includegraphics[width=1.\linewidth]{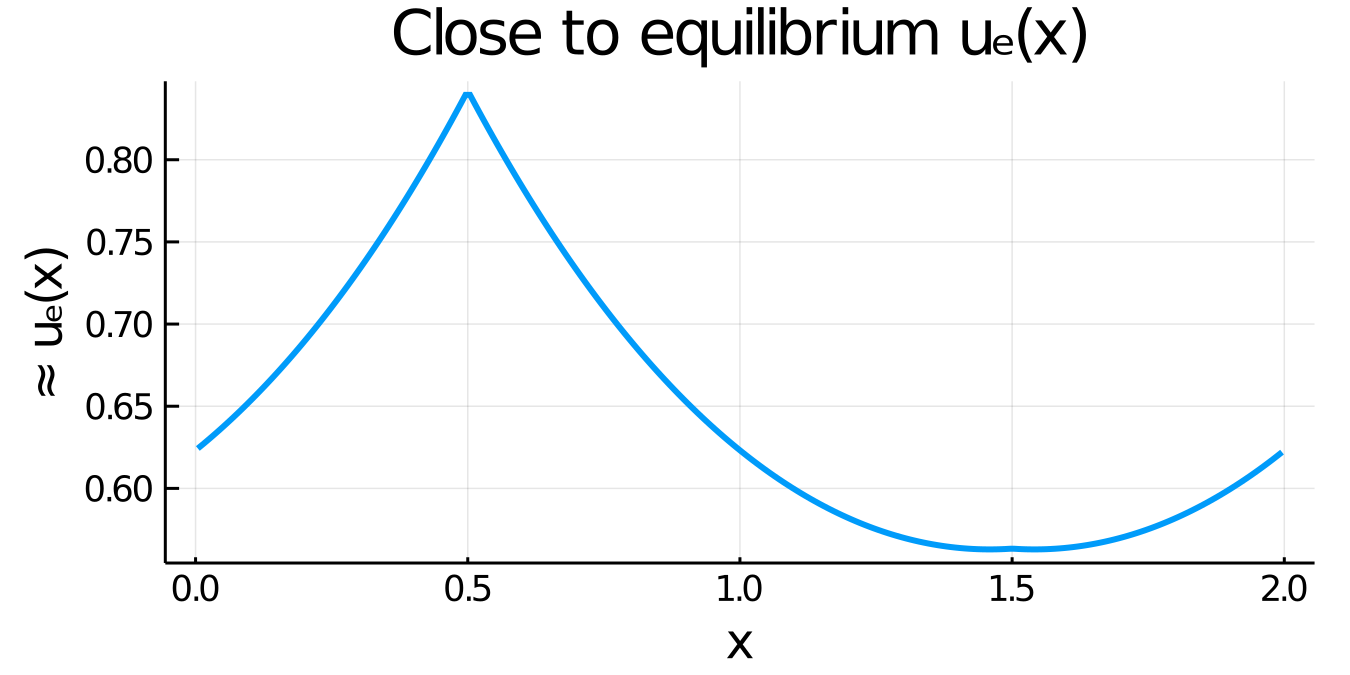}  
		\end{subfigure}
	\end{subfigure}
	\caption{GM kinetics with $n=2$. Full numerical time-dependent results
          from the BE-RK4 IMEX scheme. Left: convergence to a
          stable asymmetric branch when $\rho=9$ starting close to the
          stable asymmetric branch. Right: convergence to a stable
          asymmetric branch when $\rho=25$ starting close to the unstable
          symmetric branch. Remaining parameter values are as in
          the caption to Fig.~\ref{fig:per:gm_bif}.}
	\label{fig:asymmhysteresisring}
\end{figure}

From the numerical results given in Table \ref{tab:ring-cell_1}, we observe
that asymmetric branches do bifurcate from the symmetric branch even when
${D_v/D_u<1}$, and that the range of $\rho$ where hysteresis occurs is larger
as the ratio $D_v/D_u$ decreases. In Table \ref{tab:ring-cell_2} we show
for $D_u=D_v=1$ that as $\beta_u$ is increased beyond $\beta_u=0.7$ the
pitchfork bifurcation switches from sub- to super-critical. The overall
trends are qualitatively similar to those given in
Tables~\ref{tab:2-cell GM hysteresis} and \ref{tab:2-cell GM hysteresis iota}
for the case of reactive domain boundaries.

\begin{table}[ht]
\centering
\begin{tabular}{|c||c|c|c|c|c|c|c|}
    \hline
    $\frac{D_v}{D_u}$   & 0.4   & 0.5       & 0.6       & 0.7       & 0.8       & 0.9       & 1\\
    \hline\hline
    $\rho_p$            & $>100$& 53.17080  & 35.32088  & 28.23715  & 24.43531  & 22.06333  & 20.44213 \\
    \hline
    $\mu_e$             &       & 8.21230   & 7.81358   & 7.53645   & 7.33272   & 7.17669   & 7.05336 \\
    \hline\hline
    $\rho_f$            &       & 6.25939   & 5.66710   & 5.29081   & 5.03058   & 4.83989   & 4.69415 \\
    $\mu_1^e$           &       & 4.45524   & 4.23893   & 4.08858   & 3.97806   & 3.89341   & 3.82650 \\
    $\mu_2^e$           &       & 0.37314   & 0.35503   & 0.34243   & 0.33318   & 0.32609  & 0.32048 \\
    \hline
\end{tabular}

\begin{tabular}{|c||c|c|c|c|c|c|c|}
    \hline
    $\frac{D_v}{D_u}$   & 1.1       & 1.2       & 1.3       & 1.4       & 1.5       & 1.6 \\
    \hline\hline
    $\rho_p$            & 19.26390  & 18.36889  & 17.66592  & 17.09916  & 16.63252  & ...  \\
    \hline
    $\mu_e$             & 6.95344   & 6.87085   & 6.80144   & 6.74229   & 6.69129   & \\
    \hline\hline
    $\rho_f$            & 4.57915   & 4.48608   & 4.40922   & 4.34468   & 4.28971   &  \\
    $\mu_1^e$           & 3.77230   & 3.72749   & 3.68983   & 3.65775   & 3.63008   &  \\
    $\mu_2^e$           & 0.31594   & 0.31219   & 0.30904   & 0.30635   & 0.30403   &  \\
    \hline
\end{tabular}
\caption{Numerical values (rounded to 5th decimal place) of the
  subcritical pitchfork bifurcation point $\rho_p$ for the symmetric
  branch and the fold points $\rho_f$ along the asymmetric branches.
  Bifurcation values for the symmetric ($\mu_e$) and one of the
  asymmetric ($\mu_1^e$, $\mu_2^e$) solution branches are listed. As
  ${D_v/D_u}$ decreases, the range of $\rho$ where hysteresis
  occurs increases. Parameters: $D_u=1, \sigma_u = \sigma_v=1, L=1$,
  and $\beta_u=0.1$.}\label{tab:ring-cell_1}
\end{table}

\begin{table}[ht]
\centering
\begin{tabular}{|c||c|c|c|c|c|c|c|c|}
    \hline
    $\beta_u$             & 0.05      & 0.1       & 0.15      & 0.2       & 0.25      & 0.3       & 0.35      & 0.4  \\
    \hline\hline
    $\rho_p$            & 36.28648  & 20.44213  & 15.39744  & 13.09728  & 11.94362  & 11.42015  & 11.32738  & 11.59667 \\
    \hline
    $\mu_e$             & 12.90881  & 7.05336   & 5.11479   & 4.15511   & 3.58673   & 3.21380   & 2.95240   & 2.76056  \\
    \hline\hline
    $\rho_f$            & 4.32660   & 4.69415   & 5.11246   & 5.59447   & 6.15768   & 6.82631   & 7.63502   & 8.63523  \\
    $\mu_1^e$           & 3.91408   & 3.82650   & 3.73728   & 3.64635   & 3.55352   & 3.45845   & 3.36053   & 3.25882  \\
    $\mu_2^e$           & 0.16492   & 0.32048   & 0.46828   & 0.60972   & 0.74617   & 0.87895   & 1.00951   & 1.13950  \\
    \hline
\end{tabular}

\begin{tabular}{|c||c|c|c|c|c|c|}
    \hline
    $\beta_u$             & 0.45      & 0.5       & 0.55      & 0.6       & 0.65      & 0.7\\
    \hline\hline
    $\rho_p$            & 12.23745  & 13.33412  & 15.08439  & 17.91935  & 22.88248  & 33.18867\\
    \hline
    $\mu_e$             & 2.61497   & 2.50167   & 2.41175   & 2.33930   & 2.28021   & 2.231556 \\
    \hline\hline
    $\rho_f$            & 9.90668   & 11.58010  & 13.88586  & 17.27125  & 22.73469  & - \\
    $\mu_1^e$           & 3.15185   & 3.03717   & 2.91034   & 2.76171   & 2.55994   & - \\
    $\mu_2^e$           & 1.27099   & 1.40694   & 1.55225   & 1.71692   & 1.93262   & - \\
    \hline
\end{tabular}
\caption{Similar caption as in Table \ref{tab:ring-cell_1} except that
  the bifurcation quantities for $\rho={\beta_v/\beta_u}$ are
  tabulated for various values of $\beta_u$ when $D_v=D_u=1$. The
  range where hysteresis occurs decreases as $\beta_u$ increases, and
  when $\beta_u\approx 0.7$ the pitchfork bifurcation switches from
  sub- to super critical. Other parameters are as in the caption of Table
  \ref{tab:ring-cell_1}.}
\label{tab:ring-cell_2}
\end{table}

Although we have not provided a linear stability theory for the asymmetric
steady-state solution branches in Fig.~\ref{fig:per:gm_bif}, the full PDE
numerical results shown in
Figs.~\ref{fig:symmhysteresisring}--\ref{fig:asymmhysteresisring}, as
computed using the BE-RK4 IMEX scheme of Appendix \ref{num:ring},
confirm the hysteresis structure of Fig.~\ref{fig:per:gm_bif} and
the prediction of stable asymmetric patterns.

\subsection{Rauch-Millonas kinetics}

\begin{figure}[htbp]
    \centering
    \includegraphics[width=0.48\textwidth]{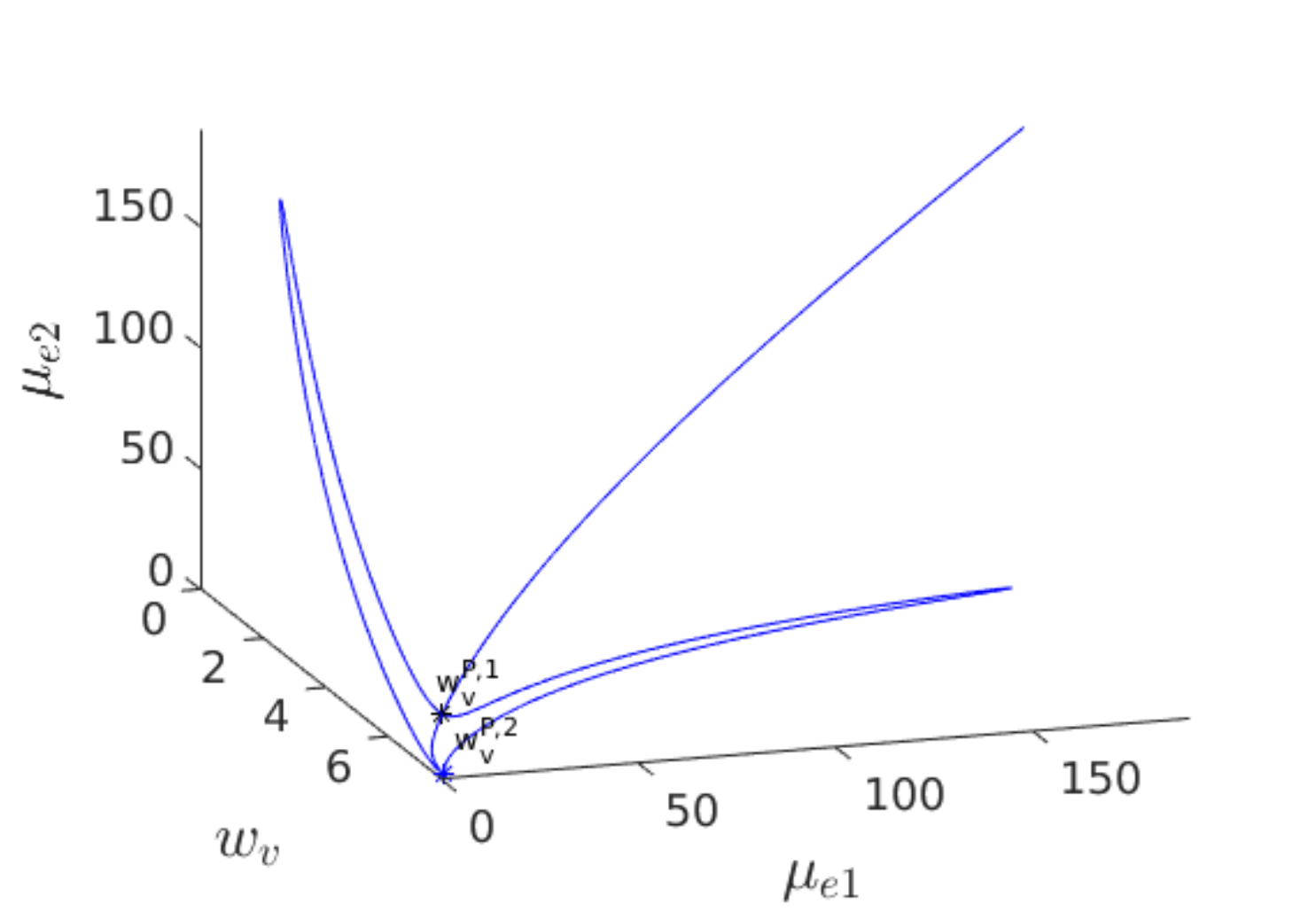}
    \includegraphics[width=0.48\textwidth]{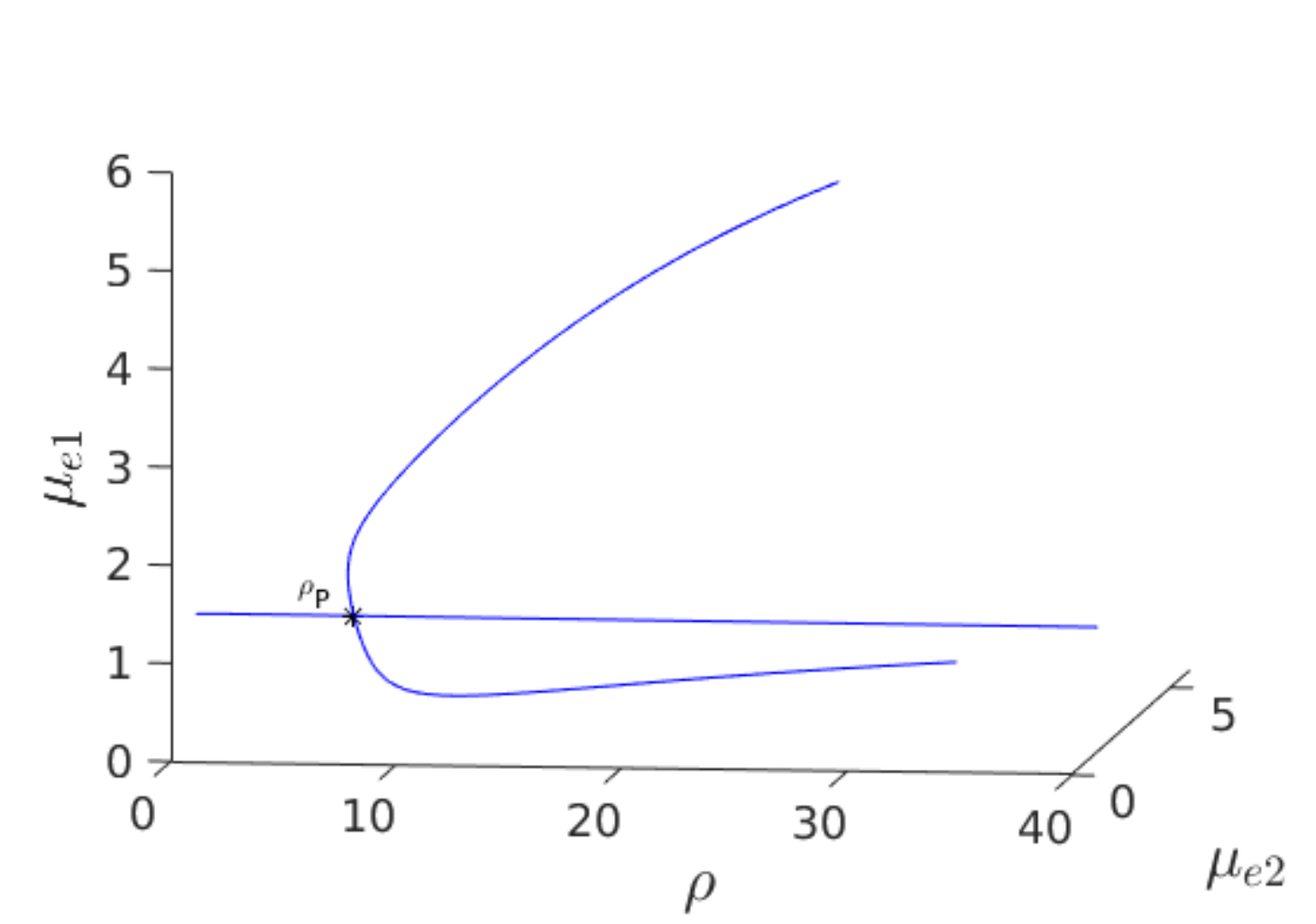}
    \caption{{3-D Bifurcation diagrams for Rauch-Millonas kinetics
      (\ref{ss:rauch}) with $n=2$ computed from (\ref{per2:eq}) using
      MatCont \cite{matcont}.}  Left: Plot of $\mu_1^{e}$ showing that
      asymmetric steady-states occur inside a degenerate pitchfork
      bubble bounded by $w_v^{P,1}\approx 6.34518$ and
      $w_v^{P,2}\approx 7.64062$ when $\rho={\beta_v/\beta_u}=
      7$. Right: Supercritical pitchfork bifurcation in $\rho$ from
      the symmetric branch occurs when $w_v=w_v^{P,2}$. Stable
      asymmetric branches occur past this threshold in
      $\rho$. Parameters:
      $D_u=D_v=1, \sigma_u=\sigma_v=0.01,c_u=c_v=1,q_u={1/100}, q_v=7,
      \alpha_1^u=600,\alpha_2^u=6,
      \gamma_1^u=100,\gamma_2^u={1/10}$, and
      $\beta_u=0.3$.} \label{fig:pitchforkbubbleandrhobifRMring}
\end{figure}

\begin{figure}[htbp]
	\begin{subfigure}[b]{.49\textwidth}
	    \begin{subfigure}[b]{1.\textwidth}
    		\centering
			\includegraphics[width=1.\linewidth]{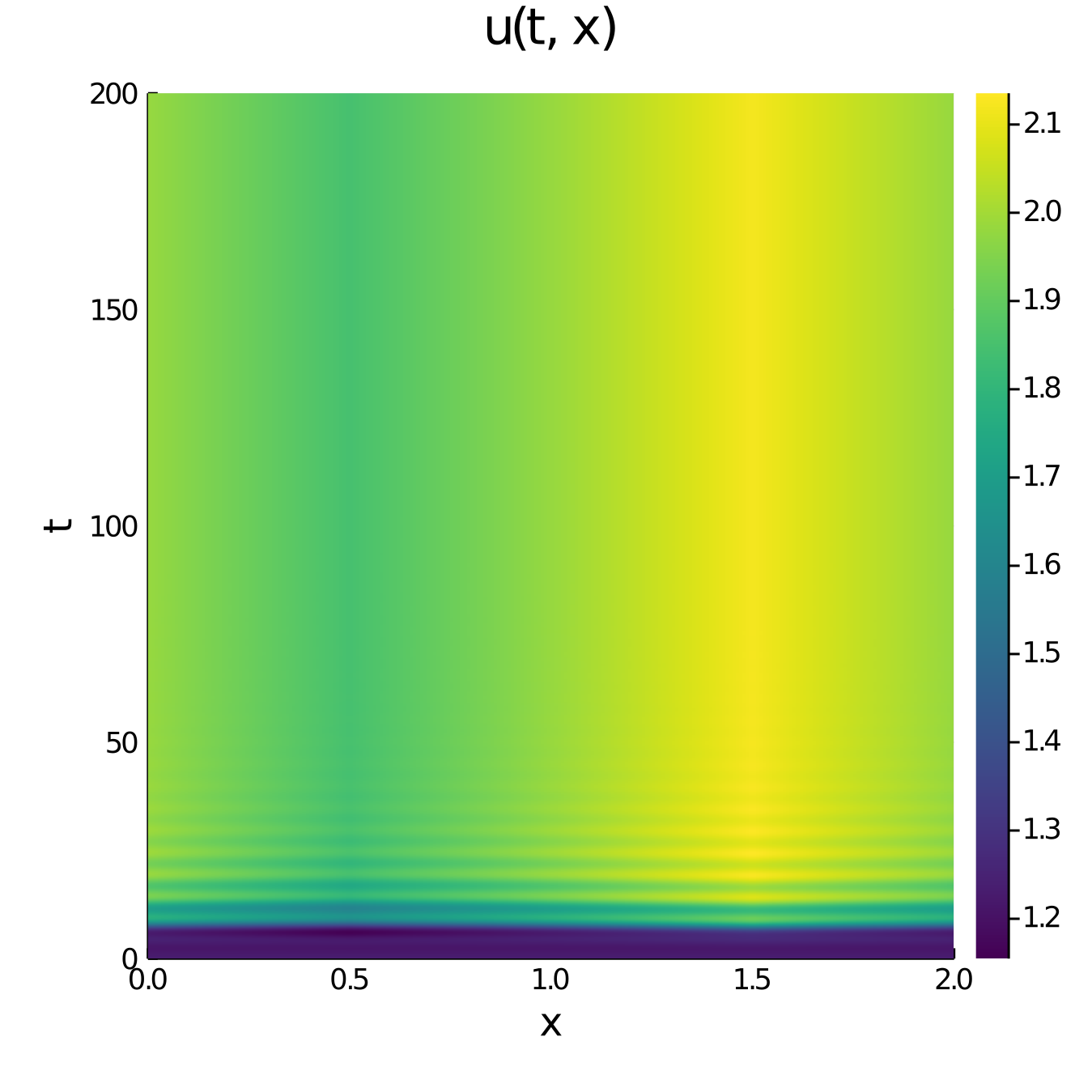}  
		\end{subfigure}
		\begin{subfigure}[b]{1.\textwidth}
    		\centering
			\includegraphics[width=1.\linewidth]{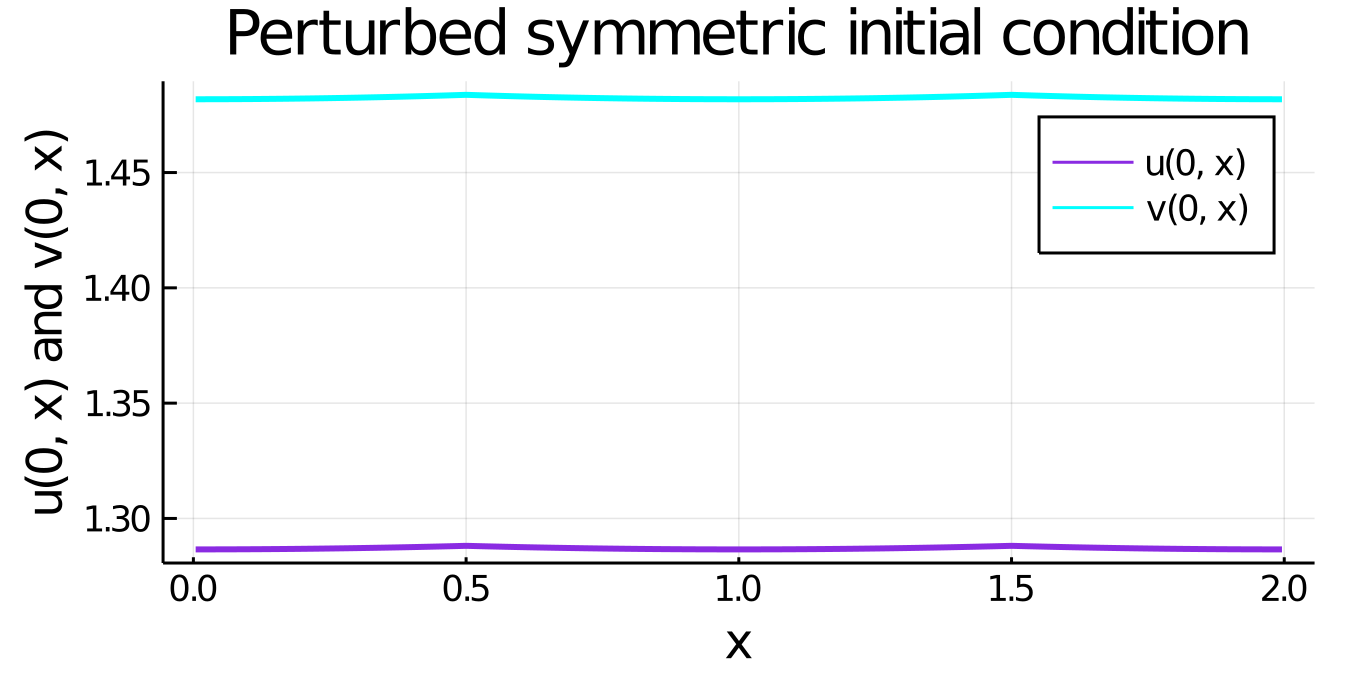}  
		\end{subfigure}
	\end{subfigure}
	\begin{subfigure}[b]{.50\textwidth}
		\begin{subfigure}[b]{1.\textwidth}
    		\centering
			\includegraphics[width=1.\linewidth]{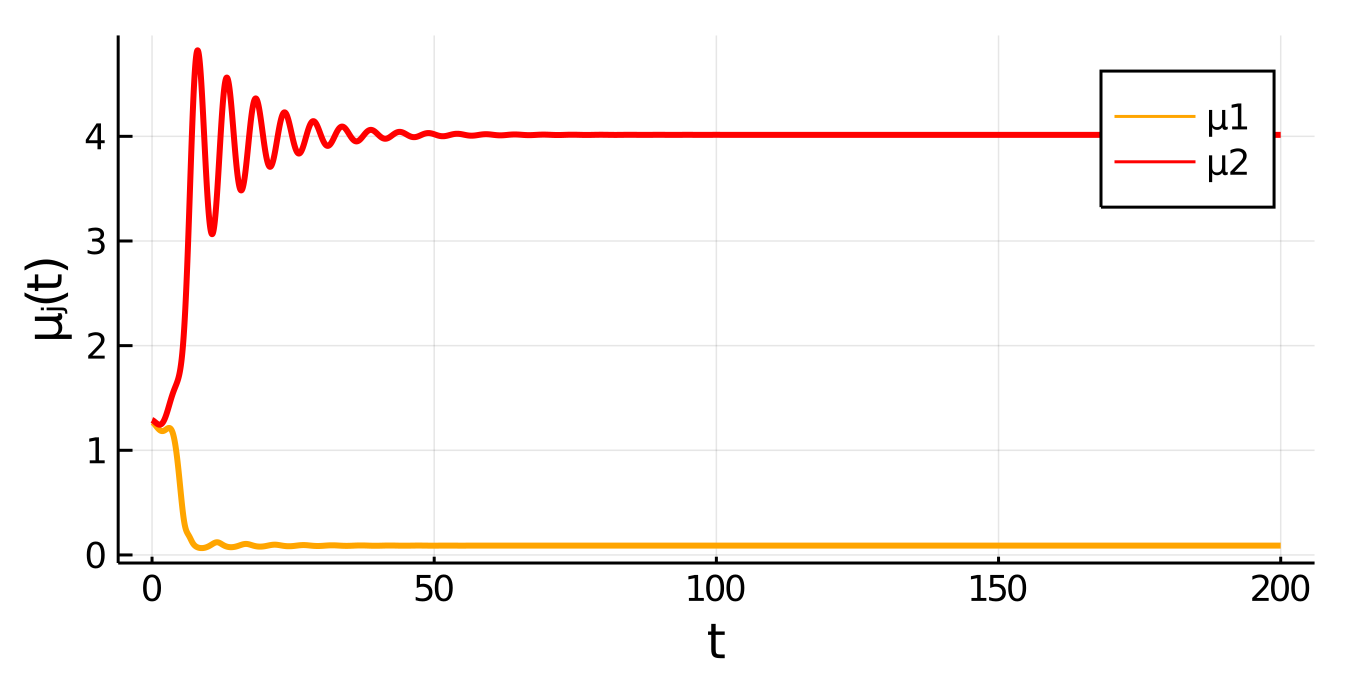}  
		\end{subfigure}
		\begin{subfigure}[b]{1.\textwidth}
			\centering
			\includegraphics[width=1.\linewidth]{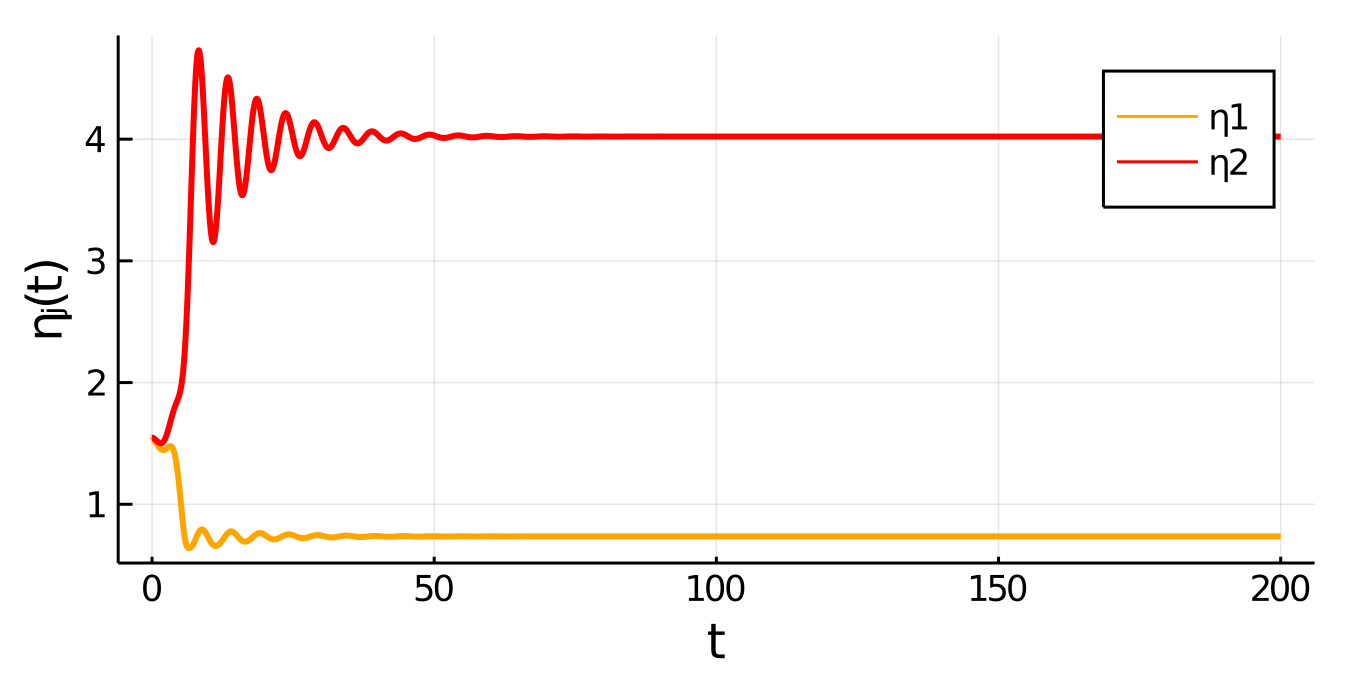}  
		\end{subfigure}
		\begin{subfigure}[b]{1.\textwidth}
			\centering
			\includegraphics[width=1.\linewidth]{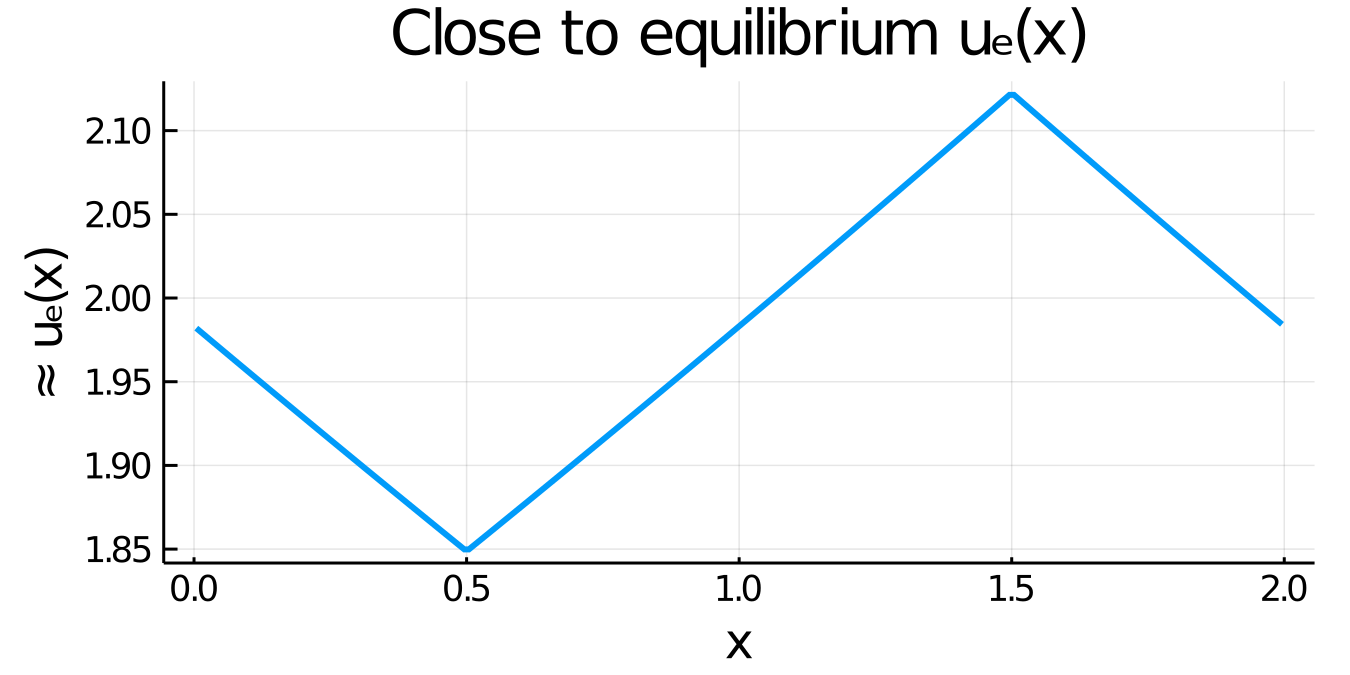}  
		\end{subfigure}
	\end{subfigure}
	\caption{Rauch-Millonas kinetics with $n=2$. For an initial
          condition near the unstable symmetric branch, and for
          $\rho=15$ and $w_v=w_v^{P,2}$, the full time-dependent
          solution computed using the BE-RK4-IMEX scheme of Appendix
          \ref{num:ring} converges to a stable asymmetric
          steady-state. The results are consistent with the prediction
          from the bifurcation diagram in
          Fig.~\ref{fig:pitchforkbubbleandrhobifRMring}(right).  The
          remaining parameter values are as in the caption of
          Fig.~\ref{fig:pitchforkbubbleandrhobifRMring}. Observe that
          $u_e$ is nearly piecewise linear owing to the fact that the
           bulk degradation $\sigma_u$ is very small, i.e.
          $\sigma_u=0.01$. }\label{fig:RMringcell}
\end{figure}

Finally, we briefly show that stable asymmetric patterns can also
occur with Rauch-Millonas kinetics (\ref{ss:rauch}) for $n=2$
compartments when $D_u=D_v=1$. Upon solving (\ref{per2:eq}) with
(\ref{ss:rauch}) using MatCont \cite{matcont}, in the left and right
panels of Fig.~\ref{fig:pitchforkbubbleandrhobifRMring} we plot
bifurcation diagrams of $\mu_{1}^{e}$ versus either $w_v$ or $\rho$,
respectively.  In Fig.~\ref{fig:pitchforkbubbleandrhobifRMring}(left)
we observe that asymmetric steady-states occur in a $w_v$-pitchfork
bubble, while from
Fig.~\ref{fig:pitchforkbubbleandrhobifRMring}(right) we observe  that
asymmetric steady-states emerge from a supercritical $\rho$-pitchfork
bifurcation that opens as $\rho$ increases above a threshold. The
full-numerical simulations shown in Fig.~\ref{fig:RMringcell} for
$\rho=15$ and $w_v=w_v^{P,2}$ are consistent with the prediction of
Fig.~\ref{fig:pitchforkbubbleandrhobifRMring}(right) of stable
asymmetric patterns.

\section{Discussion}\label{sec:disc}

We have formulated and analyzed a new class of 1-D
compartmental-reaction diffusion system in which a two-component
nonlinear reaction kinetics is restricted to the domain boundaries and
where two bulk diffusing species effectively couple the spatially
localized reactions. For the case of a common reaction kinetics, and
when one of the kinetic species has a linear dependence, we have
derived a nonlinear algebraic system characterizing the steady-states
of the model.  Through a bifurcation analysis and from path-following
numerical methods we have shown that symmetry-breaking pitchfork
bifurcations can occur even when the two bulk diffusing species have
identical or very comparable diffusivities. The linear stability
properties of the symmetric steady-state was investigated numerically
by deriving a nonlinear matrix eigenvalue problem characterizing both
in-phase and anti-phase perturbations of a symmetric
steady-state. Through numerical computations of branches of
steady-state solutions together with the numerical solution of the
nonlinear matrix eigenvalue problem, the theory was illustrated for
either FitzHugh-Nagumo, Gierer-Meinhardt, or Rauch-Millonas reaction
kinetics. The theoretical prediction of symmetry-breaking leading to
stable asymmetric patterns in these models was confirmed through full
time-dependent PDE computations. Overall, it was shown that it is the
ratio of the two binding rates to the compartments that controls
whether stable asymmetric patterns can occur. This is in contrast to
the case of a traditional activator-inhibitor RD system where stable
spatial patterns exist only for a sufficiently large, but typically
biologically unrealistic, diffusivity ratio of the two diffusing
species. For FitzHugh-Nagumo and Rauch-Millonas kinetics we have also
shown that stable asymmetric patterns can emerge from a symmetric
pattern at a fixed binding rate ratio when a control parameter in the
intracellular kinetics is varied. Extensions of the analytical theory
to treat a periodic chain of reaction compartments was made, and the
results confirmed through full PDE simulations.

{We remark that our findings are qualitatively similar to those
  associated with a postulated diffusive transport mechanism for
  biological morphogens, referred to as a {\bf binding-mediated
    hindrance diffusion process} \cite{morphogen}, in which a
  differential binding rate ratio on the cell boundaries for two
  morphogen species with comparable diffusivities was suggested to
  provide the large {\bf effective} diffusivity ratio that is required
  for pattern formation and symmetry-breaking in tissues. However,
  since our condition (\ref{ss:odd_FN}) for the existence of
  symmetry-breaking bifurcations is highly implicit in the bulk
  diffusivities, it seems rather intractable analytically to isolate
  via a simple scaling analysis an {\bf effective diffusivity}
  bifurcation parameter for the bulk species that incorporates the
  membrane binding rate ratio and/or the ratio of bulk degradation
  rates. An effective diffusivity ratio condition was also derived
  analytically in \cite{korv} for a $2+1$ RD system with an immobile
  substrate. It is an open problem to isolate an effective diffusivity
  ratio parameter for our compartmental-reaction diffusion system that
  encapsulates analytically how the symmetry-breaking bifurcation point
  depends on the membrane binding rate ratio and the ratio of bulk
  degradation rates.}

In the context of our 1-D framework, there are four main additional open
problems. {Firstly, it would be worthwhile to formulate a linear
stability problem for asymmetric patterns of the 1-D
compartmental-reaction diffusion system, while allowing for arbitrary
reaction kinetics.} Secondly, it would be interesting to extend the
weakly nonlinear analysis of \cite{paquin_1d}, which dealt with
one-bulk diffusing species, to our compartmental-reaction diffusion
setting.  With this extension, amplitude equations near the
symmetry-breaking bifurcation would be valuable so as to classify
whether the bifurcation is sub- or super-critical, and thereby predict
whether the emerging asymmetric patterns are linearly stable or
unstable. Finally, it would be interesting to analytically represent
the time-dependent bulk diffusion fields in terms of a time-dependent
Green function so as to reduce the bulk-cell system to an
integro-differential system of ODEs with memory. In this setting,
large amplitude solutions, including time-periodic solutions, would
represent a new frontier to be investigated. A further extension of
our 1-D model would be to investigate how the symmetry-breaking
bifurcations are influenced by random noise arising from the
compartmental reaction kinetics. In \cite{goldstein} it was shown that
including stochasticity in RD systems can lower the
threshold for the onset of pattern formation (see \cite{erban} for an
overview of stochastic modeling in RD systems).

An important open direction is to analyze symmetry-breaking behavior
associated with a similar modeling framework in a 2-D spatial setting
where intra-compartmental reactions occur only within a disjoint
collection of small circular disks within a bounded 2-D domain. In
this context, two bulk diffusing species with comparable diffusivities
can effectively be transported across the compartment boundaries as
regulated by certain binding rates. This exchange between the
compartments and the bulk medium can either trigger or modulate
the intra-compartmental reactions. Similar bulk-cell models in 2-D, but
with only one diffusing bulk species, have been analyzed recently to model
quorum-sensing behavior (cf.~\cite{gou2d}, \cite{smjw_diff},
\cite{ridgway}, \cite{q_survey}).

Another possible extension of our modeling framework is to consider a
compartmental-reaction diffusion system on a network. In this
graph-theoretic setting, the reaction kinetics are restricted to the
nodes of the graph, while two bulk species are taken to diffuse along
the edges. Related network models of this type have been analyzed in
\cite{macron}, \cite{macron1} and \cite{landge} for general
multi-component systems, and in \cite{faye} for a spatially-extended
SIR epidemic model.

\begin{appendix}
    \section{CN-RK4 IMEX method on a non-centered grid} \label{CN-RK4 IMEX}
    
    In this appendix we provide details on the numerical method used to
    solve the 2-cell system of \S \ref{2-cell system}. For illustration,
    we focus on the FN kinetics as in \S \ref{2-cell FN system}. Let
    $M \in \bN\backslash\{0,1\}, N\in\bN\backslash\{0,1\}$, and let
    $\Delta x := \frac{L}{N-1}$ and $\Delta t:= \frac{t_f}{M-1}$ be,
    respectively, the spatial step and the time step for some final
    time $t_f$. Let then $\{x_n\}_n$ with $x_n:=(n-1)\Delta x$,
    $n\in\{1,...,N\}$, be the spatial grid and $\{t_m\}_m$ with
    $t_m:=(m-1)\Delta t$, $m\in\{1,...,M\}$, be the temporal grid.
    
    \begin{center}
	    \def\svgwidth{0.7\textwidth}
        %% Creator: Inkscape inkscape 0.92.3, www.inkscape.org
%% PDF/EPS/PS + LaTeX output extension by Johan Engelen, 2010
%% Accompanies image file 'noncentredgrid.pdf' (pdf, eps, ps)
%%
%% To include the image in your LaTeX document, write
%%   \input{<filename>.pdf_tex}
%%  instead of
%%   \includegraphics{<filename>.pdf}
%% To scale the image, write
%%   \def\svgwidth{<desired width>}
%%   \input{<filename>.pdf_tex}
%%  instead of
%%   \includegraphics[width=<desired width>]{<filename>.pdf}
%%
%% Images with a different path to the parent latex file can
%% be accessed with the `import' package (which may need to be
%% installed) using
%%   \usepackage{import}
%% in the preamble, and then including the image with
%%   \import{<path to file>}{<filename>.pdf_tex}
%% Alternatively, one can specify
%%   \graphicspath{{<path to file>/}}
%% 
%% For more information, please see info/svg-inkscape on CTAN:
%%   http://tug.ctan.org/tex-archive/info/svg-inkscape
%%
\begingroup%
  \makeatletter%
  \providecommand\color[2][]{%
    \errmessage{(Inkscape) Color is used for the text in Inkscape, but the package 'color.sty' is not loaded}%
    \renewcommand\color[2][]{}%
  }%
  \providecommand\transparent[1]{%
    \errmessage{(Inkscape) Transparency is used (non-zero) for the text in Inkscape, but the package 'transparent.sty' is not loaded}%
    \renewcommand\transparent[1]{}%
  }%
  \providecommand\rotatebox[2]{#2}%
  \newcommand*\fsize{\dimexpr\f@size pt\relax}%
  \newcommand*\lineheight[1]{\fontsize{\fsize}{#1\fsize}\selectfont}%
  \ifx\svgwidth\undefined%
    \setlength{\unitlength}{326.48030064bp}%
    \ifx\svgscale\undefined%
      \relax%
    \else%
      \setlength{\unitlength}{\unitlength * \real{\svgscale}}%
    \fi%
  \else%
    \setlength{\unitlength}{\svgwidth}%
  \fi%
  \global\let\svgwidth\undefined%
  \global\let\svgscale\undefined%
  \makeatother%
  \begin{picture}(1,0.15481655)%
    \lineheight{1}%
    \setlength\tabcolsep{0pt}%
    \put(0,0){\includegraphics[width=\unitlength,page=1]{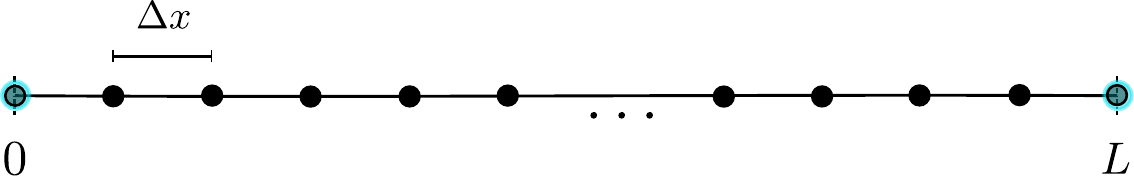}}%
  \end{picture}%
\endgroup%

        %\caption{}
    \end{center}
    
    We discretize the 1-D Laplacian by the centered scheme
    \begin{equation*}
      \partial_{xx} u(t,x_n) =  \frac{u(t,x_{n+1})-2u(t,x_n)+
        u(t,x_{n-1})}{(\Delta x)^2} + \O(\Delta x^2)\,,
    \end{equation*}
    for all points away from the boundary, i.e., for
    $n\in\{2,3,...,N-1\}$. For $n=1$, use the Robin boundary
    conditions of our system, so that
    \begin{equation*}
      \partial_{xx} u(t,0) = \frac{\frac{u(t,\Delta x)-u(t,0)}{\Delta x} -
        \partial_x u(t,0)}{\frac{1}{2}\Delta x} + \O(\Delta x) =
      \frac{u(t,\Delta x)-u(t,0)}{\frac{1}{2} (\Delta x)^2} -
      \frac{\beta_u}{D_u} \frac{u(t,0)-\mu_1(t)}{\frac{1}{2}\Delta x} +
      \O(\Delta x)\,.
    \end{equation*}
    Similarly, for $x=N$, we have
    \begin{equation*}
        \partial_{xx} u(t,L) = \frac{\partial_x u(t,L) -
        \frac{u(t,L)-u(t,L-\Delta x)}{\Delta x}}{\frac{1}{2}\Delta x} +
      \O(\Delta x) = \frac{\beta_u}{D_u} \frac{\mu_2(t)-u(L,t)}
      {\frac{1}{2}\Delta x} -
      \frac{u(t,L)-u(t,L-\Delta x)}{\frac{1}{2}(\Delta x)^2} + \O(\Delta x)\,.
    \end{equation*}
    This yields the discretization matrix defined by
    \begin{equation*}
        L_{\Delta x}^u :=
        \begin{pmatrix}
            -2 - 2\frac{\beta_u}{D_u} \Delta x &2         &0         &...         &       &0 \\
            1                                  &-2        &1         &0           &...    &0 \\
                                               &\ddots    &\ddots    &\ddots      &       &  \\
                                               &          &1         &-2          &1      & \\
                                               &          &          &\ddots      &\ddots &\ddots \\
            0                                  &...       &          &0           &2      &-2 - \frac{\beta_u}{D_u}\Delta x
        \end{pmatrix}\,,
    \end{equation*}
    so that to $\O(\Delta x)$ we have
    \begin{equation*}
        \frac{d}{dt} 
        \begin{pmatrix}
            u(t,x_1)  \\
            u(t,x_2)  \\
            \vdots \\
            u(t,x_{N-1}) \\
            u(t,x_N)
        \end{pmatrix}
        = \frac{D_u}{(\Delta x)^2} L_{\Delta x}^u
        \begin{pmatrix}
            u(t,x_1)  \\
            u(t,x_2)  \\
            \vdots \\
            u(t,x_{N-1}) \\
            u(t,x_N)
        \end{pmatrix}
        + \frac{1}{\Delta x}
        \begin{pmatrix}
            2\beta_u\mu_1(t) \\
            0  \\
            \vdots  \\
            0  \\
            2\beta_u\mu_2(t)
        \end{pmatrix}
        - \sigma_u
        \begin{pmatrix}
            u(t,x_1)  \\
            u(t,x_2)  \\
            \vdots \\
            u(t,x_{N-1}) \\
            u(t,x_N)
        \end{pmatrix}\,.
    \end{equation*}
The discretization matrix $L_{\Delta x}^v$ can be obtained analogously from
the equations for $v$.\\

Then, upon including the intracellular FN kinetics of (\ref{s:FN}), we
obtain the lumped system
\begin{equation}
  \frac{d}{dt} w = \cL w + \cN[w] + \vartheta\,,
\end{equation}
where
\begin{equation*}
       w :=
        \begin{pmatrix}
            u(t,x_1)  \\
            \vdots \\
            u(t,x_N)  \\
            v(t,x_1)  \\
            \vdots \\
            v(t,x_N) \\
            \mu_1(t) \\
            \eta_1(t) \\
            \mu_2(t) \\
            \eta_2(t)
        \end{pmatrix} \qquad
\cN[w] :=  \begin{pmatrix}
            0 \\
            \vdots \\
            0 \\
            0 \\
            \vdots \\
            0 \\
            -q(\mu_1-2)^3 \\
            0 \\
            -q(\mu_2-2)^3 \\
            0
        \end{pmatrix}
        \,, \qquad
 \vartheta := \begin{pmatrix}
            0 \\
            \vdots \\
            0 \\
            0 \\
            \vdots \\
            0 \\
            4 \\
            0 \\
            4 \\
            0
        \end{pmatrix}\,,
\end{equation*}
and where the linear matrix operator $\cL$ is defined as
\begin{equation*}
 \cL :=
        \begin{pmatrix}
                &                                                   & && &&2\frac{\beta_u}{\Delta x} &0       &0  &0       \\
                & \frac{D_u}{(\Delta x)^2} L_{\Delta x}^u - \sigma_u I&  && &&                          &\vdots  &  &\vdots  \\
                &                                                   & && &&0 &0       &2\frac{\beta_u}{\Delta x}  &0       \\
                & &                                                  & && &0 &2\frac{\beta_v}{\Delta x}       &0  &0       \\
                &&  &&\frac{D_v}{(\Delta x)^2} L_{\Delta x}^v - \sigma_v I& &\vdots                     &  &\vdots  &        \\
                &&                                                    && &&0           &0       &0  &2\frac{\beta_v}{\Delta x}       \\
                \beta_u & &0 &0  &  &0                                &1-\beta_u   &-1      &0  &0 \\
                0 & &0 &\beta_v  &  &0                                &\varepsilon z   &-\varepsilon-\beta_v      &0  &0 \\
                0& &\beta_u &0  &  &0                                 &0   &0      &1-\beta_u  &-1 \\
                0 & &0 &0  &  &\beta_v                                &0   &0      &\varepsilon z  &-\varepsilon - \beta_v \\
              \end{pmatrix}\,.
\end{equation*}

As inspired by the operator-splitting approach of \cite{cross}, 
we consider the split equations
\begin{equation*}
        \frac{d}{dt} w_L = \cL w + \vartheta\,, \qquad
        \frac{d}{dt} w_N = \cN[w] \,,
\end{equation*}
for a \emph{single} time step of length $\Delta t$, and note that
$w(t) = w_L(t)+w_N(t)$. The affine equation can easily be integrated
implicitly with the Crank-Nicholson scheme (CN) yielding an
$\O(\Delta t^2)$-error and the nonlinear equation can be approximated
with the classical Runge-Kutta scheme (RK4) yielding an
$\O(\Delta t^5)$ error. All combined, this gives the scheme for the
full numerical solution at time step $m+1$. Denoting the numerical
solution at time step $t_m$ by $w_m$, we have
    \begin{equation*}
        \begin{array}{lrcl}
          & \frac{w_{m+1} - w_m}{\Delta t} &=& \frac{1}{2} (\cL w_{m+1} +
 \cL w_m) + \frac{1}{6} (k_{1,m} + 2 k_{2,m} + 2 k_{3,m} + k_{4,m}) + \vartheta \\
          \Leftrightarrow & w_{m+1} &=& (I-\frac{\Delta t}{2} \cL)^{-1}
     ((I+\frac{\Delta t}{2} \cL) w_m + \frac{\Delta t}{6}
      (k_{1,m} + 2 k_{2,m} + 2 k_{3,m} + k_{4,m}) + \Delta t \vartheta)
        \end{array}
    \end{equation*}
    where the weights for RK4 are given by $k_{1,m} := \cN[w_m]$,
    $k_{2,m} := \cN[w_m + \frac{\Delta t}{2} k_{1,m}]$,
    $k_{3,m} := \cN[w_m + \frac{\Delta t}{2} k_{2,m}]$, and
    $k_{4,m} := \cN[w_m + \Delta t k_{3,m}]$.
    
\section{BE-RK4 IMEX method on a centered grid}\label{num:ring}
    
In this appendix we provide details on the numerical method used to
solve the ring-cell system of \S \ref{ring-cell system}. For
illustration, we again use the FN kinetics (\ref{s:FN}) as in \S
\ref{ring-cell FN system}.

Let $N\in\bN\backslash\{0,1\}$ be even and denote the number of grid
points per fundamental domain with centered cell. Let
$M \in \bN\backslash\{0,1\}$ and label $\Delta x := \frac{L}{N}$ and
$\Delta t:= \frac{t_f}{M-1}$, respectively, as the spatial step and
the time step for some final time $t_f$. We let $\{x_k\}_k$ with
$x_k:=(k-1)\Delta x$, $k\in\{1,...,N\}$, be the spatial grid while
$\{t_m\}_m$ with $t_m:=(m-1)\Delta t$, $m\in\{1,...,M\}$ is the
temporal grid. We note that $x_{nN}= nL - \frac{\Delta x}{2}$ and that
the $j$-th cell is centered exactly in between $x_{\frac{2j-1}{2}N}$
and $x_{\frac{2j-1}{2}N+1}$ for $j\in\{1,...,n\}$.
    
We discretize the 1-D Laplacian by the centered scheme
\begin{equation*}
      \Delta u(t,x_k) =  \frac{u(t,x_{k+1})-2u(t,x_k)+u(t,x_{k-1})}
      {(\Delta x)^2} + \O(\Delta x^2) \,,
\end{equation*}
for all points away from the cells, i.e., for
$k\in\{1,...,nN\}\backslash\bigcup\{\frac{2j-1}{2}N, \frac{2j-1}{2}N+1\}$.
\begin{center}
	    \def\svgwidth{0.7\textwidth}
        %% Creator: Inkscape inkscape 0.92.3, www.inkscape.org
%% PDF/EPS/PS + LaTeX output extension by Johan Engelen, 2010
%% Accompanies image file 'centredgrid.pdf' (pdf, eps, ps)
%%
%% To include the image in your LaTeX document, write
%%   \input{<filename>.pdf_tex}
%%  instead of
%%   \includegraphics{<filename>.pdf}
%% To scale the image, write
%%   \def\svgwidth{<desired width>}
%%   \input{<filename>.pdf_tex}
%%  instead of
%%   \includegraphics[width=<desired width>]{<filename>.pdf}
%%
%% Images with a different path to the parent latex file can
%% be accessed with the `import' package (which may need to be
%% installed) using
%%   \usepackage{import}
%% in the preamble, and then including the image with
%%   \import{<path to file>}{<filename>.pdf_tex}
%% Alternatively, one can specify
%%   \graphicspath{{<path to file>/}}
%% 
%% For more information, please see info/svg-inkscape on CTAN:
%%   http://tug.ctan.org/tex-archive/info/svg-inkscape
%%
\begingroup%
  \makeatletter%
  \providecommand\color[2][]{%
    \errmessage{(Inkscape) Color is used for the text in Inkscape, but the package 'color.sty' is not loaded}%
    \renewcommand\color[2][]{}%
  }%
  \providecommand\transparent[1]{%
    \errmessage{(Inkscape) Transparency is used (non-zero) for the text in Inkscape, but the package 'transparent.sty' is not loaded}%
    \renewcommand\transparent[1]{}%
  }%
  \providecommand\rotatebox[2]{#2}%
  \newcommand*\fsize{\dimexpr\f@size pt\relax}%
  \newcommand*\lineheight[1]{\fontsize{\fsize}{#1\fsize}\selectfont}%
  \ifx\svgwidth\undefined%
    \setlength{\unitlength}{327.97666376bp}%
    \ifx\svgscale\undefined%
      \relax%
    \else%
      \setlength{\unitlength}{\unitlength * \real{\svgscale}}%
    \fi%
  \else%
    \setlength{\unitlength}{\svgwidth}%
  \fi%
  \global\let\svgwidth\undefined%
  \global\let\svgscale\undefined%
  \makeatother%
  \begin{picture}(1,0.14424085)%
    \lineheight{1}%
    \setlength\tabcolsep{0pt}%
    \put(0,0){\includegraphics[width=\unitlength,page=1]{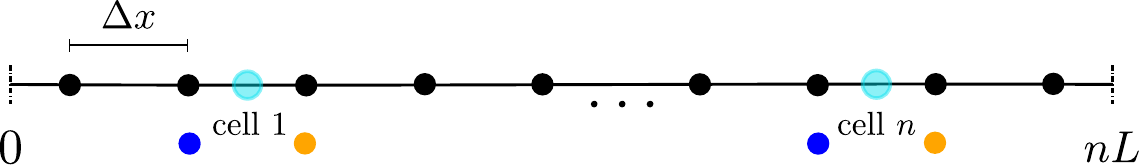}}%
  \end{picture}%
\endgroup%

        %\caption{}
\end{center} 
For $k_j:=\frac{2j-1}{2}N$, which labels the index of the grid point to the left
of the $j$-th cell, we discretize the 1-D Laplacian by
    \begin{align*}
      \partial_{xx} u(t,x_{k_j}) &= \frac{\orange{\tilde{u}_{k_j+1}(t)} -
     2u(t,x_{k_j}) + u(t,x_{k_j-1})}{(\Delta x)^2} + \O((\Delta x)^2)\,, \\
      \partial_{xx}u(t,x_{k_j+1}) &= \frac{u(t,x_{k_j+2}) - 2u(t,x_{k_j+1}) +
       \blue{\tilde{u}_{k_j}(t)}}{(\Delta x)^2} + \O((\Delta x)^2)\,,
    \end{align*}
where the coloured points are values at the ghost points depicted in the
schematic figure above and are obtained from
    \begin{align*}
      \frac{u(t,x_{k_j+1}) + \blue{\tilde{u}_{k_j}(t)}}{2} &=
   \frac{\orange{\tilde{u}_{k_j+1}(t)} + u(t,x_{k_j})}{2}\,, \quad
        \text{(continuity condition)} \\
      \frac{u(t,x_{k_j+1}) - \blue{\tilde{u}_{k_j}(t)}}{\Delta x} -
      \frac{\orange{\tilde{u}_{k_j+1}(t)} - u(t,x_{k_j})}{\Delta x} &=
       \frac{\beta_u}{D_u} \left(\frac{u(t,x_{k_j+1}) + u(t,x_{k_j})}{2} -
       \mu_j(t)\right)\,, \quad \text{(jump condition)} \,.
    \end{align*}
    Upon solving for the ghost point values we get
    \begin{align*}
      \blue{\tilde{u}_{k_j}(t)} &= \frac{1}{2} (2-\frac{\beta_u \Delta x}{2D_u})
   u(t,x_{k_j}) + \frac{\beta_u \Delta x}{2D_u} \mu_j(t) -
       \frac{\beta_u\Delta x}{4 D_u} u(t,x_{k_j+1})\,, \\
      \orange{\tilde{u}_{k_j+1}(t)} &= -\frac{\beta_u \Delta x}{4D_u}
    u(t,x_{k_j}) + \frac{\beta_u \Delta x}{2D_u} \mu_j(t) +
   \frac{1}{2} (2- \frac{\beta_u\Delta x}{2 D_u}) u(t,x_{k_j+1})\,.
    \end{align*}
    This yields the discretization matrix defined by
    \begin{equation*}
        L_{\Delta x}^u :=
        \begin{pmatrix}
            -2 &1         &0         &...         &0      &\quad &1 \\
            1  &-2        &1         &0           &...      &\quad &0 \\
               &    &\ddots    &      &           &\quad         &  \\
               &          &1         &-2-\frac{\beta_u\Delta x}{4D_u}          &1-\frac{\beta_u\Delta x}{4 D_u}      &\quad &     \\
               &          &1-\frac{\beta_u\Delta x}{4 D_u}          &-2-\frac{\beta_u\Delta x}{4D_u}    &1    & & \\
               &          &          &    &\ddots      &\quad &  \\
            1  &0       &...          &\quad           &0    &-2       &1
        \end{pmatrix}
    \end{equation*}
    satisfying, to $\O(\Delta x)$,
    \begin{equation*}
        \frac{d}{dt} 
        \begin{pmatrix}
            u(t,x_1)  \\
            u(t,x_2)  \\
            \vdots \\
            u(t,x_{k_j}) \\
            u(t,x_{k_j+1}) \\
            \vdots\\
            u(t,x_{N-1}) \\
            u(t,x_N)
        \end{pmatrix}
        = \frac{D_u}{(\Delta x)^2} L_{\Delta x}^u
        \begin{pmatrix}
            u(t,x_1)  \\
            u(t,x_2)  \\
            \vdots \\
            u(t,x_{k_j}) \\
            u(t,x_{k_j+1}) \\
            \vdots \\
            u(t,x_{N-1}) \\
            u(t,x_N)
        \end{pmatrix}
        + \frac{1}{\Delta x}
        \begin{pmatrix}
            0 \\
            \vdots  \\
            0  \\
            \frac{\beta_u}{2}\mu_j(t)  \\
            \frac{\beta_u}{2}\mu_j(t)  \\
            \\
            0
        \end{pmatrix}
        - \sigma_u
        \begin{pmatrix}
            u(t,x_1)  \\
            u(t,x_2)  \\
            \vdots \\
            u(t,x_{k_j}) \\
            u(t,x_{k_j+1}) \\
            \vdots \\
            u(t,x_{N-1}) \\
            u(t,x_N)
        \end{pmatrix}.
\end{equation*}
      
Note the abuse of notation for the entries indexed by $j$. Read here
that such entries are in $L_{\Delta x}^u$ for each
$j\in\{1,...,n\}$. The discretization matrix $L_{\Delta x}^v$ can be
obtained analogously from the equations for $v$.

Next, the intracellular FN reaction kinetics (\ref{s:FN}) are approximated by
    \begin{align*}
      \frac{d}{dt} \mu_j(t) &= (1-\beta_u) \mu_j(t) - q(\mu_j(t)-2)^3 + 4 -
    \eta_j(t) + \beta_u \left(\frac{u(t,x_{k_j}) + u(t,x_{k_j+1})}{2}\right)\,, \\
      \frac{d}{dt} \eta_j(t) &= \varepsilon z \mu_j(t) +
     (-\varepsilon - \beta_v) \eta_j(t) + \beta_v \left(
      \frac{v(t,x_{k_j}) + v(t,x_{k_j+1})}{2}\right)\,.
    \end{align*}

Then, upon including the intracellular FN kinetics, we
obtain the lumped system
\begin{equation}
  \frac{d}{dt} w = \cL w + \cN[w] + \vartheta\,,
\end{equation}
where now we define
\begin{equation*}
  w := \begin{pmatrix}
            u(t,x_1)  \\
            \vdots \\
            u(t,x_{k_j}) \\
            u(t,x_{k_j+1}) \\
            \vdots \\
            u(t,x_N)  \\
            v(t,x_1)  \\
            \vdots \\
            v(t,x_{k_j}) \\
            v(t,x_{k_j+1}) \\
            \vdots \\
            v(t,x_N) \\
            \mu_j(t) \\
            \eta_j(t) \\
          \end{pmatrix}  \,, \qquad
 \cN[w] :=  \begin{pmatrix}
            0  \\
            \vdots \\
            0 \\
            0 \\
            \vdots \\
            0  \\
            0  \\
            \vdots \\
            0 \\
            0 \\
            \vdots \\
            0 \\
            -q(\mu_j-2)^3 \\
            0
        \end{pmatrix}\,, \qquad
\vartheta :=
        \begin{pmatrix}
            0  \\
            \vdots \\
            0 \\
            0 \\
            \vdots \\
            0  \\
            0  \\
            \vdots \\
            0 \\
            0 \\
            \vdots \\
            0 \\
            4 \\
            0
        \end{pmatrix}
        \,,
\end{equation*}
and where the linear matrix operator $\cL$ is now defined by
\begin{equation*}
      \cL :=
        \begin{pmatrix}
                & & & & & & & & & & &                                                  &0  &0   \\
                & & & & & & & & & & &                                                  &   &\vdots   \\
                & &\frac{D_u}{(\Delta x)^2} L_{\Delta x}^u - \sigma_u I & & & & & &0 & & &  &\frac{\beta_u}{2\Delta x}  &0   \\
                & & & & & & & & & & &                                                  &\frac{\beta_u}{2\Delta x}  &0   \\
                & & & & & & & & & & &                                                  &   &\vdots   \\
                & & & & & & & & & & &                                                  &0  &0   \\
                & & & & & & & & & & &                                                  &0  &0   \\
                & & & & & & & & & & &                                                  &   &\vdots   \\
                & &0 & & & & & &\frac{D_v}{(\Delta x)^2} L_{\Delta x}^v - \sigma_v I & & &   &  &\frac{\beta_v}{2\Delta x}   \\
                & & & & & & & & & & &                                                      &0 &\frac{\beta_v}{2\Delta x}   \\
                & & & & & & & & & & &                                                  &   &\vdots   \\
                & & & & & & & & & & &                                                  &0  &0   \\
                0 & &\frac{\beta_u}{2} &\frac{\beta_u}{2}  &  &0 &0   & &                  &                  & &0 &1-\beta_u &-1 \\
                0 & &                  &                   &  &0 &0   & &\frac{\beta_v}{2} &\frac{\beta_v}{2} & &0 &\varepsilon z  &-\varepsilon - \beta_v \\
        \end{pmatrix}\,.
\end{equation*}
Again note that there is an abuse of notation for the $j$-indexed rows,
including the reaction kinetics rows, as it refers to all $j\in\{1,...,n\}$.

With the operator-splitting approach inspired by \cite{cross}, we again
consider the split equations
\begin{equation*}
        \frac{d}{dt} w_L = \cL w + \vartheta\,, \qquad
        \frac{d}{dt} w_N = \cN[w] 
\end{equation*}
for a \emph{single} time step of length $\Delta t$, and note that
$w(t) = w_L(t)+w_N(t)$. The affine equation is approximated
implicitly with the Backward-Euler scheme (BE) yielding an
$\O(\Delta t)$-error, while the nonlinear equation is integrated
explicitly using the classical Runge-Kutta scheme (RK4) yielding an
$\O(\Delta t^5)$ error. All combined, this gives the scheme for the
full numerical solution at time step $m+1$. Denoting the numerical
solution at time step $t_m$ by $w_m$, we have
\begin{equation*}
        \begin{array}{lrcl}
          & \frac{w_{m+1} - w_m}{\Delta t} &=& \cL w_{m+1} + \frac{1}{6}
          \left(k_{1,m} + 2 k_{2,m} + 2 k_{3,m} + k_{4,m}\right) + \vartheta \\
          \Leftrightarrow & w_{m+1} &=& (I-\Delta t \cL)^{-1}
   \left(w_m + \frac{\Delta t}{6} (k_{1,m} + 2 k_{2,m} + 2 k_{3,m} + k_{4,m}
                                        \right) + \Delta t \vartheta)\,,
        \end{array}
\end{equation*}
where the weights for RK4 are given by $k_{1,m} := \cN[w_m]$,
$k_{2,m} := \cN[w_m + \frac{\Delta t}{2} k_{1,m}]$,
$k_{3,m} := \cN[w_m + \frac{\Delta t}{2} k_{2,m}]$, and
$k_{4,m} := \cN[w_m + \Delta t k_{3,m}]$.

We remark that this scheme is of order $\O((\Delta x)^2)$ in contrast to
the order $\O(\Delta x)$ scheme of Appendix \ref{CN-RK4 IMEX}. This is due
to using a centered grid with two ghost points per cell. Also
note that we have used the implicit $\O(\Delta t)$-scheme BE
instead of the implicit $\O((\Delta t)^2)$-scheme CN since the
latter can possibly generate numerical instabilities. As a result,
the scheme presented in this appendix is more somewhat robust than
that given in Appendix \ref{CN-RK4 IMEX}.
\end{appendix}
%\enlargethispage{20pt}
\vspace{0.5cm}

\begin{flushleft}
\textbf{Authors' Contributions:} The authors contributed equally to this article. \\

\vspace{0.2cm}

\textbf{Competing Interests:} There are no competing interests.\\

\vspace{0.2cm}

\textbf{Funding:} Merlin Pelz was supported by a Four-Year-Graduate-Fellowship from UBC. M. J. Ward was supported by the NSERC Discovery Grant Program. \\

\vspace{0.2cm}

\textbf{Acknowledgements:} The authors are thankful to Prof. Brian Wetton for his help with introducing ghost points for the BE-RK4 IMEX numerical scheme.
\end{flushleft}

\vspace{0.5cm}

%%%%%%%%%% Insert bibliography here %%%%%%%%%%%%%%

\vskip2pc

\bibliographystyle{plain}

\vspace*{-1.1cm}
\bibliography{1Dcoupledsingcells}

\end{document}